\documentclass[12pt]{article}
\usepackage{amsmath}

\usepackage{amssymb}
\usepackage{mathrsfs}
\usepackage{graphicx}
\usepackage{mathtools}
\usepackage[polutonikogreek,english]{babel}
\usepackage[linktocpage]{hyperref}
\usepackage{comment}

\DeclareFontFamily{U}{mathx}{}
\DeclareFontShape{U}{mathx}{m}{n}{<-> mathx10}{}
\DeclareSymbolFont{mathx}{U}{mathx}{m}{n}
\DeclareMathAccent{\widehat}{0}{mathx}{"70}
\DeclareMathAccent{\widecheck}{0}{mathx}{"71}

\renewcommand{\d}{\mathrm{d}}
\newcommand{\rl}{\ell}
\newcommand{\tilS}{\tilde{S}}
\newcommand{\bN}{\mbox{\boldmath$N$}}
\newcommand{\on}{{\overline{n}}}
\newcommand{\bp}{{\mbox{\boldmath$p$}}}
\renewcommand{\Im}{\mathop{\rm Im}\nolimits}
\newcommand{\oDelta}{{\overline{\Delta}}}
\newcommand{\bv}{\mbox{\boldmath$v$}}
\newcommand{\bl}{\mbox{\boldmath$l$}}
\newcommand{\pR}{\,^{+}\!R}
\newcommand{\pOmega}{\,^+\!\Omega}
\newcommand{\mOmega}{\,^-\!\Omega}
\newcommand{\pmOmega}{\,^\pm\!\Omega}
\newcommand{\pmR}{\,^{\pm}\!R}
\newcommand{\overOm}{\overline{\Omega}}
\newcommand{\overT}{\overline{T}}
\newcommand{\bomega}{\mbox{\boldmath$\omega$}}
\newcommand{\be}{\mbox{\boldmath$e$}}
\newcommand{\btau}{\mbox{\boldmath$\tau$}}
\newcommand{\tbomega}{{\widetilde{\bomega}}}
\renewcommand{\Re}{\mathop{\rm Re}\nolimits}
\newcommand{\D}{{\cal D}}
\newcommand{\cS}{{\widecheck{S}}}
\newcommand{\N}{{\cal N}}
\newcommand{\rv}{\mbox{\rm v}}
\newcommand{\sh}{\mathop{\rm sh}\nolimits}
\newcommand{\rnu}{{\rm \textgreek{n}}}
\newcommand{\rlambda}{{\rm \textgreek{l}}}
\newcommand{\ralpha}{{\rm \textgreek{a}}}
\newcommand{\cG}{{\widecheck{G}}}
\newcommand{\tg}{{\widetilde g}}
\newcommand{\bDelta}{\mbox{\boldmath$\Delta$}}
\newcommand{\ccM}{{\widecheck{\cal M}}}
\newcommand{\cmf}{{\widecheck {\mathfrak f}}}
\newcommand{\mm}{{\mathfrak m}}
\newcommand{\ccF}{{\widecheck{\cal F}}}
\newcommand{\crO}{{\widecheck{\mathrm O}}}
\newcommand{\cmM}{{\widecheck {\mathfrak M}}}
\newcommand{\cF}{{\cal F}}
\newcommand{\mf}{{\mathfrak f}}
\newcommand{\mM}{{\mathfrak M}}
\newcommand{\rO}{{\mathrm O}}
\newcommand{\cPhi}{{\widecheck \Phi}}
\newcommand{\ccO}{{\widecheck{\cal O}}}
\newcommand{\Det}{\mathop{\mathrm{Det}}}
\newcommand{\cO}{{\cal O}}
\newcommand{\ialpha}{{\mbox{\it\textgreek{a}}}}
\newcommand{\rmm}{\mbox{\rm m}}
\newcommand{\rgamma}{{\rm \textgreek{g}}}
\newcommand{\halpha}{{\widehat{\alpha}}}
\newcommand{\no}{<}
\newcommand{\hbeta}{{\widehat{\beta}}}
\newcommand{\cA}{{\cal A}}
\newcommand{\tdelta}{{\widetilde{\delta}}}
\newcommand{\trmm}{{\widetilde{\rmm}}}
\newcommand{\tk}{{\widetilde{k}}}
\newcommand{\tw}{{\widetilde w}}
\newcommand{\hone}{{\widehat{1}}}
\newcommand{\htwo}{{\widehat{2}}}
\newcommand{\hthr}{{\widehat{3}}}
\newcommand{\tp}{{\widetilde{p}}}
\newcommand{\bq}{\mbox{\boldmath$q$}}
\newcommand{\Tr}{\mathop{\mathrm{Tr}}}
\newcommand{\cL}{{\cal L}}
\newcommand{\hS}{{\widehat{S}}}

\allowdisplaybreaks[4]
\begin{document}

\title{Discrete gravitational diagram technique and corrections to the Newtonian potential}

\author{V.M. Khatsymovsky \\
 {\em Budker Institute of Nuclear Physics} \\ {\em of Siberian Branch Russian Academy of Sciences} \\ {\em
 Novosibirsk,
 630090,
 Russia}
\\ {\em E-mail address: khatsym@gmail.com}}
\date{}
\maketitle

\begin{abstract}
To develop a gravitational diagram technique that is ultraviolet finite, we start from Regge's simplicial gravity and use a bell-shaped form of the measure obtained via functional integration over connection. A "hypercubic" structure is considered (some variables are frozen), it is described by the metric $g_{\lambda \mu}$ at the sites. The measure is known in closed form if the timelike length scale is fixed at a low level. This is possible in the synchronous gauge (properly softened to remove nonphysical infrared singularities). This gauge also allows, with a refined finite-difference form of the action, to match the analytical properties of the propagator to the continuum case.

The metric is parameterized to make the measure Lebesgue. The linear part of this parametrization leads to a discrete form of standard Feynman diagrams that approximates finite continuum diagrams and is finite for infinite ones; the nonlinear part gives new vertices and diagrams.

The maximum of the measure is at the edge length scale $b = b_{\rm s} \sim \eta^{1/2}$, where $\eta$ defines the free factor like $(- \det \|g_{\lambda \mu}\|)^{\eta/2}$ in the measure and should be a large parameter to ensure true action upon integration over connection. For general perturbative expansion (including both that for the measure and the S matrix) to be free of increasing powers of $\eta$, its initial point must be at $b$ sufficiently close to $b_{\rm s}$; this appears to be a dynamic mechanism for establishing $b$ as an optimal initial point of the perturbative expansion.

This technique is applied to Newton's potential. Some of new diagrams, including potentially large ones, are mutually cancelled. The S matrix expansion is analyzed to consist of standard diagrams. These diagrams form series with a small parameter $\eta^{-1}$; one-loop diagrams calculated in the literature represent the leading order.
\end{abstract}


MSC classes: 83C27; 81S40

keywords: general relativity; discrete gravity; Regge calculus; Feynman diagrams; synchronous gauge; functional integral

\tableofcontents

\section{Introduction}

An advantage of the minisuperspace theory of gravity (a discrete theory), similar to the one proposed by Regge \cite{Regge} and known as “Regge calculus” (RC), is the countability of its degrees of freedom and the possibility of approximating all the degrees of freedom of the standard continuum general relativity theory (GR). In quantum theory, this is important in view of the formal non-renormalizability of GR. RC is GR on piecewise flat spacetime. Piecewise flat spacetime can approximate an arbitrary continuum Riemannian geometry in some topology with arbitrary accuracy \cite{Fein,CMS}; it is described by a countable set of edge lengths of the flat 4-simplices that make up the piecewise flat spacetime, forming the so-called simplicial complex. The countability of the set of variables is advantageous for using path integrals in calculating physical quantities like the Newtonian potential \cite{HamWil1,HamWil2,Ham1}. Additional constraints can be imposed on the simplicial complex that still allow smooth manifold approximation at large scales, but simplify the analysis of path integrals in Causal Dynamical Triangulations Theory (CDT) \cite{cdt,cdt1}. In addition to using piecewise flat spacetime as a regularization tool, there are approaches that assume that spacetime is in fact piecewise flat \cite{Mik}.

Along with using RC for analysing the Newtonian potential \cite{HamWil1}, this quantity was analyzed on the one-loop level in the ordinary (continuum) perturbation theory. On this level, the divergences in gravity were studied in Ref. \cite{hooft}. The graviton loop corrections to the Newtonian potential were analysed in \cite{don,don1,don2,don3,muz,akh,hamliu1,kk,BDH1,BDH2,kk1}. These are some of the works that consider the direct calculation of graviton loop diagrams, which is important for us, although other methods, such as functional ones, can also be used to analyze the Newtonian potential \cite{Frob,Tiber}. The Newtonian potential appears to be a natural touchstone for applications of the continuum gravity perturbation theory and its discrete form based on RC. At higher orders, the problem of diagram divergence arises and the simplicial structure of spacetime may have an impact. The general form of Feynman diagrams in simplicial gravity was considered in \cite{HamLiu}. We will now analyze the diagrammatic technique that arises within the framework of a specific mechanism for fixing an elementary length scale (a specific functional integral measure) and the true finite-difference action of the leading order and the soft synchronous gauge-fixing term attached to this mechanism under the requirement of reproducing, at non-Planckian momenta, those diagrams of continuum perturbation theory that are convergent.

As a source of ultraviolet (UV) cutoff, we consider the bell-shaped dependence of the functional measure in terms of edge lengths on the elementary length scale \cite{our1}. This dependence has a maximum at some finite nonzero length scales. Such a scale is measured in Planck units, but is proportional to some formally large parameter of the theory. At the same time, this measure provides additional graviton vertices compared to the naive discretization with the Lebesgue measure.

This measure is the result of functional integration over the connection variables. In turn, the latter provide a consistent non-singular transition to the continuous time limit and the Hamiltonian canonical formalism, the construction of which in terms of edge length variables alone encounters difficulties. Meanwhile, these are the stages of the standard construction of a functional integral. By introducing the connection variables proposed by Fr\"{o}hlich \cite{Fro}, we extend the set of originally purely edge length variables. The latter are edge vectors defined in the local frames of the 4-simplices. The connection variables are SO(3,1) matrices defined on the 3-simplices and can be decomposed into self-dual and anti-self-dual parts. We take the discrete gravity action as the sum of two terms, the individual contributions of these parts of the connection \cite{Kha}. The connection can be removed at the classical level using the equations of motion and thus the exact RC action can be obtained.

By taking the action as the sum of the self-dual and anti-self-dual contributions, we simplify the analysis. This is reminiscent of the transition to Ashtekar variables \cite{Ash}, based only on the (anti-)self-dual su(2) connection and used to formulate Loop Quantum Gravity (LQG). In contrast, we use both parts of the connection: self-dual and anti-self-dual. Moreover, in LQG the factor intended to cut off the theory at small distances and make it finite is the discreteness of the area spectrum with a quantum of the Planck scale \cite{RovSmo}; in our case, such a factor is the elementary area/length scale as a characteristic of the probability distribution of the elementary area/length whose spectrum is continuous.

The considered functional measure can be found in the extended superspace of the tetrad-connection variables in the continuous time formalism. Then it is natural to pose the problem of finding a full discrete measure that goes over into the found continuous time measure, no matter what direction is chosen, along which the edge lengths are made arbitrarily small, and some coordinate along this direction becomes the required continuous time. We can solve the close problem of finding such a measure on some extended configuration superspace and then reducing (projecting) the found measure onto the ordinary superspace, considered as a physical hypersurface in the extended one. Here, the extended configuration superspace is one in which area bivectors are generalized to independent area tensors. Projection of the measure can be accomplished by inserting a delta-function factor, ensuring that the independent area tensors are bivectors constructed from some edge vectors. It is clear that this projection operation can give rise to ambiguities, and we, taking into account symmetry requirements, define it up to some parameter $\eta$. This parameter defines the quantum version of the theory. Its change $\Delta \eta$ results in the appearance of the additional factors $V^{\Delta \eta}_{\sigma^4}$ in the measure, where $V_{\sigma^4}$ is the 4-volume of the 4-simplex $\sigma^4$. These factors are analogues of $ ( - \det \| g_{\lambda \mu} \| )^{\Delta \eta / 2}$ in the continuum theory.

Let us denote the set of variables of the tetrad type $\rl$, where these can be either genuine edge vectors or, for example, area tensors. The invariant Haar measure ${\cal D} \Omega$ on the connection $\Omega \in$ SO(3,1) (more precisely, $\Omega_{\sigma^3}$ on the 3-simplices $\sigma^3$) is the $\Omega$-defined part of the mentioned full discrete measure $\d \mu ( \rl ) {\cal D} \Omega$. The integration over ${\cal D} \Omega$ gives some effective action $\tilS_{\rm g} ( \rl )$ and a new measure $F ( \rl ) D \rl$ in the functional integral
\begin{equation}\label{int-S D-Omega=F}                                     
\int \exp [ i S_{\rm g}(\rl , \Omega  ) ] ( \cdot ) \d \mu ( \rl ) {\cal D} \Omega = \int \exp [ i \tilS_{\rm g} ( \rl ) ] ( \cdot ) F ( \rl ) D \rl .
\end{equation}

\noindent Here $S_{\rm g}(\rl , \Omega )$ is the mentioned discrete action, which reduces to the RC action $S_{\rm g}(\rl )$ on the equations of motion for $\Omega$.

To evaluate the phase $\tilS_{\rm g} ( \rl )$ in this result, it is natural to expand the original functional integral around some solution for $\Omega$, which gives a nonzero result already in the zeroth order. If $\Omega = \Omega_0 ( \rl )$ solves the equations of motion for $\Omega$, then the zeroth approximation for $\tilS_{\rm g} ( \rl )$ is $S_{\rm g}(\rl , \Omega_0 ( \rl ) )$. For $\Omega = \Omega_0 \exp \omega$, $\omega \in$ so(3,1), subsequent orders in the expansion of $S_{\rm g}(\rl , \Omega )$ in $\omega$ start with a bilinear one. The terms in $S_{\rm g}(\rl , \Omega )$ are proportional to the areas of the 2-simplices (triangles), and typical values of the integration variable $\omega$ and thus corrections to $S_{\rm g}(\rl , \Omega_0 ( \rl ) )$ are small for large areas. In particular, the spacelike areas should be large, and the scale of these areas or the spacelike elementary edge length scale $b_{\rm s}$ arises dynamically in the present approach (as a result of finding an optimal starting point of the perturbative expansion under the requirement of maximization of the measure) and depends on $\eta$ (\ref{b=l-pl-sqrt-eta}). Thus, $\eta$ should be a large parameter. Then we can use the action $S_{\rm g}(\rl , \Omega_0 ( \rl ) )$, which is exactly the RC action $S_{\rm g}(\rl )$ as already mentioned.

When evaluating $F ( \rl )$, a contribution in the zeroth order arises if we expand over some edge vectors which can be called {\it $t$-like} or {\it temporal}. This concept corresponds to the covariant world index zero in the continuum theory. Such edge vectors $l^a_\lambda$ are discrete analogues of the tetrad $e^a_\lambda$ at $\lambda = 0$ or the ADM lapse-shift functions \cite{ADM1} $(N, \bN )$. At zero order, the temporal triangles make no contribution to the action, and without this contribution, the functional integration over the connections factorizes into factors related to the individual remaining (spatial and diagonal) triangles.

Let functions $\rl = \rl ( u )$ express the variables $\rl = (l_1, \dots, l_n )$ through some new variables $u = (u_1, \dots, u_n )$ transforming the measure into the Lebesgue measure, $F ( \rl ) D \rl = D u$. Then the perturbative expansion for the functional integral (\ref{int-S D-Omega=F}) can be developed by first expanding the action $\tilS_{\rm g} ( \rl )$ around some starting point $\rl_{(0)} = \rl ( u_{(0)} )$,
\begin{eqnarray}\label{ddS}                                                 
& & \hspace{-10mm} \tilS_{\rm g} ( \rl ) = \tilS_{\rm g} ( \rl_{(0)} ) + \frac{1}{2} \sum_{j, k, l, m} \frac{\partial^2 \tilS_{\rm g} ( \rl_{(0)} )}{\partial l_j \partial l_l} \frac{\partial l_j (u_{(0)} )}{\partial u_k} \frac{\partial l_l (u_{(0)} )}{\partial u_m} \Delta u_k \Delta  u_m + \dots , ~ \Delta u = u - u_{(0)} .
\end{eqnarray}

\noindent There is no linear order here since the equations of motion (Regge's skeleton equations for $\tilS_{\rm g} ( \rl ) = S_{\rm g} ( \rl )$) are assumed to be satisfied at this point,
\begin{equation}\label{dS/dl=0}                                             
\frac{\partial \tilS_{\rm g} ( \rl_{(0)} )}{\partial l_j} = 0 .
\end{equation}

\noindent By working from the extreme point (\ref{dS/dl=0}), we minimize the corrections and eliminate the need to partially sum a subsequence of corrections to reproduce the correct result. If, within the framework of compliance with the equations of motion, it is possible to adjust the second order terms, then in this case we can speak of the condition of the extremum of the determinant of the second order form. Which, in view of the bell-shaped form of $F ( \rl )$, is reduced to a condition of the type of the condition for maximizing the measure $F ( \rl )$ (which looks quite natural physically),
\begin{equation}\label{def-l0}                                              
F (\rl_{(0)} )^{-2} \det \left \| \frac{\partial^2 \tilS_{\rm g} (\rl_{(0)} )}{\partial l_i \partial l_k} \right \| = \mbox{ minimum}.
\end{equation}

\noindent To be exact, here $\tilS_{\rm g}$ is replaced by $S$, which also includes gauge-breaking and Faddeev-Popov ghost terms.

The aforementioned functional integration over $\Omega$ in the factorization approximation \cite{Kha3} leads to an exponentially decaying factor $\exp ( - 2 \pi V_{\sigma^2} )$ in $F ( \rl )$, with the area of the triangle $V_{\sigma^2}$ (for a spatial or diagonal triangle $\sigma^2$). This can be illustrated using a simplified qualitative example of an integral of the form
\begin{equation}                                                            
\int \exp ( i v_{ a b } R^{ a b } ) {\cal D} R
\end{equation}

\noindent ($ v_{a b}$ is the area tensor, $\frac{1}{2} \epsilon_{abcd} l^c_1 l^d_2 $ for the triangle formed by the edges with vectors $l^c_1$ and $l^d_2 $). Integrating this function of $ v_{a b}$ as independent variables with a monomial of the components $ v_{a b}$ over $\d^6 v$ (the requirement of analytic extendibility to independent $ v_{a b}$ aligns well with the aforementioned concept of an extended superspace of independent area tensors), we get the integral over ${\cal D} R$ of the multiple partial derivative of $\delta^6 ( R - \overline{R} )$. This integral turns out to be finite. For this to be true for any monomial in $ v_{a b}$, the function $\int \exp ( i v_{ a b } R^{ a b } ) {\cal D} R$ must contain exponential suppression of large areas.

$F ( \rl )$ also contains the aforementioned factor $V^\eta_{\sigma^4}$ with the 4-simplex volume $V_{\sigma^4}$ and is bell-shaped. Condition (\ref{def-l0}) leads to the following mentioned spacelike elementary edge length scale $ l = b_{\rm s} $ (Subsection \ref{initial-point}), here in usual units,
\begin{equation}\label{b=l-pl-sqrt-eta}                                     
b_{\rm s} = l_{\rm Pl} \sqrt{ \frac{ \eta - 9 }{ 2 \pi } }, ~~~ l_{\rm Pl} = \sqrt{8 \pi G} .
\end{equation}

Given the loosely fixed length scale, we can first study the very starting point of the perturbative expansion. If this point is a discrete analogue of a continuum background with singularities of the type of black holes, the presence of a definite length scale allows us to resolve these singularities and find the metric/field distribution in the main types of black holes \cite{our}. Now we analyze perturbations around a flat background.

To have an algorithmizable diagrammatic technique, we need some periodic simplicial structure, and as such it is natural to take the simplest one with a hypercubic cell consisting of 4!=24 4-simplices whose edges are cubic edges or diagonals \cite{RocWil}. We then freeze some tetrad-connection variables, leaving a sufficient fraction of them active to approximate any smooth Riemannian manifold to arbitrary accuracy.

The symmetry broken by the choice of lattice axes will be restored if all simplicial structures are taken into account, as implied by the RC strategy. However, even within the framework of the considered hypercubic lattice, the diffeomorphism symmetry is preserved in the leading order over metric variations from site to site and, thus, in configurations close to the continuum spacetime. Consequently, at least in this region of the configuration superspace, sets of arbitrarily large measure of physically almost equivalent configurations arise, as is the case in the continuum spacetime.

Correspondingly, analogously to the continuum theory, we should take into account not all configurations in the functional integral. But unlike the continuum case, the result depends on the chosen subset in the configuration superspace over which the functional integration is performed, that is, we can say, on the chosen gauge. Nevertheless, the exact value of the functional integral should be obtained by summing/averaging over all gauges. A more general approach than rigidly fixing the gauge conditions is to average over some neighborhood of such conditions by adding a gauge-fixing term to the action, that is, by introducing a weight factor into the functional integral. In the discrete case, this averaging seems even more natural, since it can be considered as part of the mentioned eventual averaging over all gauges. In our paper \cite{khat}, as such a term we take a finite-difference version of the gauge non-invariant term corresponding to the de Donder-Fock gauge.

This gauge admits variations of all metric components, so we need to know the measure when $l^a_\lambda \forall \lambda$ are not small. This is not known in closed form, so in that paper we used the simplest model of the measure, which reduces to the already known measure in the "factorization approximation" when the scale of $l^a_\lambda$ is small for some $\lambda$. Then we could calculate the degree of dependence of the magnitude of an arbitrary loop diagram on the parameter $\eta$, which, as mentioned above, is considered a large parameter. In principle, the perturbative expansion can contain increasing powers of $\eta$, as shown below by eq (\ref{(1+eta x)^(1/eta)}). However, it turns out that these increasing powers of $\eta$ are absent from the perturbative expansion, but only if the initial point of the perturbative expansion is located close enough to the point that ensures the measure maximization condition (\ref{def-l0}). At such an initial point, the scales of $l^a_\lambda \forall \lambda$ are not small. However, there is a loophole that allows us to preserve the absence of increasing powers of $\eta$ and at the same time not to leave the region of small $l^a_\lambda$ for some $\lambda$, let it be $\lambda$ equal to 0, where the measure is known. Namely, we need to prohibit the variations of the temporal components. Thus, we arrive at (a discrete version of) the synchronous gauge.

The use of the synchronous gauge requires the elimination of some difficulties that are already present in the continuum case. There the synchronous gauge in its standard form as the condition $g_{0 \lambda} = g_{0 \lambda}^{(0 ) } = ( - 1 , 0 , 0 , 0 )$ (the "hard" synchronous gauge) suffers from the presence of poles at $p_0 = 0$ (up to $p_0^{- 4}$) in the graviton propagator. A possible way out of this situation is to consider some “soft” gauge analogous to the gauge $n^\lambda A^a_\lambda = 0$ in Yang-Mills theories, where instead of singularities like $p_0^{- 1}$ in the gauge field propagator for the gauge $A^a_0 = 0$ we have $( n p )^{- 1} = (p_0 + i \varepsilon )^{- 1}$, $( \on p )^{- 1} = (p_0 - i \varepsilon )^{- 1}$ with poles shifted into the plain of complex momentum for $n^\lambda = (1, - \varepsilon (\partial^\alpha \partial_\alpha )^{- 1} \partial^\beta ) $ \cite{Ste}, thus justifying the Landshoff prescription \cite{Land}, which involves the substitution $p_0^{- 2} \Rightarrow (p_0^2 + \varepsilon^2)^{- 1}, \varepsilon \to 0$. We have analyzed such a gauge for gravity \cite{khat0}, $n^\lambda w_{\lambda \mu} = 0$, $w_{\lambda \mu} = g_{\lambda \mu} - g_{0 \lambda}^{(0 ) }$, in particular, the ghost contribution was found to be zero in the limit $\varepsilon \to 0$.

In such a soften gauge we have terms $\sim (p_0^2 + \varepsilon^2)^{- j}$ in the graviton propagator, which are non-singular themselves, but integrating them over $p_0$ can lead to negative powers of $\varepsilon$. This circumstance takes place for the non-pole part of the propagator (not containing the graviton pole at $p^2 = \bp^2 - p_0^2 = 0$), whereas, for example, at the one-loop level, the non-pole terms do not contribute to the absorptive part of the S-matrix, as considered in \cite{hooft}; therefore, passing to the limit $\varepsilon \to 0$ in this case does not encounter any difficulties. Nevertheless, with multi-loop computations in mind, we pay special attention to the possible implementation of a prescription of the type $p_0^{-j} \Rightarrow [ (p_0 + i \varepsilon )^{-j} + (p_0 - i \varepsilon )^{-j} ] / 2$ (a principal value prescription, although not in the original Cauchy sense). A convenient feature of this prescription is the separate use of the factors $(p_0 + i \varepsilon )^{-j}$ and $(p_0 - i \varepsilon )^{-j}$ in different terms in a given diagram, so that the integration path $\Im p_0 = 0$ is not clamped between the poles $p_0 = + i \varepsilon$ and $p_0 = - i \varepsilon$ and can be deformed to lie at a distance $O ( 1 )$ from the poles, thus making passing to the limit $\varepsilon \to 0$ obviously non-singular. A prescription of this type is achieved if we issue from the required graviton propagator of the form $[ G ( n, n ) + G ( \on, \on ) ] / 2$ (the graviton propagator obtained after adding the gauge-violating term bilinear in $ n^\lambda w_{\lambda \mu } $ is a Hermitian symmetric function $G ( n, \on )$ of $n, \on$, and the non-Hermitian but mutually Hermitian conjugates $G ( n, n )$ and $G ( \on, \on )$ follow from $G ( n, \on )$ by replacing here $\on \Rightarrow n$ or $n \Rightarrow \on$, respectively). We can recover the required gauge-fixing term \cite{khat0.5}. This is a function of the propagator (but only of its non-pole part). We can also find that the ghost contribution is zero in the limit $\varepsilon \to 0$.

New issues arise when passing from the continuum to the discrete case in question. There is a rare case where some changes to an action that have a non-leading order over metric variations from site to site lead to a significant change in the result of calculations with this action. Namely, the choice of a particular finite-difference form of the coordinate derivative affects the analytical properties of the propagator. If we adopt the simplest finite-difference derivative $\Delta_\lambda = \exp ( i p_\lambda ) - 1$ (in the momentum representation; $p_\lambda$ is the quasi-momentum), then the propagator will have a complex form, where, along with the factor $( \oDelta \Delta )^{- 1}$, the terms contain factors with complex analytical properties with denominators of degree greater than two with respect to $\Delta_\lambda$, $\oDelta_\lambda$ (say, six when calculated in the hard synchronous gauge). This complexity is due to the need to distinguish between $\Delta_\lambda$ and $\pm \oDelta_\lambda$ in the process of calculating the propagator.

The expression will not become more complex compared to the continuum case if we adopt the symmetrized finite-difference derivative $\Delta^{(s )}_\lambda = i \sin p_\lambda$. It is anti-Hermitian as the usual derivative $\partial_\lambda$, and the propagator can be obtained from the continuum one by replacing $\partial_\lambda \Rightarrow \Delta^{(s )}_\lambda$. However, its analytical properties qualitatively differ from those of the continuum propagator. It is sufficient to say that for small $| \bp |$ ($\sin p_\alpha \approx p_\alpha$), the poles of $( \oDelta^{(s )} \Delta^{(s )} )^{- 1} = - ( \Delta^{(s )} \Delta^{(s )} )^{- 1} $ are located at $\sin p_0 \approx \pm | \bp |$, that is, $p_0 \approx \pm | \bp |$ and $p_0 \approx \pm ( \pi - | \bp |)$ (in the period $p_0 \in ( - \pi , \pi ]$). Whereas the smaller pair of $p_0 $ does correspond to the continuum graviton pole at $p^2 = \bp^2 - p_0^2 = 0$, the larger pair $p_0 \approx \pm \pi$ does not. Taking it into account leads to the pole contribution to a given diagram approximately twice as large as that in the continuum case. At the same time, the poles of $( \oDelta \Delta )^{- 1}$ for small $| \bp |$ are given by the relation $2 \sin ( p_0 / 2 ) \approx \pm | \bp |$, that is, only $p_0 \approx \pm | \bp |$, and there is a correspondence with the continuum case.

To combine the correspondence with the continuum case and the comparative simplicity of the propagator, we can simultaneously use $\Delta_\lambda$ and $\Delta^{(s )}_\lambda$ in different terms in the action. Of course, there is a more significant cumbersomeness here compared to using only $\Delta^{(s )}_\lambda$. We have analyzed the propagator for such an action in the synchronous gauge in the principal value type prescription \cite{khat1}. A one-parameter family of possible such actions is considered, differing by some terms of non-leading orders over metric variations from site to site, and the one for which there is a complete correspondence of the propagator to the continuum case is identified (favorably, it turns out to be the simplest among this family). We have analyzed the required gauge-fixing term and its finiteness (which is important when we further use the mentioned parametrization of the metric $\rl = \rl ( u )$ here, and the gauge-fixing term becomes a source of vertices). It was also found that the ghost contribution is zero in the limit $\varepsilon \to 0$.

Here we consider the vertices arising from the gauge-fixing term using the mentioned metric parametrization $\rl = \rl ( u )$ and find that their contribution to the diagrams tends to zero at $\varepsilon \to 0$.

The aforementioned situation, where the perturbative expansion contains increasing powers of $\eta$, can arise quite easily. Let us assume for a moment that there is no exponential suppression of large areas or we consider small areas for which the exponential is $\approx 1$, and the above power volume factor in the measure of the type of $ ( - \det \| g_{\lambda \mu} \| )^{\eta / 2 }$ is used in the hope of suppressing small distances (the UV contribution). For a variable $y$ having the meaning of some length, we model the corresponding part of the measure as $y^{\eta - 1} \d y$ and parameterize $y = y ( x )$ to obtain the Lebesgue measure $C \d x$. Let's take $C = y_0^\eta$ and expand $y = y ( x )$ near some point $y = y_0$, $x = 0$:
\begin{equation}\label{(1+eta x)^(1/eta)}                                   
y_0^{- 1} y = (1 + \eta x)^{1 / \eta} = 1 + x + \frac{1}{2} ( 1 - \eta ) x^2 + \dots + O ( \eta^{j - 1} ) \cdot x^j + \dots .
\end{equation}

\noindent The bilinear part of the action is conditioned by the linear term in $x$ here and dynamically ensures that the typical value of $x$ is $O ( 1 )$ relative to $\eta$. The subsequent terms define new vertices of interaction. The coefficients of these terms are proportional to $\eta$ and higher powers of $\eta$, so that the new vertices have similar coefficients. Thus, powers of $\eta$ appear in the perturbative expansion.

But in the presence of exponential suppression of large areas in the measure, the situation turns out to be quite different. (Instead of the example measure given in the previous paragraph we have something like $y^{\eta - 1} \exp ( - y ) \d y$, where $y$ is some area.) We analyze the dependence of the perturbative expansion on the large parameter $\eta$ and find that it does not contain increasing powers of $\eta$, but only if we choose the starting point of this expansion sufficiently close to the maximum point of the measure. This looks as a dynamic mechanism of establishing this starting point, including the elementary length scale.

If the linear terms in the metric parametrization $\rl = \rl ( u )$ are considered, we get a finite-difference form of the standard perturbative expansion of the continuum GR. In particular, if continuum diagrams or some of their structures converge (as is the case, for example, for the one-loop corrections to the Newtonian potential), then this discrete approach simply gives a lattice approximation for them, and for ordinary external momenta or distances much larger than the typical edge length scale $b_{\rm s}$, this approximation works with high accuracy.

If continuum diagrams diverge, their discrete versions, of course, remain finite; they even form a series with a small parameter $b_{\rm s}^{- 1}$ or $\eta^{- 1}$. The latter happens because the lattice provides an effective UV cutoff $\sim b_{\rm s}^{- 1} \sim ( l_{\rm Pl } \eta^{1 / 2} )^{- 1}$, and powers of this cutoff cancel out with the corresponding powers of the gravitational coupling, which is just $l_{\rm Pl }$, while the powers of $\eta^{- 1}$ remain as factors in the diagram values.

In addition to the discrete analogues of the continuum vertices and diagrams, there are new vertices due to the bilinear and higher order terms in the metric parametrization. These vertices lead to new diagrams.

This is further examined using the example of corrections to the Newtonian potential. Some new diagrams cancel each other out. In the diagram technique, there are anti-Hermitian local imaginary contributions to the effective action from the new vertices. In the continuum case, these imaginary contributions would also be divergent, like $\delta^4(0)$. However, similar contributions also come from standard vertices and diagrams. This may create difficulties in constructing the unitary S matrix.

The S matrix is defined in continuum theory; accordingly, we consider the continuum analogue of the present discrete formulation, and conversely, for continuum consequences we consider their discrete analogues.

The general expression for the S matrix is considered which is defined not only by the interaction Lagrangian density but also by an infinite sequence of quasi-local operators \cite{BogShir}. The theoretical a priori uncertainty of these operators is connected with the uncertainty of the T-ordered products of field operator expressions at coincident points. These operators can also be included as counterterms in the full effective interaction Lagrangian density. Then in renormalizable theories, T-product singularities, if we absorb them into these operators, can thus be included in the renormalization of the constants of the theory, which can then be viewed as finite observables. In the considered non-renormalizable theory, we use a restricted version of the renormalization procedure, which consists in keeping the imaginary part of the Lagrangian density equal to zero. As the new vertices contribute to the imaginary part rather than the real part of the Lagrangian density, an appropriate choice of counterterms should cancel the new vertices.

As for the imaginary contribution to the effective Lagrangian density from the standard vertices, in the continuum theory such a divergent contribution arises for the actions containing interaction terms with multiple derivatives. These terms affect the canonical structure of the theory; Refs. \cite{HonMee,GJLW} demonstrated, using the example of such Lagrangians (chiral Lagrangians), that a careful application of the canonical quantization procedure leads to S matrix expansion rules that allow one to avoid the leading divergences characteristic of naive Feynman diagrams. In particular, when employing such an analysis and accounting for terms describing the interaction of gravity with matter fields, it was shown in Ref. \cite{don4} that the divergent contributions from field loop diagrams to the cosmological term cancel out.

In our case, to pass to the discrete version, it is expedient to employ the approach proposed in \cite{FraVil}; this approach involves defining the functional measure based on the requirement that it exactly compensates for the imaginary part of the effective action when the measure is rewritten as an imaginary counterterm to the action. In our case, we reserve a part of the measure (relatively small for large $\eta$) for this purpose. We then note that the action with such a counterterm is equivalent to the action without this counterterm if we use a certain prescription that discards non-pole contributions when integrating over loop (quasi-)momenta. This prescription consists of extending the standard path of integration over $p_0$ along the real number axis to a closed contour in the plane of complex $p_0$. (In habitual theories, where the interaction is less than quadratic in derivatives, this prescription changes nothing.) Then we note that this prescription also nullify the contribution of new vertices and makes the above procedure of their cancellation unnecessary.

This approach also considers the normal-ordered form of the action, in which field lines starting and ending at the same vertex (so-called "tadpoles") are excluded.

Although imaginary non-pole contributions to the effective action are discarded when calculating the S matrix elements, the full perturbative expansion that preserves these contributions also makes sense. Here the difference between the discrete and continuous formulations appears. In the continuum case, these non-pole contributions are infinite; in the discrete case, they are finite and form an expansion for an object such as the measure (including radiative corrections), which is obviously not a physical amplitude, but defines a framework (including the discretization length or elementary length) in which the physical amplitudes themselves can be found.

As a result, the physical amplitude is determined by the (discrete form) of the standard Feynman diagrams.
(In more detail, the prescription is used which singles out the pole contribution from each loop quasi-momentum integration.) Finite continuum diagrams are reproduced by such discrete analogues for ordinary or non-Planckian external momenta with high accuracy. In addition, as mentioned above, such an expansion for higher loops goes in positive powers of the small parameter $\eta^{- 1}$. Corrections to the Newtonian potential in the continuum formulation turn out to be finite and have been found in the literature at the one-loop level. Thus, in the present approach, these results represent the leading order in the small parameter $\eta^{- 1}$ of the perturbative expansion of the coefficient of the considered dependence $r^{- 3}$ to be added to the zero order potential $\propto r^{- 1}$.

In this case, we rely on the gauge independence of the result in the continuum theory, since we use the synchronous gauge (the softened discrete version) and compare discrete diagrams with their continuum counterparts again in the synchronous gauge (the softened continuum version), while the calculations in the literature were carried out in other gauges (usually in the Feynman one).

The paper is organized as follows. In Section \ref{starting}, we specify the starting functional integral in terms of the metric variables. In Subsection \ref{action} we consider the action; namely, in Subsection \ref{action=phase} it follows as a phase from the functional integration, in Subsection \ref{accuracy-of-action} we consider the range of values of the elementary length scales (timelike and spacelike) favorable for the accuracy of such a reproduction of the action, in Subsection \ref{suitable-action} we consider the finite-difference form of the action with both the usual and symmetrized form of the derivative, suitable for the correspondence of the propagator to the continuum case. Subsection \ref{measure} discusses the functional measure. In Subsection \ref{gauge}, the gauge-fixing term is considered. In Section \ref{expansion}, the perturbative expansion is constructed. In Subsection \ref{transform}, we rewrite the measure partially in terms of the areas over which the bell-shaped dependence is observed, also taking into account the soft synchronous gauge at $\varepsilon \to 0$. In Subsection \ref{initial-point}, we find the elementary edge length scale (spacelike) as an optimal starting point of the perturbative expansion from the measure maximization condition. In Subsection \ref{scale}, we estimate the scale of the graviton perturbations arising in diagrams depending on the length scales. In Subsection \ref{parameter}, the metric variables are parameterized, and before the actual parametrization in Subsection \ref{replace}, some simplifying replacement of the functional variables $g_{0 \lambda}$ is made; then, in Subsection \ref{parameterize}, the parametrization is found in the form of a general series, and for the sake of generality, it is assumed that the starting point does not necessarily coincide with the point of maximization of the measure. In Subsection \ref{large-parameter}, this series is used to show that the perturbative expansion does not contain increasing powers of $\eta$ for the initial point chosen in a certain neighbourhood of the maximum point of the measure, and does contain increasing powers of $\eta$ for the initial point outside this neighbourhood. In Subsection \ref{concrete-parameter}, a concrete form of metric parametrization up to bilinear terms is given for the starting point coinciding with the point of maximization of the measure. In Subsection \ref{vertex-scales}, the scales of interaction vertices and diagrams are considered; namely, in Subsection \ref{diagr-from-gauge-fix} the diagrams with the vertices arising from the gauge-fixing term are found to vanish at $\varepsilon \to 0$, in Subsection \ref{scales} we consider effective values of interaction terms when substituted into diagrams, in Subsection \ref{continuum-correspond} the correspondence with the continuum diagrams is considered, in Subsection \ref{new-vertex} new vertices are considered. In Section \ref{Newton} we consider corrections to the Newtonian potential as a typical effect for which graviton loop diagrams can be analyzed and used. In Subsection \ref{cancel} the mutual cancellation of some diagrams is analyzed. Subsection \ref{contrib-new} describes the contribution of the new vertices. Subsection \ref{Smatrix} analyzes the S-matrix; in Subsection \ref{normal}, this analysis is generalized to the normal-ordered form of the action. Then the Conclusion follows.

\section{Functional integral}\label{starting}

\subsection{Gravity action}\label{action}

\subsubsection{The action as a phase from functional integration}\label{action=phase}

We start with a gravitational system described by variables of both tetrad and connection types. The role of the tetrad type variables should be played by the edge vectors. They enter the gravitational action in the form of area tensors $v$. The area tensor of a triangle $\sigma^2$ formed by two edges $\sigma^1_1 $, $\sigma^1_2$ with edge vectors $l_{\sigma^1_1}$, $l_{\sigma^1_2}$ is $v_{\sigma^2}^{ab} = \frac{1}{2}\epsilon^{ab}{}_{cd}l^c_{\sigma^1_1}l^d_{\sigma^1_2}$. Instead of an antisymmetric tensor described by six components, a complex 3-vector can be used, $2 \bv_{\sigma^2} = i \bl_{\sigma^1_1} \times \bl_{\sigma^1_2} - \bl_{\sigma^1_1} l^0_{\sigma^1_2} + \bl_{\sigma^1_2} l^0_{\sigma^1_1}$.

Such vector and tensor quantities are defined in a certain local pseudo-Euclidean frame of reference, usually assigned to that 4-simplex $\sigma^4$ to which the considered geometric objects belong as subsets. In this aspect, the more detailed notation for these quantities would be $v_{\sigma^2 | \sigma^4}^{ab}$, $l_{\sigma^1_1 | \sigma^4}^a$, etc. Specifying $l_{\sigma^1_1 | \sigma^4}^a$ in a certain $\sigma^4 \supset \sigma^1$, this vector in any other $\sigma^{\prime 4} \supset \sigma^1$ can be found by some transformations of local pseudo-Euclidean reference frames; if we try to maximally extend the local frame of $\sigma^4$, this vector can enter the expression for $v_{\sigma^2}^{ab}$ for some $\sigma^2 \supset \sigma^1$ being transformed by the curvature matrices rotating by the defect angles at some triangles having common points with $\sigma^1$. Thus, the area tensors of some triangles may depend on the curvature matrices at the nearby triangles.

Another way of specifying the tetrad variables might be to set the values of the vector or tensor of a given geometric object in the frames of all 4-simplices containing that object. Thus we extend the set of characteristics of the system in the configuration superspace, but at the same time some geometrical constraints should be introduced. These constraints should ensure that the characteristics of the same considered objects in different frames should be related by some transformations of local pseudo-Euclidean reference frames. As such constraints, we could take, for example, the continuity of the scalar products of edge vectors or area vectors of any 3-simplex face when passing between the local frames of two 4-simplices sharing this face.

However, below we will consider a simplifying ansatz in which some of the degrees of freedom are frozen and at the same time are capable of reproducing all the degrees of freedom of the continuum theory. In this ansatz, the edge vectors or area tensors on which the gravitational action depends, can be attributed to the sites of a 4-cubic lattice, and such objects attributed to any given site can be considered as defined in some 4-simplex frame also attributed to this site. In other words, there will be no dependence of the area tensors involved on the curvature matrices (which here in the general case could be the intrinsic curvature matrices of a 4-cube).

The connection variables are SO(3,1) matrices $\Omega_{\sigma^3}$, which are set on the 3-simplices $\sigma^3$ (tetrahedra). The connection-related curvature matrices set on the 2-sim\-plices $\sigma^2$ (triangles) are the holonomy of $\Omega$, $R_{\sigma^2} ( \Omega ) = \prod_{{\sigma^3} \supset {\sigma^2}} \Omega^{\pm 1}_{\sigma^3}$. An SO(3,1) matrix can be factorized into self-dual and anti-self-dual matrix factors, each of which is an element of an instance of the group SO(3,C), since SO(3,1) embeds as a subgroup of SO(3,C) $\times$ SO(3,C): $\Omega = \pOmega \mOmega , \mOmega = ( \pOmega )^* , \pR ( \Omega ) = R ( \pOmega )$. Then $\pmOmega$ and $\pmR$ are 3 $\times$ 3 matrices. The gravity action in terms of connection-tetrad variables $S_{\rm g}(\rl , \Omega )$ or here in terms of complex area 3-vectors $\bv$ and connection $\Omega$ is
\begin{eqnarray}\label{S-simplicial}                                        
& & \hspace{-10mm} S_{\rm g} [ v, \Omega ] = \frac{1}{2} \sum_{\sigma^2} \left ( 1 + \frac{i}{\gamma } \right ) \sqrt{ \bv^2_{\sigma^2} } \arcsin \frac{\bv_{\sigma^2} * \pR_{\sigma^2} ( \Omega )}{\sqrt{ \bv^2_{\sigma^2}}} + \mbox{ c.c.} , ~ \bv * R \stackrel{\rm def}{=} \frac{1}{2}v^i R^{kl} \epsilon_{ikl} .
\end{eqnarray}

\noindent Here "c.c." means "complex conjugate", $\gamma$ is the Barbero-Immirzi parameter.

The aforementioned ansatz for the simplicial complex includes using the simplest periodic one with the cubic cell divided by diagonals into $4! = 24$ 4-simplices. Besides that, we set the following:

1) $\Omega_{\sigma^3} = 1$ for the internal $\sigma^3$s in the 4-cube,

2) $\Omega_{\sigma^3} = \Omega_\lambda$ for each of the six $\sigma^3$ into which the 3D face (the 3-cube orthogonal to the edge along $x^\lambda$) is divided,

3) the area tensors of two triangles constituting a quadrangle in the $x^\lambda, x^\mu$ plane are the same.

\noindent Then we have a mini-superspace simplicial action:
\begin{eqnarray}\label{S-cube}                                              
& & S_{\rm g} [ v, \Omega ] = \frac{1 }{4 } \sum_{\stackrel{\scriptstyle \lambda \mu \nu \rho}{\rm sites}} \left ( 1 + \frac{i}{\gamma } \right ) \epsilon^{\lambda \mu \nu \rho} \sqrt{ \bv^2_{\lambda \mu} } \arcsin \frac{\bv_{\lambda \mu} * \pR_{\nu \rho} ( \Omega )}{\sqrt{ \bv^2_{\lambda \mu}}} + {\rm c. c. } .
\end{eqnarray}

\noindent Here
\begin{eqnarray}\label{v=ll-R=wwww}                                        
& & 2 \bv_{\lambda\mu} = i \bl_\lambda \times \bl_\mu + l^0_\lambda \bl_\mu - l^0_\mu \bl_\lambda, \quad R_{\lambda\mu} (\Omega ) = \overOm_\lambda (\overT_\lambda \overOm_\mu) (\overT_\mu \Omega_\lambda) \Omega_\mu ,
\end{eqnarray}

\noindent where $T_\lambda$ acts on a function $f ( x )$ as $T_\lambda f(\dots , x^\lambda , \dots ) = f(\dots , x^\lambda + 1 , \dots )$ (the shift operator); the overlining in $\overline{T}_\lambda$, $\overline{\Omega}_\lambda$ means the Hermitian conjugation. $\epsilon^{0123} = +1$.

In the leading approximation, for small variations of the vectors $l^a_\lambda$ from site to site, the action $S_{\rm g} [ v, \Omega ]$ (\ref{S-cube}) in the aspect of excluding $\Omega$ using the equations of motion, as we have found in \cite{our1}, is equivalent to the action
\begin{eqnarray}\label{v-D-omega}                                          
& & \hspace{-15mm} - \frac{1}{4} \sum_{\rm sites} \left ( 1 + \frac{i}{\gamma } \right ) \epsilon^{\lambda \mu \nu \rho} \bv_{\lambda \mu} \cdot (\Delta_\nu \bomega_\rho - \Delta_\rho \bomega_\nu + \bomega_\nu \times \bomega_\rho) + {\rm c. c. } , ~ \epsilon_{123} = +1 , ~ \Delta_\lambda = T_\lambda - 1 ,
\end{eqnarray}

\noindent where $\omega^k_\lambda = - \frac{1}{4} \epsilon^k{}_{l m} \omega^{l m}_\lambda + \frac{i}{2} \omega^{0 k}_\lambda$ is the 3-vector whose (complex) direction and length $\sqrt{ \bomega^2 }$ mean the axis and the angle of some SO(3,C) rotation generated by the self-dual part of the generator $\omega \in so(3,1)$ of $\Omega$ ($\Omega_\lambda = \exp \omega_\lambda$), an independent variable in (\ref{v-D-omega}).

It can be noted that (\ref{v-D-omega}) is a finite-difference form of the tetrad-connection representation \cite{Holst,Fat} of the Einstein action $\frac{1}{2} \int R \sqrt{-g} \d^4 x$ in the continuum GR,
\begin{eqnarray}\label{Cartan}                                             
& & \hspace{-10mm} S_{\rm g} [ e, \omega ] = - \frac{1}{8}\int{(\epsilon_{abcd}e^a_{\lambda}e^b_{\mu} + \frac{2}{ \gamma}e_{\lambda c}e_{\mu d }) \epsilon^{\lambda\mu\nu\rho}} [\partial_{\nu} + \omega_{\nu}, \partial_{\rho} + \omega_{\rho}]^{cd}{\rm d}^4x \nonumber \\ & & \hspace{-15mm} = - \frac{1}{8} \epsilon^{\lambda\mu\nu\rho} \left ( 1 + \frac{i}{\gamma } \right ) \int (i \be_\lambda \times \be_\mu + e^0_\lambda \be_\mu - e^0_\mu \be_\lambda) \cdot (\partial_\nu \bomega_\rho - \partial_\rho \bomega_\nu + \bomega_\nu \times \bomega_\rho) \d^4 x + {\rm c. c. } ,
\end{eqnarray}

\noindent if we take $e^a_\lambda = l^a_\lambda$ for the tetrad and $\Delta x^\lambda = 1$ for the neighbouring sites along $x^\lambda$.

\subsubsection{Accuracy of recovering the action as a phase}\label{accuracy-of-action}

It is convenient to make a multiplicative shift of the connection variables $\Omega = \Omega_0 \exp \omega$, $\omega \in$ so(3,1), where $\Omega = \Omega_0 ( v )$ solves the equations of motion for $\Omega$. Then $S_{\rm g} (v, \Omega )$ has no term linear in $\omega$, and $S_{\rm g} (v, \Omega_0 (v ) )$ is the RC action $S_{\rm g} (\rl )$ (a function of the edge lengths $\rl$) by construction of the connection representation $S_{\rm g} (v , \Omega )$. The invariant Haar measure ${\cal D} \Omega$ reduces to some ${\cal D} \omega$. The functional integral over ${\cal D} \Omega$ takes the form
\begin{eqnarray}\label{stationary}                                         
& & \hspace{-15mm} \int \exp [ i ( S_{\rm g} (v, \Omega_0 ) + (\omega B \omega) + O( \omega^3 ) ) ] {\cal D} \omega = \exp [ i ( S_{\rm g} ( \rl )] \int \exp [ i (\omega B \omega) ] ( 1 + O( \omega^3 ) ) {\cal D} \omega ,
\end{eqnarray}

\noindent where the bilinear form $(\omega B \omega)$ in the expansion of $S_{\rm g} (v , \Omega )$ over $\omega$ is singled out.

The form $(\omega B \omega)$ consists of the terms proportional to
\begin{equation}\label{GoGoG*v}                                            
\left( \Gamma_1 \omega_{\sigma^3_1} \Gamma_0 \omega_{\sigma^3_2} \Gamma_2 \right) * \bv_{\sigma^2}
\end{equation}

\noindent (plus the complex conjugate ones). Here $ \Gamma_0 $, $ \Gamma_1 $, $ \Gamma_2 $ are SO(3,C) matrices, functions of edge vectors, depending on $\sigma^2$ and on the two 3-simplices $\sigma^3_1$, $\sigma^3_2$, containing $\sigma^2$, some products of matrices $\Omega_{0 \sigma^3}$. In the case of a flat spacetime, we can extend the local frame of any 4-simplex to the overall spacetime and set $\Omega_0 = 1$. That is, we cannot choose $\Omega_0 = 1$ only if $\exists \sigma^2$ such that the corresponding curvature $R_{0 \sigma^2} \neq 1$. Then $\Omega_0$ is a function of $R_0$ (certain products of $R_{0 \sigma^2}$s). Extremely large curvature corresponds to rotations $R_{0 \sigma^2}$ by defect angles $\sim 1$. If defect angles are small, less than 1, $R_0$ and thus $\Omega_0$ and the matrices $ \Gamma_0 $, $ \Gamma_1 $, $ \Gamma_2 $ are close to unity. Then the terms (\ref{GoGoG*v}) constituting the bilinear form $(\omega B \omega)$ are close to forms proportional to
\begin{equation}\label{o*o*v}                                              
\bomega_{\sigma^3_1} \times \bomega_{\sigma^3_2} \cdot \bv_{\sigma^2} .
\end{equation}

\noindent Here the simplices $\sigma^3_1$, $\sigma^3_2$, $\sigma^2$, on which the complex vectors $\bomega$, $\bv$ are taken, are conveniently classified in the case of some regular simplicial structure, a special case of which is the hypercubic lattice. This structure consists of 3-dimensional leaves of the foliation, which are themselves the 3-dimensional simplicial complexes. These complexes have the same structure and are labelled by integer values of some coordinate $t$. Corresponding vertices in any two neighboring leaves are connected by edges. These edges can be called {\it $t$-like} or {\it temporal}. The edges of any leaf itself are naturally called {\it leaf} edges. A quadrilateral formed by two analogous leaf edges of neighboring leaves and two $t$-like edges connecting the analogous ends of these leaf edges is divided into two triangles by a {\it diagonal} edge. Apart from the edges, a simplex can be a $t$-like simplex if it contains a $t$-like edge, a leaf simplex if it is a simplex of a leaf, and a diagonal simplex in other cases. The leaves divide the whole spacetime into regions. Each such region is limited by two neighboring leaves and consists of 4-dimensional prisms whose bases are 3-simplices of these leaves and lateral 3-dimensional surface consists of $t$-like 3-simplices. Each 4-prism consists of four 4-simplices. It is natural to consider $t$ as a time, then the $t$-like edge vectors are discrete counterparts of the ADM (Arnowitt-Deser-Misner) lapse-shift functions.

In this structure, we denote the vectors of the $t$-like triangles $\sigma^2$ as $\btau_{\sigma^2}$, while $\bv_{\sigma^2}$ itself denotes the vectors of the leaf/diagonal triangles. Now, in $\bomega_{\sigma^3_1} \times \bomega_{\sigma^3_2} \cdot \btau_{\sigma^2}$, the 3-simplices $\sigma^3_1$ and $\sigma^3_2$ are $t$-like. In the hypercubic model, this is $\bomega_\alpha \times \bomega_\beta \cdot \btau_\gamma \epsilon^{\alpha \beta \gamma} $, $\btau_\alpha \equiv \bv_{0 \alpha}$. For a leaf/diagonal $\sigma^2$, the expression $\bomega_{\sigma^3_1} \times \bomega_{\sigma^3_2} \cdot \bv_{\sigma^2}$ accepts all three combinations of $\sigma^3_1$ and $\sigma^3_2$ being both $t$-like or both leaf/diagonal or $t$-like and leaf/diagonal. In the hypercubic model, the first two possibilities would be $\overline{T}_0 \bomega_\alpha \times \bomega_\alpha \cdot \bv_\alpha $, $\bv_\alpha \equiv \epsilon_{\alpha \beta \gamma} \bv_{\beta \gamma} / 2$ and $\overline{T}_\alpha \bomega_0 \times \bomega_0 \cdot \bv_\alpha $, respectively. Of course, these are discrete artifacts vanishing in the continuum limit, but, for example, the determinant of the matrix of the quadratic form $\overline{T}_0 \bomega_\alpha \times \bomega_\alpha \cdot \bv_\alpha $ of $\bomega_\alpha$ for a given $\alpha$ is not identically zero and the same can be said about the form $\overline{T}_\alpha \bomega_0 \times \bomega_0 \cdot \bv_\alpha $ of $\bomega_0$.

In overall, we cannot reduce $\det B$ to a product of local factors, and there is no reason for its being identically zero at $\btau = 0$; $\det B (\btau = 0) \neq 0$ for a random configuration. Therefore, in (\ref{stationary}), there will be an expansion in $| \bv |^{- 1 / 2}$, or, by parity, in $| \bv |^{- 1 }$ at small $| \btau |$ in the most part of the configurations. Here $| \bv |$, $| \btau |$ are typical values of $| \bv_{\sigma^2} |$, $| \btau_{\sigma^2} |$. The corrections to the action will be small for $| \bv | \gg 1$. However, the coefficients will be singular for some configurations. For example, the simplest such case is when all $\bv_{\sigma^2}$ are collinear, then $(\omega B \omega)$ does not change if each $\bomega_{\sigma^2}$ changes by a value proportional to $\bv_{\sigma^2}$. In such cases, the terms $\bomega_{\sigma^3_1} \times \bomega_{\sigma^3_2} \cdot \btau_{\sigma^2}$ play a regularizing role. Then, to provide the corrections to the action small for all configurations, we should consider the case $| \btau | \gg 1$. Then the expansion also occurs in powers of $| \btau |^{- 1 }$.

This can be illustrated by the finite-difference form (\ref{v-D-omega}) of the hypercubic action. For the initial values of $l^a_\lambda$, $l^{(0 ) a}_\lambda = {\rm diag} (b_{\rm t}, b_{\rm s}, b_{\rm s}, b_{\rm s})$ (\ref{g^(0)l^(0)}), the form $(\omega B \omega)$ (per site) is proportional to
\begin{eqnarray}                                                           
& & \epsilon^{\alpha \beta \gamma} \left[ i | \bv | \left( \be^{(0 ) }_\alpha \times \be^{(0 ) }_\beta \right) \cdot \left( \bomega_0 \times \bomega_\gamma \right) + | \btau | \be^{(0 ) }_\alpha \cdot \left( \bomega_\beta \times \bomega_\gamma \right) + \mbox{ c.c.} \right] , \nonumber \\ & & e^{(0 ) i}_\alpha = \delta^i_\alpha , ~ 2 | \bv | = b_{\rm s}^2 , ~ 2 | \btau | = b_{\rm s} b_{\rm t} .
\end{eqnarray}

\noindent We scale the variables $\bomega_\lambda$ to $\tbomega_\lambda$ so that the form remains the same when replacing the letters $\bomega_\lambda$ with $\tbomega_\lambda$, $| \bv | $ with 1, $| \btau | $ with 1.
\begin{equation}                                                           
\bomega_\alpha = \frac{ 1 }{ \sqrt{| \btau |}} \tbomega_\alpha , \quad \bomega_0 = \frac{ \sqrt{| \btau |}}{| \bv |} \tbomega_0 .
\end{equation}

\noindent When integrating over ${\cal D} \omega$ in (\ref{stationary}), the largest contribution is made by typical values of $\tbomega_\lambda$ of order up to $O ( 1 )$. Therefore, the expansion in (\ref{stationary}) occurs in terms of $| \btau |^{- 1 / 2}$, $| \btau |^{1 / 2} | \bv |^{- 1}$ or, by parity, in terms of $| \bv |^{- 1}$ and $| \btau |^{- 1}$, but positive powers of $| \btau |$ also occur. With taking into account the relation $|\btau | \ll | \bv |$ (\ref{t<<v}) (required for the validity of the factorization approximation when calculating the functional measure by functional integration over connection), the considered expansion is majorized by an expansion over $| \bv |^{- 1}$ and $| \btau |^{- 1}$. In overall, it is possible to limit ourselves to the leading terms in the expansions at
\begin{equation}                                                           
| \bv | \gg | \btau | \gg 1 \mbox{ or } b_{\rm s}^{- 1} \ll b_{\rm t} \ll b_{\rm s} .
\end{equation}

\noindent (The latter automatically means $b_{\rm s} \gg 1$.)

To be precise, in (\ref{stationary}) there are two more deviations from the Gaussian type. First, in the integration limits. The vector $\bomega$ parameterizes an SO(3,C) rotation by the complex angle $\sqrt{\bomega^2} \mbox{mod} (2 \pi )$. Therefore, the actual integration should be performed in the band $0 \leq \Re \sqrt{\bomega^2} < 2 \pi$ of the complex plane of $\sqrt{\bomega^2}$. However, the contribution of configurations with $\Re \sqrt{\bomega^2} \geq 2 \pi$ is suppressed by the oscillating exponential at large $| \bv |$, $| \btau |$, and the integration limits can be safely extended to infinity.

Second, the Haar measure ${\cal D} \omega$ deviates from the Lebesgue one. It can be written as $\D \bomega \D \bomega^*$ where $\D \bomega = ( 4 \pi^2 \bomega^2 )^{- 1} \sin^2 ( \sqrt{ \bomega^2 } / 2 ) \d^3 \bomega$ and the same for $\bomega \Rightarrow \bomega^*$ (the product of two Haar measures on SO(3,C)), where it is understood that $\d^3 \bomega \d^3 \bomega^* \equiv 2^3 \d^3 \Re \bomega \d^3 \Im \bomega$. Again, the modifying factor in front of the Lebesgue measure takes the form $1 + O ( \omega^2 )$ and is close to 1, since typical values of $\omega^2$ are small for large area tensors.

Finally, above we assumed that the actual defect angles are much smaller than 1 (to go from (\ref{GoGoG*v}) to (\ref{o*o*v})). However, it is clear that for the evaluation purposes considered here, this condition can be extended to say that the defect angles must be $\lesssim 1$, less than, or of the order of 1.

\subsubsection{Suitable finite-difference form of the action}\label{suitable-action}

Excluding $\omega$ from the action (\ref{v-D-omega}) with the help of the equations of motion, we operate with finite differences in the leading order over themselves in the same way as with ordinary derivatives, and thus come to a finite-difference form of the Einstein action $\frac{1}{2} \int R \sqrt{-g} \d^4 x$. Again, in the leading order, different forms of the finite-difference derivative coincide, and we first choose the symmetric such form, $\Delta^{(s )}_\lambda = ( T_\lambda - \overT_\lambda ) / 2$. This form has the anti-Hermitian property $\oDelta^{(s )} = - \Delta^{(s )}$ just like the ordinary derivative. This makes it easier to find the propagator and simplifies its analytical properties, bringing them closer to the properties of the propagator in the continuum theory. Thus, the action reads
\begin{eqnarray}\label{Sg}                                                 
& & \tilS_{\rm g} [ g ] = \frac{1}{ 8 } \sum_{\rm sites} g^{\lambda \rho} g^{\mu \sigma} g^{\nu \tau} \left[ 2 \left( \Delta^{(s )}_\lambda g_{\mu \nu} \right) \left( \Delta^{(s )}_\tau g_{\sigma \rho} \right) - \left( \Delta^{(s )}_\nu g_{\mu \lambda} \right) \left( \Delta^{(s )}_\tau g_{\sigma \rho} \right) \right. \nonumber \\ & & \left. - 2 \left( \Delta^{(s )}_\lambda g_{\mu \rho} \right) \left( \Delta^{(s )}_\sigma g_{\nu \tau} \right) + \left( \Delta^{(s )}_\nu g_{\lambda \rho} \right) \left( \Delta^{(s )}_\tau g_{\mu \sigma} \right) \right] \sqrt{- g} .
\end{eqnarray}

\noindent This action should be combined with a certain term fixing the (approximate) gauge, i.e. the subset of the configuration superspace (with a weighting factor) over which the functional integral is calculated (in the long run, some averaging over all gauges must be performed). Then it becomes possible to determine the propagator. In the simplest case of the initial metric $g^{(0 ) }_{\lambda \mu} = \eta_{\lambda \mu} \stackrel{\rm def}{=} \mathrm{diag} (-1, 1, 1, 1)$, around which we expand $g_{\lambda \mu}$ in small variations, $g_{\lambda \mu} = g^{(0 ) }_{\lambda \mu} + w_{\lambda \mu}$, the graviton propagator has a pole part $\propto ( \Delta^{(s ) 2} + i 0 )^{- 1} = ( \sin^2 p_0 - \sum^3_{j = 1} \sin^2 p_j + i 0 )^{- 1}$. Here $\Delta^{(s )}_\lambda = i \sin p_\lambda$, $p_\lambda$ is the quasi-momentum. The (anti)symmetry of $\Delta^{(s )}_\lambda$ with respect to $p_\lambda \to p_\lambda + \pi$ leads to the appearance of quasi-momentum configurations in the period $p_\lambda \in (- \pi , \pi ]$ related by this relation and contributing equally to any given diagram. In particular, if there is a pole at some point $p_0$, then there are the poles at the points $- p_0$ and $\pm ( \pi - | p_0 | )$. Then the pole contribution to the diagram when integrating over $\d p_0$ for small $\bp$ and other quasi-momenta is approximately twice as large as that obtained with such integration in the analogous continuum diagram.

This doubling of poles is absent if the symmetrized finite-difference derivative $\Delta^{(s )}_\lambda$ is replaced by the usual $\Delta_\lambda = \exp ( i p_\lambda ) - 1$. However, the anti-Hermitian property of $\Delta^{(s )}_\lambda$ does not extend to operator $\Delta_\lambda$, but this is precisely the property that allows the discrete propagator to follow from the continuous one by replacing the derivative $\partial_\lambda$ with the symmetrized finite difference $\Delta^{(s )}_\lambda$. When using the usual finite difference $\Delta_\lambda$ in the action, there are terms with factors whose denominators are polynomials of $\Delta_\lambda$, $\oDelta_\lambda$ of order higher than two with respect to $\Delta_\lambda$, $\oDelta_\lambda$ (namely, six in the case of the hard synchronous gauge) and complex analytical properties. Fortunately, it is sufficient to replace $\Delta^{(s )}_\lambda$ with $\Delta_\lambda$ only in the terms where two derivatives are contracted with each other, and even then not in all of them. In $\tilS_{\rm g} [ g ]$ (\ref{Sg}) $\Delta^{(s )}_\lambda$ should be replaced by $\Delta_\lambda$ in the second and possibly fourth terms. In \cite{khat1} we have considered the general case where $\Delta^{(s )}_\lambda$ is replaced by $\Delta_\lambda$ in the second term and in the $k$ part of the fourth term, while the $1 - k$ part of this term remains with $\Delta^{(s )}_\lambda$ (a priori, $k$ can be any real number, including, for example, $k > 1$). We found that it is precisely for $k = 1$ that gauge fixing in the form of such a principal value type prescription does work since the considered nonphysical poles of each term in the propagator are located on one side of the integration path $\Im p_0 = 0$ (whereas for $k \neq 1$ this path is clamped between the poles of the same term) and, for small quasi-momenta, smoothly passes to such regularization in the continuum case, when quasi-momenta become momenta. Thus, we take the value of $k$ to be 1 when $\Delta^{(s )}_\lambda$ is replaced by $\Delta_\lambda$ in both the second and fourth terms, and the action is
\begin{eqnarray}\label{cSg}                                                
& & \cS_{\rm g} = \frac{1}{ 8 } \sum_{\rm sites} g^{\lambda \rho} g^{\mu \sigma} g^{\nu \tau} \left[ 2 \left( \Delta^{(s )}_\lambda g_{\mu \nu} \right) \left( \Delta^{(s )}_\tau g_{\sigma \rho} \right) - \left( \Delta_\nu g_{\mu \lambda} \right) \left( \Delta_\tau g_{\sigma \rho} \right) \right. \nonumber \\ & & \left. - 2 \left( \Delta^{(s )}_\lambda g_{\mu \rho} \right) \left( \Delta^{(s )}_\sigma g_{\nu \tau} \right) + \left( \Delta_\nu g_{\lambda \rho} \right) \left( \Delta_\tau g_{\mu \sigma} \right) \right] \sqrt{- g} .
\end{eqnarray}

\subsection{Functional measure}\label{measure}

We issue from the system of area tensors and SO(3,1) connection matrices assigned to the 3D faces between 4-simplices. As usual in a discrete system, the Jacobian of the Poisson brackets of the constraints is singular at zero fields, in this case at a flat metric. This is typical for discrete gravity due to the lack of diffeomorphism invariance, as is reviewed, eg, in Ref.~\cite{Loll}. To correctly approach the quantization of this discrete system, we can start from the extended configuration superspace of independent area tensors $v^{ab}_{\lambda\mu}$. In this superspace, the system can be quantized canonically. The physical hypersurface of interest is singled out by some conditions ensuring existence of certain edge vectors spanning the given area tensors in each 4-simplex. Then we can project the functional measure from the extended superspace to the physical hypersurface. We take this projection to be defined up to a factor of the type of volume to some power $V^\eta$. The projection is performed by inserting some $\delta$-function factor $\delta_{\rm metric}$ under the functional integral sign which enforces the conditions at each site which single out the physical hypersurface in the superspace,
\begin{equation}\label{delta(vv)}                                          
\delta_{\rm metric} ( v ) = \int V^\eta \delta^{21} \left ( \epsilon_{abcd}v^{ab}_{\lambda\mu}v^{cd}_{\nu\rho}
- V \epsilon_{\lambda\mu\nu\rho} \right ) \d V .
\end{equation}

\noindent This reduces the measure over independent area tensors to the product of
\begin{equation}\label{int-d36v-delta-metric}                              
\int \d^{36} v^{a b}_{\lambda \mu} \delta_{\rm metric} ( v ) \sim ( - g )^\frac{ \eta - 7 }{ 2 } \d^{10} g_{\lambda \mu}
\end{equation}

\noindent over the 4-cubes or sites.

Besides that, in the connection sector, we have the invariant (Haar) measure. Integration over the discrete connection is an analogue of the Gaussian integration over continuum connection, which allows to pass from the functional integral with the Palatini form of the action to the functional integral with the Einstein-Hilbert action in terms of purely metric variables. The situation simplifies when the discrete analogues of the ADM lapse-shift functions have a small scale and their contribution in the action can be disregarded. This is exactly what is favored by the synchronous gauge we are considering in this paper, when we fix the scale of discrete ADM lapse-shift functions at some low level. In this case, the integral over connection factorizes over spatial 2-simplices (triangles). Each hypercube per site consisting of 24 4-simplices contains 6 spatial triangles of which essentially different in our hypercubic model are 3 areas with the area vectors $\bv_\alpha \equiv \epsilon_\alpha{}^{\beta \gamma} \bv_{\beta \gamma} / 2$ ($\alpha , \beta , \gamma , ... = 1, 2, 3$) corresponding to 3 faces of the spatial 3-cubic face. As a result of integration over connections, we obtain \cite{Kha3} the factor
\begin{equation}\label{N0}                                                 
\N_0 (\rv ) = \left | \frac{1}{\frac{1}{4} \left ( \frac{1}{\gamma} - i \right )^2 \rv^2 + 1} \frac{\frac{1}{4} \left ( \frac{1}{\gamma} - i \right ) \rv}{ \sh \left [ \frac{\pi}{2} \left ( \frac{1}{\gamma} - i \right ) \rv \right ]} \right |^2
\end{equation}

\noindent per spatial triangle (with area vector $\bv$), which should be introduced into the above measure (\ref{int-d36v-delta-metric}). Here $\rv = \sqrt{\bv^2}$. There we have also considered master integrals with the $n$-th order monomials of curvature variables (generators) inserted, and these behave as $| \bv |^{- n}$ relative to $\N_0$ at large $| \bv |$ and define the $n$-th order correction $\sim (|\btau | |\bv |^{- 1})^n$. This reveals the meaning of the term "low level", at which we should fix the scale of discrete ADM lapse-shift functions as the relation
\begin{equation}\label{t<<v}                                               
|\btau | \ll | \bv | \mbox{ or } b_{\rm t} \ll b_{\rm s}
\end{equation}

\noindent between the $t$-like $|\btau |$ and leaf/diagonal $| \bv |$ triangle area scales or the $t$-like $b_{\rm t}$ and leaf/diago\-nal $b_{\rm s}$ edge length scales.

The resulting measure reads
\begin{equation}\label{N0-v-alpha}                                         
\prod_{\rm sites} \left ( \prod_\alpha \N_0 (2 \rv_\alpha ) \right ) ( - g )^\frac{ \eta - 7 }{ 2 } \d^{10} g_{\lambda \mu}.
\end{equation}

\subsection{Gauge-fixing term}\label{gauge}

Despite that the gauge (diffeomorphism) symmetry is violated by discreteness, there are degrees of freedom close to the gauge ones of the continuum theory, when the metric variations from 4-simplex to 4-simplex are small. To eliminate a set of physically almost equivalent configurations of infinite measure from the functional integral, we can impose a gauge which means that we integrate only over a subset of all the configurations in the superspace. In contrast to the continuum theory, where the gauge symmetry is exact, in the discrete theory the result of calculating amplitudes generally depends on the gauge. For estimates, we can use a certain gauge. A certain averaging over all gauges with a certain weight factor such that it is equivalent to functional integration over all (nonintersecting) subsets must give the functional integral over the total configuration superspace and final value for the considered amplitude.

To fix the gauge, we can also introduce an averaging factor into the functional integral, which is equivalent to adding a gauge-fixing term to the action. Such a term for averaging around the synchronous gauge would be proportional to $(\rnu^\mu w_{\mu \lambda}) \rlambda^{\lambda \sigma } (\rnu^\tau w_{\tau \sigma})$ per site where $\rnu^\lambda = (1, 0, 0, 0)$. There are the gauge singularities at $p_0 = \rnu p = 0$. To "soften" the gauge singularities for the Yang-Mills field in the analogous temporal gauge with the gauge-fixing term $(\rnu^\lambda A^a_\lambda )^2$, it is proposed in Ref. \cite{Ste} to replace $\rnu^\lambda$ here by $n^\lambda$, where $n^\lambda$ is some differential operator infinitely close to $\rnu^\lambda$. This confirms the earlier proposed prescription by Landshoff \cite{Land} that involves the replacement $p_0^{- 2} \Rightarrow (p_0^2 + \varepsilon^2)^{- 1}, \varepsilon \to 0$ in the propagator. We aim at a prescription of the type of a principal value prescription (not in the standard Cauchy sense), which amounts, roughly speaking, to replacing singularities similarly to (the discrete analogue of) $p_0^{-j} \Rightarrow [ (p_0 + i \varepsilon )^{-j} + (p_0 - i \varepsilon )^{-j} ] / 2$. The corresponding gauge-fixing term is being constructed in two steps. In the first step we consider the gravity action with the synchronous gauge-fixing term "softened" by the replacement $\rnu^\lambda \Rightarrow n^\lambda$. Also with a source term it takes the form
\begin{eqnarray}\label{cS_g'}                                              
& & \cS_{\rm g}^\prime [ g , J ] = \cS_{\rm g} - \sum_{\rm sites} \left[ J^{\lambda \mu} w_{\lambda \mu} + \frac{1}{4} (n^\mu w_{\mu \lambda}) \rlambda^{\lambda \sigma } (n^\tau w_{\tau \sigma}) \right] , \quad g_{\lambda \mu} = g^{(0 )}_{\lambda \mu} + w_{\lambda \mu} , \nonumber \\ & & w_{\lambda \mu} = \cG_{\lambda \mu \sigma \tau} ( n, \on ) J^{\sigma \tau} , \quad  ( \| \rlambda^{\lambda \mu} \|^{- 1} )_{\sigma \tau} \stackrel{\rm def}{=} \ralpha_{\sigma \tau} , \quad \eta_{\lambda \mu } = {\rm diag} (-1, 1, 1, 1) , \nonumber \\ & & n^\lambda = \rnu^\lambda - \varepsilon \frac{ \Delta^{(s ) \lambda}_\perp }{ \Delta^{(s ) 2}_\perp } , \quad \Delta^{(s ) \lambda}_\perp = \Delta^{(s ) \lambda} - \frac{ \rnu^\lambda }{ \rnu^2 } (\rnu \Delta^{(s )} ) .
\end{eqnarray}

\noindent Assuming that spacetime is asymptotically flat, we can set
\begin{equation}                                                           
g^{(0 )}_{\lambda \mu} = g_{\lambda \mu} ( x ) |_{x \to \infty}.
\end{equation}

The above choice of the discrete derivative as directly the finite difference $\Delta^{(s )}_\lambda$ or $\Delta_\lambda$ fixes the coordinate differences between the sites $\Delta x^\lambda = 1$. Then $g^{(0 )}_{\lambda \mu}$ is defined just by the elementary edge lengths, spacelike $b_{\rm s}$ and timelike $b_{\rm t}$ ones, and follows partially (for spatial directions) from the maximization condition for the measure and partially (for temporal and temporal-spatial directions) via fixing the discrete analogues of the ADM lapse-shift functions as a gauge in the considered strategy. The corresponding background metric reads
\begin{eqnarray}\label{g^(0)l^(0)}                                         
& & \hspace{-10mm} g^{(0 )}_{\lambda \mu} = {\rm diag} (-b_{\rm t}^2, b_{\rm s}^2, b_{\rm s}^2, b_{\rm s}^2) = l^{(0) a}_\lambda \eta_{a b} l^{(0) b}_\mu , \quad l^{(0) a}_\lambda = {\rm diag} (b_{\rm t}, b_{\rm s}, b_{\rm s}, b_{\rm s}) , \quad l^{(0) a}_\mu l^{(0) \lambda}_a \equiv \delta^\lambda_\mu .
\end{eqnarray}

For the considered $g^{(0 )}_{\lambda \mu} \neq \eta_{\lambda \mu } = {\rm diag} (-1, 1, 1, 1)$ we can pass to the scaled metric tensor variable $\tg_{a b} = l^{(0) \lambda}_a g_{\lambda \mu} l^{(0) \mu}_b$, which has $\tg^{(0 )}_{a b} = \eta_{a b}$. The propagator of the type $\cG_{\lambda \mu \sigma \tau} ( n, \on )$ is a function of the initial metric $g^{(0 )}_{\lambda \mu}$, the set of the finite differences $\Delta^{(s )}_\lambda$ and $\Delta^{(s )}_\lambda$ and the gauge parameters $\ralpha_{\lambda \mu}$ and $n^\lambda$. We write this as $\cG_{\lambda \mu \sigma \tau} ( n, \on | \{ g^{(0 )}_{\lambda \mu} \} , \{ \Delta^{(s )}_\lambda \} , \{ \Delta_\lambda \} , \{ \ralpha_{\lambda \mu} \} , \{ n^\lambda \} )$. In terms of $\tg_{a b}$, the action and hence the corresponding propagator contain the corresponding scaled finite differences and gauge parameters. Besides that, the bilinear form of the action in terms of $\tg_{a b}$ has the constant factor $( - g^{(0 )})^{1 / 2}$; to bring the action plus the gauge-fixing term into a familiar form, we can either additionally include $( - g^{(0 )})^{1 / 4}$ in the scaled finite differences or include $( - g^{(0 )})^{1 / 2}$ in the scaled $\ralpha_{\lambda \mu}$ thus making $( - g^{(0 )})^{1 / 2}$ an overall factor for the action summed with the gauge-fixing term. In the latter case, the latter factor results in the overall factor $( - g^{(0 )})^{- 1 / 2}$ for the propagator. We then inversely scale the propagator components by $l^{(0) a}_\lambda$ to return to the original coordinates,
\begin{eqnarray}                                                        
& & \cG_{\lambda \mu \sigma \tau} ( n, \on | \{ g^{(0 )}_{\lambda \mu} \} , \{ \Delta^{(s )}_\lambda \} , \{ \Delta_\lambda \} , \{ \ralpha_{\lambda \mu} \} , \{ n^\lambda \} ) \nonumber \\ & & = l^{(0) a}_\lambda l^{(0) b}_\mu l^{(0) e}_\sigma l^{(0) f}_\tau \cG_{a b e f} ( n, \on | \{ \eta_{a b} \} , \{ ( - g^{(0 )})^{1 / 4} l^{(0) \lambda}_a \Delta^{(s )}_\lambda \} , \{ ( - g^{(0 )})^{1 / 4} l^{(0) \lambda}_a \Delta_\lambda \} , \nonumber \\ & & \{ l^{(0) \lambda}_a \ralpha_{\lambda \mu} l^{(0) \mu}_b \} , \{ l^{(0) a}_\lambda n^\lambda \} ) \\ & & = l^{(0) a}_\lambda l^{(0) b}_\mu l^{(0) e}_\sigma l^{(0) f}_\tau \cG_{a b e f} ( n, \on | \{ \eta_{a b} \} , \{ l^{(0) \lambda}_a \Delta^{(s )}_\lambda \} , \{ l^{(0) \lambda}_a \Delta_\lambda \} , \nonumber \\ & & \{ ( - g^{(0 )})^{1 / 2} l^{(0) \lambda}_a \ralpha_{\lambda \mu} l^{(0) \mu}_b \} , \{ l^{(0) a}_\lambda n^\lambda \} ) ( - g^{(0 )})^{- 1 / 2} .
\end{eqnarray}

\noindent Thus, $g^{(0 )}_{\lambda \mu} \neq \eta_{\lambda \mu }$ can be taken into account by scaling the finite differences and gauge parameters. The scaled finite differences are true discrete derivatives $l^{(0) \lambda}_a \Delta^{(s )}_\lambda = (b^{- 1}_{\rm t} \Delta^{(s )}_0, b^{- 1}_{\rm s} \bDelta^{(s )} )$, $l^{(0) \lambda}_a \Delta_\lambda = (b^{- 1}_{\rm t} \Delta_0, b^{- 1}_{\rm s} \bDelta )$. To avoid an excessive bulkiness, we use for illustration the case $g^{(0 )}_{\lambda \mu} = \eta_{\lambda \mu }$, introducing $b_{\rm t} \neq 1$ and/or $b_{\rm s} \neq 1$ only if necessary.

In (\ref{cS_g'}), $\cG_{\lambda \mu \sigma \tau} ( n, \on )$ is the propagator, "softened" by replacements of the type $p_0^{- 2} \Rightarrow (p_0^2 + \varepsilon^2)^{- 1}, \varepsilon \to 0$. But we are interested in the non-Hermitian operators $\cG_{\lambda \mu \sigma \tau} ( n, n )$ and $\cG_{\lambda \mu \sigma \tau} ( \on, \on )$ formally obtained from $\cG_{\lambda \mu \sigma \tau} ( n, \on )$ by the replacements $\on \Rightarrow n$ or $n \Rightarrow \on$ or from the original "hard" synchronous gauge propagator $\cG_{\lambda \mu \sigma \tau} ( \rnu, \rnu )$ by analytic continuation (in the momentum representation) from $\rnu$ to $n$ or $\on$. In the second step we form a Hermitian operator,
\begin{equation}\label{Gnn+Gn*n*}                                          
\cG_{\lambda \mu \sigma \tau} = \frac{1}{2} \cG_{\lambda \mu \sigma \tau} ( n, n ) + \frac{1}{2} \cG_{\lambda \mu \sigma \tau} ( \on, \on ) .
\end{equation}

\noindent Up to terms that vanish as $\varepsilon$ tends to zero, the prescription (\ref{Gnn+Gn*n*}) amounts to replacing singularities like (the discrete analogue of) $p_0^{-j} \Rightarrow [ (p_0 + i \varepsilon )^{-j} + (p_0 - i \varepsilon )^{-j} ] / 2$, but more importantly, the poles in each term in (\ref{Gnn+Gn*n*}) are on one side of the integration path $\Im p_0 = 0$, unlike $\cG_{\lambda \mu \sigma \tau} ( n, \on )$, where this path is clamped between the poles. Therefore, the integration path can be deformed so that it would pass at distances $O(1)$ from all singularities, and the result of integration over $\d p_0$ will be $O(1)$.

In the actual discrete system, the nonphysical singularities themselves (for the "hard" synchronous gauge, $n = \rnu$) and their softening (introducing $\varepsilon \neq 0$) look like
\begin{eqnarray}\label{p0^2-pk^4}                                          
\left[ \sin^2 \frac{p_0}{2} - \frac{b_{\rm t}^2}{b_{\rm s}^2} \sum_{\alpha = 1}^3 \sin^4 \frac{ p_\alpha }{ 2 } + O ( \varepsilon^2 ) \right]^{- j} \Rightarrow \frac{ 1 }{ 2 } \left[ \sin^2 \frac{p_0 + i \varepsilon }{2} - \frac{b_{\rm t}^2}{b_{\rm s}^2} \sum_{\alpha = 1}^3 \sin^4 \frac{ p_\alpha }{ 2 } + O ( \varepsilon^2 ) \right]^{- j} & & \nonumber \\ + \frac{ 1 }{ 2 } \left[ \sin^2 \frac{p_0 - i \varepsilon }{2} - \frac{b_{\rm t}^2}{b_{\rm s}^2} \sum_{\alpha = 1}^3 \sin^4 \frac{ p_\alpha }{ 2 } + O ( \varepsilon^2 ) \right]^{- j} . & &
\end{eqnarray}

\noindent If a continuum diagram is convergent, it is contributed by non-Planckian momenta (for habitual external momenta or distances much larger than the typical length scale). The powers of the corresponding quasi-momenta are suppressed as powers of the typical length scale. In particular, $p_\lambda$ in (\ref{p0^2-pk^4}) are small and $p_l^4$ can be neglected. Thus we smoothly pass to the prescription for $p_0^{- 2 j}$ in the continuum case.

These are not the only singular terms in the propagator, these can be multiplied by the singular factor $( \rnu \Delta^{(s )} )^{- 1}$, which, e. g., in $\cG ( n, n )$ corresponds to $( n \Delta^{(s )} )^{- 1} = ( \rnu \Delta^{(s )} - \varepsilon )^{- 1} \propto ( \sin p_0 + i \varepsilon )^{- 1}$. Its poles are located at $p_0 \approx - i \varepsilon$ and $p_0 \approx \pm \pi + i \varepsilon$. Despite the number of poles of this factor two per period is twice as large as that of the corresponding continuum factor $( p_0 + i \varepsilon )^{- 1}$, these poles are located on different sides of the integration path, so only one of them is effective, and the above-mentioned doubling of the integration result compared to the case of the continuum is absent.

In the leading order over metric variations from site to site or for small quasi-momenta (we can limit ourselves to such momenta if, say, the continuum counterpart of the calculated diagram converges, and the external fields have habitual non-Planckian momenta), in the propagator $\cG$ we can keep only the actual denominator $\oDelta \Delta$ (more accurately, $\oDelta \Delta - i 0$) and set $\oDelta \Delta \approx - \Delta^{(s ) 2}$ in the rest of the propagator, where this approximation is not crucial. One can see that this amounts to calculating the propagator $G$ for the ordinary finite-difference gravity action $\tilS_{\rm g}$ (\ref{Sg}) with symmetrized finite-difference derivatives $\Delta^{(s )}$ and then replacing the denominator $\Delta^{(s ) 2}$ in it with $- \oDelta \Delta$, which gives some effective propagator $G^{\rm eff}$,
\begin{eqnarray}\label{Geff}                                               
& & \frac{1}{2} G^{\rm eff}_{\lambda \mu \sigma \tau} ( n, n ) = \frac{- 1}{\oDelta \Delta } [ L_{\lambda \sigma} ( n, n ) L_{\mu \tau} ( n, n ) + L_{\mu \sigma} ( n, n ) L_{\lambda \tau} ( n, n ) - L_{\lambda \mu} ( n, n ) L_{\sigma \tau} ( n, n ) ] \nonumber \\ & & - \frac{ ( \ralpha_{\lambda \sigma} \Delta^{(s) }_\tau + \ralpha_{\lambda \tau} \Delta^{(s) }_\sigma) \Delta^{(s) }_\mu + ( \ralpha_{\mu \sigma} \Delta^{(s) }_\tau + \ralpha_{\mu \tau} \Delta^{(s) }_\sigma) \Delta^{(s) }_\lambda }{(n \Delta^{(s) })^2} + \Delta^{(s) }_\lambda \Delta^{(s) }_\mu \frac{ n^\nu \ralpha_{\nu \sigma} \Delta^{(s) }_\tau + n^\nu \ralpha_{\nu \tau} \Delta^{(s) }_\sigma }{(n \Delta^{(s) })^3 } \nonumber \\ & & + \frac{ \ralpha_{\lambda \nu} n^\nu \Delta^{(s) }_\mu + \ralpha_{\mu \nu} n^\nu \Delta^{(s) }_\lambda }{(n \Delta^{(s) })^3 } \Delta^{(s) }_\sigma \Delta^{(s) }_\tau - \frac{ n^\nu \ralpha_{\nu \rho} n^\rho }{(n \Delta^{(s) })^4 } \Delta^{(s) }_\lambda \Delta^{(s) }_\mu \Delta^{(s) }_\sigma \Delta^{(s) }_\tau , \nonumber \\ & & L_{\lambda \mu} ( n, n ) \stackrel{\rm def }{=} \eta_{\lambda \mu} - \Delta^{(s) }_\lambda \frac{n_\mu }{ n \Delta^{(s) } } - \frac{n_\lambda }{ n \Delta^{(s) } } \Delta^{(s) }_\mu + \frac{n^2 \Delta^{(s) }_\lambda \Delta^{(s) }_\mu}{(n \Delta^{(s) } )^2} ,
\end{eqnarray}

\noindent (for $g^{(0 )}_{\lambda \mu} = \eta_{\lambda \mu }$) and analogously for $G^{\rm eff} ( \on, \on )$ by replacing $n \to \on$. Now the only nonphysical singularities in $G^{\rm eff(0)} = G^{\rm eff } (\rnu , \rnu )$ are $( \rnu \Delta^{(s )} )^{- j}$, and the factors $( n \Delta^{(s )} )^{- j}$ in terms in $G^{\rm eff} ( n, n )$ or $( \on \Delta^{(s )} )^{- j}$ in $G^{\rm eff} ( \on, \on )$ regularize them.

We can express $\cG$ (\ref{Gnn+Gn*n*}) in terms of Hermitian propagators and restore the bilinear form $\cG^{- 1}$ from which $\cG$ follows, $\cG^{- 1} = \ccM + \Delta \ccM$, where $\ccM$ gives the bilinear form of the action and $\Delta \ccM$ gives the gauge-fixing term required for the principal value type propagator,
\begin{eqnarray}\label{ccF}                                                
& & \hspace{-10mm} \ccF [ g ] = \frac{1}{2} \sum_{\rm sites} w_{\lambda \mu} \Delta \ccM^{\lambda \mu \sigma \tau} w_{\sigma \tau} = \sum_{\rm sites} \left[ - \frac{1}{4 \varepsilon^2 } \cmf_\rho [ g ] \left( \cmM^{- 1} \right)^{\rho \kappa} \cmf_\kappa [ g ] + \frac{1}{2} w_{\lambda \mu} \mm^{\lambda \mu \sigma \tau} w_{\sigma \tau} \right] , \nonumber \\ & & \hspace{-10mm} \cmf_\rho = \crO^{\lambda \mu}_\rho w_{\lambda \mu} , ~ \crO^{\lambda \mu}_\rho = \delta^\lambda_\rho \rnu^\mu + \frac{ \varepsilon^2 }{ 2 } \frac{ \Delta^{( s ) \nu }_\perp }{ \Delta^{( s ) 2 }_\perp } \cG^{(0)}_{\rho \nu \pi \zeta} \rnu^\pi \rlambda^{\zeta \lambda} \frac{ \Delta^{( s ) \mu }_\perp }{ \Delta^{( s ) 2 }_\perp } , \nonumber \\ & & \hspace{-10mm} \mm^{\lambda \mu \sigma \tau} = - \frac{ \varepsilon^2 }{2} \frac{ \Delta_{ \perp}^{(s) \lambda } }{ \Delta_{ \perp}^{(s) 2 } } \left( \rlambda^{\mu \sigma} + \frac{1}{2} \rlambda^{\mu \zeta } \rnu^\pi \cG^{(0)}_{\zeta \pi \chi \psi} \rnu^\chi \rlambda^{\psi \sigma} \right) \frac{ \Delta_{ \perp}^{(s) \tau } }{ \Delta_{ \perp}^{(s) 2 } } , ~ \cmM_{\lambda \tau} = \frac{\ralpha_{\lambda \tau}}{\varepsilon^2 }  - \frac{1}{2} \frac{ \Delta_{ \perp}^{(s) \mu } }{ \Delta_{ \perp}^{(s) 2 } } \cG^{(0)}_{\lambda \mu \sigma \tau} \frac{ \Delta_{ \perp}^{(s) \sigma } }{ \Delta_{ \perp}^{(s) 2 } }, \nonumber \\ & & \hspace{-10mm} \cG^{(0)}_{\zeta \pi \chi \psi} = \left[ \left\| \ccM^{\nu \rho \kappa \varphi} - \frac{1}{2} \rnu^{( \nu } \rlambda^{\rho ) ( \kappa } \rnu^{\varphi )} \right\|^{-1} \right]_{\zeta \pi \chi \psi} = \cG_{\zeta \pi \chi \psi} ( \rnu , \rnu ) .
\end{eqnarray}

If we use the effective (in the leading order over metric variations from site to site) propagator $G^{\rm eff}$ (\ref{Geff}), we can write a similar expression for the corresponding gauge-fixing term $\cF [ g ]$ by replacing in (\ref{ccF}) $\cG^{(0 )}$ by $G^{\rm eff (0)}$ and most of the symbols marked with a check mark at the top by unmarked symbols, $\ccF \Rightarrow \cF$, $\crO \Rightarrow \rO $, $\cmM \Rightarrow \mM $, $\cmf \Rightarrow \mf $, and taking into account that in this case $\mm = 0$.

$\ccF [ g ]$ (\ref{ccF}) (or $\cF [ g ]$) uses a "hard" synchronous gauge propagator $\cG^{(0)}$ (or $G^{\rm eff (0)}$), which is singular; therefore, we should check whether the resulting gauge-fixing term suffers from these singularities. In fact, we find that the gauge-fixing term is defined in a finite way.

The action in the continuum GR is invariant with respect to the diffeomorphism group $\Xi$ generated by infinitesimal coordinate transformations $\delta x^\lambda = \xi^\lambda ( x )$, under which the metric $g_{\lambda \mu}$ is transformed as
\begin{equation}                                                           
(g_{\lambda \mu})^\Xi - g_{\lambda \mu} = \delta g_{\lambda \mu} = - g_{\lambda \nu } \partial_\mu \xi^\nu - g_{\mu \nu } \partial_\lambda \xi^\nu - \xi^\nu \partial_\nu g_{\lambda \mu } .
\end{equation}

\noindent If we consider the RC action, obtained as a phase in zero order in the functional integral via the functional integration over connection, the RC action in the leading order over metric variations from site to site is a finite-difference form of the continuum Einstein action \cite{our2}. It is described in terms of $g_{\lambda \mu}$ at the sites only in this order, in non-leading ones, it depends also on additional lattice-specific simplicial components. In the considered hypercubic model with the connection excluded, a finite-difference form of the continuum GR action arise also only in the leading order over metric variations. Therefore, the transformations of the metric should be implied in this order.

The found $\exp ( i \ccF [ g ] )$ can be treated as a weight functional factor introduced to separate out degrees of freedom close to the gauge degrees of freedom of the continuum theory (to cancel infinite integration over (approximate) gauge group $\Xi$ in the functional integral). With such treatment and target transformations of the metric in the leading order over metric variations, the corresponding normalization factor $\cPhi [ g ]$ should be introduced simultaneously,
\begin{eqnarray}\label{1/Phi=int-exp-F}                                    
& & \cPhi [ g ]^{- 1} = \int \exp{( i \ccF [ g^\Xi ] )} \prod_{\rm sites} \d \Xi , \quad \d \Xi = \prod_\lambda \d \xi^\lambda , \nonumber \\ & & \delta^\Xi g_{\lambda \mu} = g^\Xi_{\lambda \mu} - g_{\lambda \mu} = - \Delta^{(s) }_\mu \xi_\lambda - \Delta^{(s) }_\lambda \xi_\mu + 2 \Gamma^\nu_{\lambda \mu} \xi_\nu + O ( (\xi )^2 ), \nonumber \\ & & \Gamma^\nu_{\lambda \mu} = \frac{1}{2} g^{\nu \rho} ( \Delta^{(s )}_\mu g_{\rho \lambda} + \Delta^{(s )}_\lambda g_{\rho \mu} - \Delta^{(s )}_\rho g_{\lambda \mu} ) , \nonumber \\ & & \ccO_\rho{}^\nu \xi_\nu \stackrel{\rm def }{ = } \crO^{\lambda \mu}_\rho \left( \Delta^{(s) }_\mu \xi_\lambda + \Delta^{(s) }_\lambda \xi_\mu - 2 \Gamma^\nu_{\lambda \mu} \xi_\nu \right) .
\end{eqnarray}

\noindent The term $O ( (\xi )^2 )$ in $\delta^\Xi g_{\lambda \mu}$ means that a priori the transformations in (\ref{1/Phi=int-exp-F}) are not necessarily infinitesimal.

If a contribution to the effective ghost action $-i \ln \cPhi$ turns out to be of order $O( \varepsilon^2 )$ at $\varepsilon \to 0$, it can be disregarded in the limit $\varepsilon \to 0$ (the subtlety lies in the need for intermediate regularization of the estimate of a certain integral value of the type of a certain action by introducing a finite number of points in the temporal direction $N$ depending on $\varepsilon$, as we consider in \cite{khat1}, so it is essential that the contribution be of order $O( \varepsilon^2 )$ and not just $O( \varepsilon )$).

Estimating $\cPhi$ as $\varepsilon$ tends to 0, we find that we can neglect $\mm = O ( \varepsilon^2 )$, limit ourselves to the infinitesimal $\xi$ and arrive at an expression for $\cPhi$ equal to $\Det \ccO$.

We can also use the above effective (in the leading order over metric variations from site to site) propagator $G^{\rm eff}$ and the corresponding gauge-fixing term $\cF [ g ]$ and write a similar expression for the corresponding normalization factor $\Phi [ g ]$. Along with the above replacement of the symbols marked with a check mark at the top with the unmarked symbols, we also replace $\ccO$ with $\cO$. Now $\mm = 0$, so we immediately get $\Phi = \Det \cO$.

$\ccO$ (\ref{1/Phi=int-exp-F}) (or $\cO$) depend on $\cG^{(0)}$ (or $G^{\rm eff (0)}$) with coefficients $\propto \varepsilon^2$, and these terms are needed for regularization in the limit $\varepsilon \to 0$. We find that both $\cPhi$ and $\Phi$ are 1 (or unessential constants) in this limit.

Of interest is also generalization of the gauge-fixing term $\ccF [ g ]$ or $\cF [ g ]$ to bilinear one in $( - g )^\ialpha w_{\lambda \mu}$ by substitution $w_{\lambda \mu} \Rightarrow ( - g )^\ialpha w_{\lambda \mu}$ in it. For $\ialpha = - 1 / 8$, it can be introduced in order to preserve the scaling property of the action (like $g_{\lambda \mu}$ or [length]$^2$) when $\ccF$ or $\cF$ is added to it. This will play its role in the estimation of the typical edge length scale using the maximization condition which uses $\det \| \partial^2 S / \partial l_i \partial l_k \| $. We have found in \cite{khat1} that the normalization factor $\cPhi$ or $\Phi$ tends to $\prod_\mathrm{sites} ( - g)^{4 \ialpha}$ when $\varepsilon$ tends to 0, the relative corrections being $O ( \varepsilon^2 )$.

Thus, the functional integral (written here as a functional $J ( \cdot )$ on functionals of the metric to be averaged) takes the form
\begin{eqnarray}\label{funct-int-full}                                     
& & \int \exp \left\{ i \left[ \cS_{\rm g} [ g ] + \frac{1}{2} \sum_{\rm sites} w_{\lambda \mu} ( - g )^{- \frac{1}{8}} \Delta \ccM^{\lambda \mu \sigma \tau} ( - g )^{- \frac{1}{8}} \right. \right. \nonumber \\ & & \left. \left. \vphantom{\frac{1}{2} \sum_{\rm sites}} \cdot w_{\sigma \tau} \right] \right\} ( \cdot ) \prod_{\rm sites} \left ( \prod_\alpha \N_0 (2 \rv_\alpha ) \right ) ( - g )^\frac{ \eta - 8 }{ 2 } \d^{10} g_{\lambda \mu} .
\end{eqnarray}

\noindent Here we have taken unto account that the ghost contribution amounts to the normalization factor $\prod_\mathrm{sites} ( - g)^{- 1 / 2}$ up to relative corrections, which are of the order of $O ( \varepsilon^2 )$ and are thus omitted; this factor changes the measure (\ref{N0-v-alpha}) by adding -1 to $\eta$ there.

\section{Constructing the perturbative expansion}\label{expansion}

\subsection{Transforming the measure}\label{transform}

In the functional measure, we single out the dependence on the spatial triangle (quadrangle) areas, whose scales are just fixed by maximizing the measure; these areas are taken as new variables,
\begin{equation}\label{m=f(g)}                                             
i \rmm_1 = 2 \rv_1 = i \sqrt{ g_{2 2} g_{3 3} - g^2_{2 3}}, ~ 2 ~ \mbox{perm}(123) .
\end{equation}

\noindent Then the factor in the measure (\ref{N0}) reads
\begin{eqnarray}\label{N0-exact}                                           
& & \hspace{0mm} \N_0 (i\rmm_\alpha ) = \frac{\gamma^2}{\gamma^2 + 1} \left | \frac{\rmm^2_\alpha}{\rmm^2_\alpha + 4 \left ( 1 + \frac{i}{\gamma} \right )^{-2} } \right |^2 \left | \frac{2}{1 - \exp \left [ - \pi \left ( 1 + \frac{i}{\gamma} \right ) \rmm_\alpha \right ] } \right |^2 \frac{ e^{ - \pi \rmm_\alpha }}{ \rmm^2_\alpha } .
\end{eqnarray}

\noindent For $\rmm_\alpha$ large compared to 1,
\begin{equation}\label{prod-N-0}                                           
\prod_\alpha \N_0 (i\rmm_\alpha ) \propto \prod_\alpha \frac{1}{ \rmm^2_\alpha } \exp \left ( - \pi \rmm_\alpha \right ) .
\end{equation}

Expressing the determinant $g$ in terms of the determinant $\rgamma = \det \| \rgamma_{\alpha \beta} \|$ of the 3-dimensional metric $\rgamma_{\alpha \beta} = g_{\alpha \beta}$, we have for the volume degree type factor in the measure:
\begin{eqnarray}\label{(-g)^(e/2)}                                         
& & \hspace{-10mm} ( - g )^\frac{\eta - 8}{2} = \left( - g_{0 0}^{(0 )} - w_{0 0} + w_{0 \alpha} \rgamma^{\alpha \beta} w_{\beta 0} \right)^\frac{\eta - 8}{2} \rgamma^\frac{\eta - 8}{2} = \left[ - g_{0 0}^{(0 )} + O ( \varepsilon^2 ) \right]^\frac{\eta - 8}{2} \rgamma^\frac{\eta - 8}{2} \propto \rgamma^\frac{\eta - 8}{2}
\end{eqnarray}

\noindent up to relative corrections of the order of $O ( \varepsilon^2 )$, since, as we consider in \cite{khat1}, $w_{0 0} = O ( \varepsilon^2 )$, $w_{0 \alpha} = O ( \varepsilon )$ in the sense of their correlators with any components of the metric. Here $\rgamma^{\alpha \beta}$ is the reciprocal to $\rgamma_{\alpha \beta}$.

The remaining three variables, besides $g_{0 \lambda}$, are chosen to be of angular type, namely, the cosines of the angles between the spatial axes. The latter are three reduced to unit scales along the axes ("physical") non-diagonal metric components
\begin{equation}                                                           
g_{\halpha \no \hbeta} : g_{\halpha \hbeta} = |g_{\alpha \alpha}|^{- 1 / 2} g_{\alpha \beta}|g_{\beta \beta}|^{- 1 / 2} , \alpha \no \beta .
\end{equation}

\noindent We can decompose $\rgamma$ (and the integration element $\d^6 g_{\alpha \beta}$) into factors that depend on $\rmm_\alpha$ and on these variables. It is simpler to pass first from the set $\{ g_{\alpha \beta} \}$ to $\{ g_{\alpha \alpha} \}$ and $\{ g_{\halpha \no \hbeta} \}$, then express $\{ g_{\alpha \alpha} \}$ in terms of $\{ \rmm_\alpha \}$ and $\{ g_{\halpha \no \hbeta} \}$. Thus we have
\begin{equation}\label{rgamma}                                             
\rgamma = \det \| g_{\halpha \hbeta} \| \prod_{\alpha \no \beta} \left( 1 - g_{\halpha \hbeta}^2 \right)^{- 1 / 2} \prod_\alpha \rmm_\alpha .
\end{equation}

\noindent Here we imply $g_{\halpha \halpha} \equiv 1$ ($\| g_{\halpha \hbeta} \|$ is a matrix with the unit diagonal).

For the integration element, we obtain
\begin{eqnarray}\label{d^(10)g_**}                                         
& & \d^{10} g_{\lambda \mu} = \d^4 g_{0 \lambda} \prod_{\alpha \no \beta} \left( 1 - g_{\halpha \hbeta}^2 \right)^{- 1} \left( \prod_\alpha \rmm_\alpha \right) \d^3 g_{\halpha \no \hbeta} \d^3 \rmm_\alpha .
\end{eqnarray}

Taking into account (\ref{prod-N-0}), (\ref{(-g)^(e/2)}), (\ref{rgamma}), (\ref{d^(10)g_**}), we obtain the functional measure (\ref{N0-v-alpha}) in terms of the variables $\{ g_{0 \lambda} \}$, $\{ g_{\halpha \no \hbeta} \}$, $\{ \rmm_\alpha \}$ to be
\begin{eqnarray}\label{fd4g0ld3gabdma}                                     
& & \d^4 g_{0 \lambda} \left( \det \| g_{\halpha \hbeta} \| \right)^\frac{\eta - 8}{ 2 } \left[ \prod_{\alpha \no \beta} \left( 1 - g_{\halpha \hbeta}^2 \right)^\frac{ 4 - \eta }{ 4 } \right] \d^3 g_{\halpha \no \hbeta} \prod_\alpha \exp \left( - \pi \rmm_\alpha \right) \rmm_\alpha^\frac{ \eta - 10 }{ 2 } \d \rmm_\alpha .
\end{eqnarray}

\subsection{Optimal initial point of the perturbative expansion}\label{initial-point}

Here we introduce the scale along the coordinate axes $\{ \rmm_\alpha^{(0)} \}$ defined by the measure maximization condition (\ref{def-l0}). This condition depends on the action $S$ via $\det \left \| \partial^2 S (\rl_{(0)} ) / (\partial l_i \partial l_k) \right \|$, in particular on the scaling properties of $S$. Now $S = \cS_{\rm g} + \ccF [ \{ w_{\lambda \mu} \} ( - g )^\ialpha ]$, where $\cS_{\rm g}$ is proportional to the metric scale and therefore it is quadratic in the length scale, $\ccF [ \{ w_{\lambda \mu} \} ( - g )^\ialpha ]$ obeys the same property with our choice $\ialpha = - 1 / 8$. Then the factor $\det \left \| \partial^2 S (\rl_{(0)} ) / (\partial l_i \partial l_k) \right \|$ does not depend on the length scale. Here we assume that $\partial l_k$ are indeed lengths. But then we should also write the considered measure in terms of lengths. We temporarily move from the variables $\{ \rmm_\alpha \}$ to $\{ \sqrt{g_{\alpha \alpha} } \}$ in the measure using
\begin{equation}                                                           
\d^3 \rmm_\alpha = 2 \prod_{\alpha \no \beta} \left( 1 - g_{\halpha \hbeta}^2 \right)^{ 1 / 4} \left( \prod_\alpha \rmm_\alpha^{1 / 2} \right) \d^3 \sqrt{g_{\alpha \alpha} } .
\end{equation}

\noindent Then the measure (\ref{fd4g0ld3gabdma}) takes the form
\begin{eqnarray}\label{fd4g0ld3gabdsqrt(gaa)}                              
& & 2 \d^4 g_{0 \lambda} \left( \det \| g_{\halpha \hbeta} \| \right)^\frac{\eta - 8}{ 2 } \prod_{\alpha \no \beta} \left( 1 - g_{\halpha \hbeta}^2 \right)^\frac{ 5 - \eta }{ 4 } \d^3 g_{\halpha \no \hbeta} \nonumber \\ & & \cdot \left( \prod_\alpha \exp \left( - \pi \rmm_\alpha \right) \rmm_\alpha^\frac{ \eta - 9 }{ 2 } \right) \d^3 \sqrt{g_{\alpha \alpha} } .
\end{eqnarray}

The scale of $\rmm_\alpha$ providing the maximum of this measure is
\begin{equation}\label{m0alpha}                                            
\rmm_\alpha^{(0)} = \frac{\eta - 9}{2 \pi} ~~ \forall \alpha , \mbox{ and this is also } b_{\rm s}^2 .
\end{equation}

\subsection{Scale of the graviton perturbations}\label{scale}

If we try to perform a perturbative expansion in the presence of the potentially large parameter $\eta$ appearing as an exponent, it is important to estimate the scale of the metric function that is raised to this power. The latter depends on the elementary length scale. Now we have the two above mentioned length scales, $b_{\rm s}$ and $b_{\rm t}$, of which $b_{\rm s}$ is determined by $\eta$ dynamically, i.e. through the maximization of the functional measure, and the definition of $b_{\rm t}$ is of a gauge nature, and $b_{\rm s} \gg b_{\rm t}$.

It is natural to ascribe to the quantum field $w_{\alpha \beta}$ its scale in terms of $b_{\rm s}$ and $b_{\rm t}$ issuing from the value of its correlator with the different components of itself, that is, the propagator $- i \langle w_{\alpha \beta} w_{\gamma \delta} \rangle$. As this propagator we can take the one restricted to the spatial-spatial metric components, where we neglect $\ralpha = O ( \varepsilon^2 )$ and, generally speaking, quantities of order $O ( \varepsilon^2 )$, but take into account the first order in $\varepsilon$ from $ n \Delta^{(s )} = \rnu \Delta^{(s )} - \varepsilon$, which allows us to bypass the singularity,
\begin{eqnarray}\label{Gabgd}                                              
& & \hspace{-10mm} \cG_{\alpha \beta \gamma \delta} ( n , n ) = ( - g^{(0 )})^{- 1 / 2} \frac{- 2}{ \oDelta \Delta } [ L_{\alpha \gamma} ( n, n ) L_{\beta \delta} ( n, n ) + L_{\beta \gamma} ( n, n ) L_{\alpha \delta} ( n, n ) \nonumber \\ & & \hspace{-15mm} - L_{\alpha \beta} ( n, n ) L_{\gamma \delta} ( n, n ) ] , ~ L_{\alpha \beta} ( n, n ) \stackrel{\rm def }{=} g^{(0 )}_{\alpha \beta} + \frac{ n^2 }{ ( n \Delta^{(s )} )^2 + \cA n^2 } \Delta^{(s )}_\alpha \Delta^{(s )}_\beta , ~ \cA \stackrel{\rm def }{=} - \oDelta \Delta - \Delta^{(s ) 2} .
\end{eqnarray}

\noindent $\cA$ is the difference between the two above finite-difference versions of the d'Alembertian. Here we have for the orders over $b_{\rm s}$ and $b_{\rm t}$ of the values entering $\cG_{\alpha \beta \gamma \delta} ( n , n )$:
\begin{eqnarray}                                                           
& & \hspace{-10mm} g^{(0 )}_{\alpha \beta} = b_{\rm s}^2 \delta_{\alpha \beta} , ~ n^2 = - b_{\rm t}^2 ( \mbox{up to } O ( \varepsilon^2 ) ) , ~ g^{(0 )} = - b_{\rm t}^2 b_{\rm s}^6 , ~ \oDelta \Delta = b_{\rm s}^{- 2} \sum_{\alpha = 1}^3 \oDelta_\alpha \Delta_\alpha - b_{\rm t}^{- 2} \oDelta_0 \Delta_0 , \nonumber \\ & & \hspace{-10mm} \cA = b_{\rm t}^{- 2} ( \oDelta_0 \Delta_0 + \Delta_0^{(s ) 2} ) - b_{\rm s}^{- 2} \sum_{\alpha = 1}^3 ( \oDelta_\alpha \Delta_\alpha + \Delta_\alpha^{(s ) 2} ) , ~ n \Delta^{(s )} = \Delta_0^{(s )} - \varepsilon = O ( 1 ) .
\end{eqnarray}

\noindent And in the considered order in $\varepsilon$ we have
\begin{eqnarray}\label{Gabgd-explicit}                                     
\cG_{\alpha \beta \gamma \delta} ( n , n ) & = & \frac{ b_{\rm s} }{ 2 b_{\rm t} } \frac{ \tdelta_{\alpha \gamma} \tdelta_{\beta \delta} + \tdelta_{\alpha \delta} \tdelta_{\beta \gamma} - \tdelta_{\alpha \beta} \tdelta_{\gamma \delta} }{ b_{\rm t}^{- 2} \sin^2 \frac{p_0 }{2 } - b_{\rm s}^{- 2} \sum_{\alpha = 1}^3 \sin^2 \frac{p_\alpha }{2 } + i 0 }, \nonumber \\ \tdelta_{\alpha \beta} & = & \delta_{\alpha \beta} - \frac{ 1 }{ 4} \frac{ \sin p_\alpha \sin p_\beta }{ b_{\rm t}^{- 2} b_{\rm s}^2 \sin^2 \frac{p_0 + i \varepsilon }{2 } - \sum_{\alpha = 1}^3 \sin^4 \frac{p_\alpha }{2 } } .
\end{eqnarray}

There is a factor $b_{\rm t}^{- 1}$ in $\cG_{\alpha \beta \gamma \delta} ( n , n )$, but almost everywhere in momentum space there is no singularity at $b_{\rm t} \to 0$, due to the dependence of the graviton pole factor $( b_{\rm t}^{- 2} \sin^2 \frac{p_0 }{2 } - b_{\rm s}^{- 2} \sum_{\alpha = 1}^3 \sin^2 \frac{p_\alpha }{2 } + i 0 )^{- 1}$ (and possibly the nonphysical pole factor $( b_{\rm t}^{- 2} b_{\rm s}^2 \sin^2 \frac{p_0 + i \varepsilon }{2 } - \sum_{\alpha = 1}^3$ $\sin^4 \frac{p_\alpha }{2 } )^{- 1}$) on $b_{\rm t}$. In this context, we can also analyze this in an integral sense.

We can integrate $\cG_{\alpha \beta \gamma \delta} ( n , n )$ over $\d p_0$ in product with other factors in a given diagram, or in product with $\exp (i p_\lambda x^\lambda)$ when returning from the momentum representation to the correlator itself at given (integer) points $x^\lambda$. In this integration, the limit of $b_{\rm t} \to 0$ in $\cG_{\alpha \beta \gamma \delta} ( n , n )$ does not lead to a singularity. Taking into account the residues at the poles of this propagator (for $b_{\rm t}/b_{\rm s} << 1$ they are at $p_0 \sim b_{\rm t}/b_{\rm s}$) in $\int \d p_0 \cG_{\alpha \beta \gamma \delta} ( n , n ) \dots$, this propagator effectively appears as a quantity of order $b_{\rm s}^2$.

This can also be seen by looking at different regions of integration. In this integration, the value of the propagator can be large, but in a limited interval: the value of $\cG_{\alpha \beta \gamma \delta} ( n , n )$ has scale $b_{\rm s}^3/b_{\rm t}$ if $| p_0 | \lesssim b_{\rm t}/b_{\rm s}$. (Again, taking into account the integrable pole singularity does not actually change this scale under the integral sign.) Accordingly, when calculating this integral, this scale is included in the result with weight $b_{\rm t}/b_{\rm s}$. This gives an effective propagator scale of order $b_{\rm s}^2$.

If this integration is performed away from the graviton pole, then, generally speaking, the denominator is dominated by the term $b_{\rm t}^{- 2}$. The corresponding contribution to the propagator is of the order of $b_{\rm t} b_{\rm s} $.

In other words, there is no singularities at $b_{\rm t} \to 0$. This is consistent with the fact that in this limit we simply get a theory with continuous time and discrete space, in which the role of the space-time lattice step begins to be played by $b_{\rm s}$ instead of $b_{\rm t} = b_{\rm s} = b$ in the full discrete theory.

Thus, the scale of the correlator $ \langle w_{\alpha \beta} w_{\gamma \delta} \rangle$ is $b_{\rm s}^2$; therefore, the scale of the graviton perturbations $w_{\alpha \beta}$ is $b_{\rm s}$.
\begin{equation}\label{ww=O(bb)}                                           
\langle w_{\alpha \beta} w_{\gamma \delta} \rangle = O ( b_{\rm s}^2 ) , \quad w_{\alpha \beta} = O ( b_{\rm s} ) .
\end{equation}

\subsection{Parameterizing the metric}\label{parameter}

\subsubsection{Replacing metric \texorpdfstring{$0 \lambda$}{0λ} components}\label{replace}

Now we would like to transform the measure to the Lebesgue form. But before doing so, we should take into account a certain contribution to the measure. We are talking about a diagrammatic effect caused by the vertices arising due to the difference between the gauge-fixing term $\ccF [ \{ w_{\lambda \mu} \} ( - g )^\ialpha ]$ and the bilinear one when $\ialpha \neq 0$. This contribution can be taken into account with the help of some change of field variables. Due to the latter, the interaction terms in the gauge-fixing term begin to bring smallness to diagrams at $\varepsilon \to 0$, and the measure receives some contribution. Recall that when estimating the optimal initial metric scale in subsection \ref{initial-point}, we retained the original gauge-fixing term with $\ialpha = - 1 / 8 \neq 0$ to ensure its scaling property is the same as that of the gravity action, since this simplifies this estimation.

The aforementioned change of variables takes the form
\begin{equation}\label{w=w'/g^a}                                           
w_{0 \lambda} = \rgamma^{- \ialpha} w^\prime_{0 \lambda} , \quad w_{\alpha \beta} = w^\prime_{\alpha \beta} .
\end{equation}

\noindent The matrices of the bilinear forms of the action plus the gauge-fixing term, and hence the propagators, are the same for $w_{\lambda \mu}$ and for $w^\prime_{\lambda \mu}$. Then a diagram describing any correlator of $\rnu^\mu w_{\lambda \mu}$ with any components $w_{\sigma \tau}$ or a loop correction to it already contains the factor $\varepsilon$ and is equal to zero in the limit $\varepsilon \to 0$, as well as the diagram for the correlator where all these fields are replaced by primed ones. It is convenient to omit the prime. We usually concentrate on the correlators of only the physical (spatial) components $w_{\alpha \beta}$, when $w^\prime_{0 \lambda}$ can only appear at an inner vertex in a loop correction, i.e. as a dummy variable, and omitting the prime is also convenient.

In $\crO^{\lambda \mu}_\rho w_{\lambda \mu}$, the value
\begin{equation}                                                           
\rnu^\mu w_{\rho \mu} ( - g )^\ialpha \Rightarrow \rnu^\mu w_{\rho \mu} \left( - \frac{ g }{ \rgamma } \right)^\ialpha = \rnu^\mu w_{\rho \mu} \left[ - g_{0 0}^{(0 )} + O ( \varepsilon^2 ) \right]
\end{equation}

\noindent enters with a coefficient $O ( \varepsilon^0 ) = O ( 1 )$. The value $w_{\alpha \beta}$ enters with a coefficient $O ( \varepsilon^2 )$. The part of $\ccF [ \{ w_{\lambda \mu} \} ( - g )^\ialpha ]$ which has the largest coefficient $O ( \varepsilon^{ - 2 } )$ is a bilinear form (of $\rnu^\mu w_{\lambda \mu}$ plus $O ( \varepsilon^2 )$ times $w_{\alpha \beta}$). The components $w_{\alpha \beta}$ enter with the largest coefficient $O ( \varepsilon^0 ) = O ( 1 )$; the corresponding terms are linear in the field $\rnu^\mu w_{\lambda \mu}$; of these terms the interaction (3-linear and higher) terms have the factor $\ialpha$. Since $\rnu^\mu w_{\lambda \mu} = O ( \varepsilon )$ in the sense of its correlator with any component of the metric, such a vertex gives this smallness to any diagram containing it.

Replacing $\rnu^\mu w_{\lambda \mu} \Rightarrow \rnu^\mu w_{\lambda \mu} + ( \rgamma^{- \ialpha} - 1 ) \rnu^\mu w_{\lambda \mu}$ in $\cS_{\rm g}$, we again get new vertices containing the field $\rnu^\mu w_{\lambda \mu}$ and the factor $\ialpha$ and giving smallness $O ( \varepsilon )$ to any diagram containing such a vertex. At $\varepsilon \to 0$, the dependence on $\ialpha$ disappears.

In overall, the effect of $\ialpha$ differing from 0 is reduced to a change in the measure by a factor of
\begin{equation}                                                           
\prod_{\rm sites} \rgamma^{- 4 \ialpha } = \prod_{\rm sites} \rgamma^{ 1 / 2 }
\end{equation}

\noindent due to the transformation (\ref{w=w'/g^a}).

\subsubsection{Parameterizing metric \texorpdfstring{$\alpha \beta$}{αβ} components}\label{parameterize}

Thus, multiplying the measure (\ref{fd4g0ld3gabdma}) by the factor $\rgamma^{ 1 / 2 }$ with taking into account the expression of $\rgamma$ in terms of $\{ g_{\halpha \no \hbeta} \}$, $\{ \rmm_\alpha \}$ (\ref{rgamma}), we then equate the resulting measure to the Lebesgue one $\d^{1 0} \tg_{\lambda \mu}$, like $\d^{1 0} g_{\lambda \mu}$ (\ref{d^(10)g_**}), but in the new variables $\{ \trmm_\alpha \}$ (for the other variables, it is assumed that $g_{0 \lambda} = \tg_{0 \lambda}$, $g_{\halpha \no \hbeta} = \tg_{\halpha \no \hbeta}$).
\begin{eqnarray}\label{f1d4g0ld3gabdma}                                    
& & \d^4 g_{0 \lambda} \left( \det \| g_{\halpha \hbeta} \| \right)^\frac{\eta - 7}{ 2 } \left[ \prod_{\alpha \no \beta} \left( 1 - g_{\halpha \hbeta}^2 \right)^\frac{ 3 - \eta }{ 4 } \right] \d^3 g_{\halpha \no \hbeta} \prod_\alpha \exp \left( - \pi \rmm_\alpha \right) \rmm_\alpha^\frac{ \eta - 9 }{ 2 } \d \rmm_\alpha \nonumber \\ & & \propto \d^{1 0} \tg_{\lambda \mu} = \d^4 g_{0 \lambda} \prod_{\alpha \no \beta} \left( 1 - g_{\halpha \hbeta}^2 \right)^{- 1} \left( \prod_\alpha \trmm_\alpha \right) \d^3 g_{\halpha \no \hbeta} \d^3 \trmm_\alpha .
\end{eqnarray}

\noindent Then the formulas for changing functional variables take the form
\begin{eqnarray}\label{f(m)dm=tilde(m)}                                    
& & \exp \left( - \pi \rmm_\alpha \right) \rmm_\alpha^\frac{ \eta - 9 }{ 2 } \d \rmm_\alpha \propto f \trmm_\alpha \d \trmm_\alpha , \quad f = \left[ \left( \det \| g_{\halpha \hbeta} \| \right)^{- 2} \prod_{\alpha \no \beta} \left( 1 - g_{\halpha \hbeta}^2 \right) \right]^\frac{ \eta - 7 }{ 12 } .
\end{eqnarray}

\noindent Here $f$ is a function of $\{ g_{\halpha \hbeta} \}$, but when calculating the functional Jacobian, it behaves as a constant when considering the dependence on $\{ \rmm_\alpha \}$. In the integrated form
\begin{eqnarray}                                                           
& & \int^{\pi {\scriptstyle \rm m}}_0 \exp ( - z ) z^\tk \d z = C [ f \trmm^2 + C_1 ( \{ g_{\halpha \hbeta} \} ) ] , \quad \tk = \frac{ \eta - 9 }{ 2 } .
\end{eqnarray}

\noindent Here $\rmm$, $\trmm$ are $\rmm_\alpha$, $\trmm_\alpha$ for any $\alpha$, and a constant $C$ and a function $C_1 ( \{ g_{\halpha \hbeta} \} )$ are also for that $\alpha$. We would like to expand $\rmm$ as a function of $\trmm$ around some point. It is natural to take the above point of the maximum of the measure $\rmm^{(0)}$ for that. Above we have found $\rmm^{(0)} = \tk / \pi$ (\ref{m0alpha}). But for some analysis we consider a more general case where we expand not around this point. In this case, the notation $\rmm^{(0 )}$ means the initial point of the expansion,
\begin{equation}                                                           
\rmm^{(0 )} = \frac{ k }{ \pi } , \quad k \neq \tk \quad \mbox{in general} .
\end{equation}

To fix $C_1$, $C$, we choose $\trmm^{(0 )}$ and require the correspondence $\trmm = \trmm^{(0 )}$ $\Leftrightarrow$ $\rmm = \rmm^{(0 )}$ $\forall \{ g_{\halpha \hbeta} \}$, then $C_1$ should cancel the dependence on $\{ g_{\halpha \hbeta} \}$ for $\trmm = \trmm^{(0 )}$ so that $C_1 = - f \trmm^{(0 ) 2} + C_0$, where $C_0$ is a constant. Requiring the correspondence $\trmm = 0$ $\Leftrightarrow$ $\rmm = 0$ for $g_{\halpha \no \hbeta} = 0$, we find $C_0 = \trmm^{(0 ) 2}$. Returning to the correspondence $\trmm = \trmm^{(0 )}$ $\Leftrightarrow$ $\rmm = \rmm^{(0 )}$, we fix the proportionality coefficient $C$. Thus we have
\begin{eqnarray}\label{int^y=Cx}                                           
& & \int^y_0 \exp ( - z ) z^\tk \d z = \frac{ x }{ x_0 } \int^k_0 \exp ( - z ) z^\tk \d z , \quad y = \pi \rmm , \nonumber \\ & & x = \trmm^{(0 ) 2} + ( \trmm^2 - \trmm^{(0 ) 2} ) f , \quad x_0 = \trmm^{(0 ) 2} .
\end{eqnarray}

We invert this equation by expanding $y$ over $\Delta x = x - x_0$ in Taylor series,
\begin{equation}                                                           
y = k + \sum^\infty_{n = 1} \left. \frac{ \d^n y }{ \d x^n } \right|_{y = k} \frac{ \Delta x^n }{ n! } ,
\end{equation}

\noindent and finding the derivatives of the inverse function. The result reads
\begin{eqnarray}\label{m=sum(tilde(m))}                                    
& & \pi \rmm = y = k + \sum^\infty_{n = 1} \left( \frac{ \Delta x }{ x_0 } I \right)^n \frac{ 1 }{ n! } \left[ e^y \left( 1 + \frac{ y }{ k } \right)^{- \tk } \frac{ \d }{ \d y } \right]^{ n - 1 } \left. e^y \left( 1 + \frac{ y }{ k } \right)^{- \tk } \right|_{y = 0} , \nonumber \\ & & I = \int^k_0 e^{k - z} \left( \frac{ z }{ k } \right)^\tk \d z .
\end{eqnarray}

In the actual particular case $k = \tk$ ($= ( \eta - 9 ) / 2$) we find up to the trilinear terms
\begin{equation}                                                           
\pi \rmm = k + \sqrt{ k } \sqrt{ \frac{ \pi }{ 2 } } \frac{ \Delta x }{ x_0 } + \frac{ 1 }{ 6 } \sqrt{ k } \left( \sqrt{ \frac{ \pi }{ 2 } } \frac{ \Delta x }{ x_0 } \right)^3 + \dots .
\end{equation}

\subsection{Absence of a large parameter and specificity of the maximum point of the functional measure in the perturbative expansion}\label{large-parameter}

The theory contains the formally large aforementioned parameter $\eta$, and a priori it can show up in the perturbative expansion. First, $f ( \{ g_{\halpha \hbeta} \} )$ (\ref{f(m)dm=tilde(m)}) contains large powers of expressions of the type of $1 + O ( g_{\halpha \no \hbeta}^2 )$ which being expanded over $g_{\halpha \no \hbeta}$ have large coefficients of the binomial expansion. In the metric parametrization (\ref{f(m)dm=tilde(m)}), this will lead to vertices and diagrams with large coefficients.

However, this largeness is compensated by the earlier found parametrical values of the fields $w_{\alpha \beta} = O ( b_{\rm s} )$ (\ref{ww=O(bb)}), where $b_{\rm s} = O ( \sqrt{ \eta } )$ (\ref{m0alpha}): then $g_{\halpha \no \hbeta} = O ( b_{\rm s}^{- 1} ) = O ( \eta^{- 1 / 2} )$ and
\begin{equation}                                                           
f = \left[ 1 + O ( g_{\halpha \no \hbeta}^2 ) \right]^\frac{ \eta - 7 }{ 12 } = \left[ 1 + O ( \eta^{ - 1 } ) \right]^\frac{ \eta - 7 }{ 12 } ,
\end{equation}

\noindent whose (binomial) expansion in metric perturbations has no terms growing with $\eta$.

Further, the coefficients in the metric parametrization (\ref{m=sum(tilde(m))}) can contain increasing powers of $\eta$, in addition to the fact that through $f ( \{ g_{\halpha \hbeta} \} )$, also through $\tk$. We analyze the function included there:
\begin{eqnarray}\label{e^y(1+y/k)^(-tilde(k))}                             
& & e^y \left( 1 + \frac{ y }{ k } \right)^{ - \tk } = \exp \left[ y - \tk \ln \left( 1 + \frac{ y }{ k } \right) \right] \nonumber \\ & & = \exp \left\{ \frac{ \tk }{ k } \left[ \frac{ 1 }{ 2 } \left( \frac{ y }{ k^{1 / 2} } \right)^2 - \frac{ 1 }{ 3 } \left( \frac{ y }{ k^{2 / 3} } \right)^3 + \frac{ 1 }{ 4 } \left( \frac{ y }{ k^{3 / 4} } \right)^4 - \dots \right] + \frac{ k - \tk }{ k } y \right\} .
\end{eqnarray}

\noindent Here we analyze the case where $k \neq \tk$. $k$ and $\tk$ are large parameters, and formally mathematically we can consider the procedure of making them arbitrarily large and we set the behaviour of the parameters when approaching this limit to obey the law $k - \tk = O ( k^\delta )$, $\delta \geq 0$. We assume that $k$ and $\tk$ are of the same order, that is, $\delta \leq 1$.

In the exponent in (\ref{e^y(1+y/k)^(-tilde(k))}), the terms $\propto ( y k^{- j_n} )^n$ are arranged in order of increasing $j_n$ for the case $\delta = 0$: $j_n = ( n - 1 ) / n$ from $n = 2$ to $n = \infty$ and then $j_1 = 1$. When substituting (\ref{e^y(1+y/k)^(-tilde(k))}) into (\ref{m=sum(tilde(m))}), we expand (\ref{e^y(1+y/k)^(-tilde(k))}) in power series, repeatedly differentiate with respect to $y$, set $y = 0$ and take the product of the results of such procedures for $n - 1$ instances of the function (\ref{e^y(1+y/k)^(-tilde(k))}) with $n - 1$ being the total number of the differentiations. If a set of numbers $p_l$ partitions the number $n - 1$, $\sum_l {p_l} = n - 1$, then the terms $( y k^{- j_{p_l}} )^{p_l}$ from the exponent in (\ref{e^y(1+y/k)^(-tilde(k))}) take part as factors in forming some contribution to the $n$-th coefficient in (\ref{m=sum(tilde(m))}) (after total number $n - 1$ of the differentiations). The corresponding power of $k$ in this coefficient is $- \sum_l p_l j_{p_l}$, and the minimum of $\sum_l p_l j_{p_l}$ is of interest.

Thus, at $\delta > 1 / 2$, the maximal power of $k$ in the $n$th term in (\ref{m=sum(tilde(m))}) follows for $p_l = 1, l = 1, 2, ... , n - 1$ ($\sum_l p_l = n - 1$), that is, if $n - 1$ differentiations over $y$ in (\ref{m=sum(tilde(m))}) are applied to $n - 1$ factors $( k - \tk ) k^{- 1} y = O ( k^{ - 1 + \delta } ) y$. The $n = 1$ term $I \Delta x / x_0$ describes graviton perturbations of the metric-related quantity $\rmm$ and is governed by the bilinear form of the action and, according to the results of subsection \ref{scale}, should be $O ( \sqrt{ k } )$. The $n$th term in (\ref{m=sum(tilde(m))}) has the order in $k$
\begin{eqnarray}                                                           
& & \left[ O ( k^{ - 1 + \delta } ) \right]^{n - 1} \left( I \Delta x / x_0 \right)^n = O \left( k^{ ( -1 + \delta ) ( n - 1 ) + n / 2 } \right) = O \left( k^{ ( \delta - 1 / 2 ) ( n - 1 ) } \right) \cdot \sqrt{k} .
\end{eqnarray}

\noindent The additional power of $k$ compared to the $n = 1$ term is positive and grows with $n$. If we divide a diagram into blocks being neighborhoods of the vertices, each block can be characterized by its scale being the product of the scale of the coefficient of the vertex and square roots of the adjacent propagator scales. Substituting the considered parametrization of the metric into the action will lead to the fact that the scales of the blocks will be multiplied by $k$, and consequently by $\eta$ in positive powers, if new lines appear in them. This will lead to diagrams that make up an expansion in increasing powers of the large parameter $\eta$.

Also we can formulate this as incapability of the bilinear form of the action for $\{ g_{\lambda \mu} \}$ taken in the first order over $\{ \tg_{\lambda \mu} \}$ to ensure that graviton perturbations of $\{ g_{\lambda \mu} \}$ were in the limits given by the bilinear form for the total $\{ g_{\lambda \mu} \}$ in the aspect of the power dependency on $\eta$. This can be viewed as a certain inconsistency of the perturbation theory when we expand too far from the point of the maximum of the measure.

On the contrary, at $\delta \leq 1 / 2$, the maximal power of $k$ in the $n$th term in (\ref{m=sum(tilde(m))}) is achieved for $p_l = 2, l = 1, 2, ... , q$, when $n - 1 = 2 q$ or, if additionally $p_{q + 1} = 1$, when $n - 1 = 2 q + 1$. That is, $n - 1$ differentiations are applied to $[ ( n - 1 ) / 2 ]$ factors $\propto ( k^{- 1 / 2} y )^2$; if $n - 1$ is odd, there is also the $([ ( n - 1 ) / 2 ] + 1)$-th factor $( k - \tk ) k^{- 1} y$. The $n$th term in (\ref{m=sum(tilde(m))}) has the order in $k$
\begin{equation}                                                           
O \left( k^{ \frac{ 1 }{ 2 } ( \delta - 1 / 2 ) [ 1 + ( - 1 )^n ] } \right) \cdot \sqrt{k} .
\end{equation}

\noindent The additional power of $k$ compared to the $n = 1$ term is not greater than zero. There is no increase in the powers of the large parameter $\eta$ in the perturbative expansion.

\begin{figure}[h]
	\centering
	\includegraphics[scale=1]{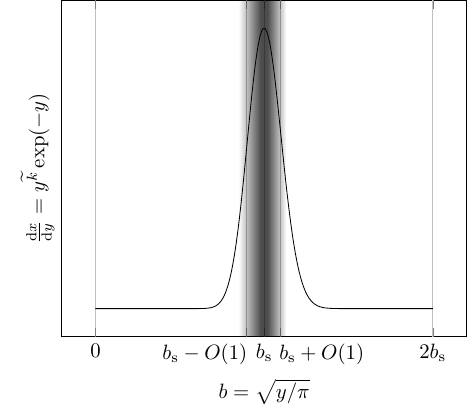}
	\caption{A typical dependence of the measure on the edge length scale $b$ if its maximum is at $b = b_{\rm s}$. An interval of the possible initial $b$ for which there is no a large parameter in the perturbative expansion. The fuzziness of its boundaries means the order-of-magnitude accuracy of estimating their location $O ( 1 )$ relative to $b_{\rm s}$.}
	\label{f1}
\end{figure}

Thus, we have an interval of the initial length scale, Fig.~\ref{f1}, for which
the large parameter $\eta$ is effectively absent in the perturbative expansion if the initial point of the perturbative expansion $k$ is sufficiently close to the maximum point of the functional measure, here $\tk$. When these parameters are formally considered as arbitrarily large, their behaviour when approaching this limit should obey the condition
\begin{equation}                                                           
| k - \tk | \lesssim O ( \sqrt{ k } ) \quad \mbox{ or } \quad | \sqrt{k} - \sqrt{\tk } | \lesssim O ( 1 )
\end{equation}

\noindent for that. $\sqrt{\tk / \pi } = b_{\rm s}$, and $\sqrt{ k / \pi } \stackrel{\rm def}{=} b$ is a current typical edge length scale, and this condition means that we should expand around an elementary length $b$ which differs from the measure-maximizing length $b_{\rm s}$ within a value of the order of the Planck length.

At $| k - \tk | << \sqrt{ k } $, the factor $I$ at $\Delta x / x_0$ in (\ref{m=sum(tilde(m))}) is $ k \sqrt{ \pi / ( 2 \tk ) } $, and at $| k - \tk | \lesssim O ( \sqrt{ k } )$ it is $O ( \sqrt{ k } )$, and the bilinear form of the action makes $\Delta x / x_0$ to be $O ( 1 )$:
\begin{equation}                                                           
I = O ( \sqrt{ k } ), \quad \frac{ \Delta ( \trmm^2 ) }{ \trmm^{(0 ) 2} } = O \left( \frac{ \Delta x }{ x_0 } \right) = O ( 1 ) .
\end{equation}

\subsection{A concrete form of metric parametrization}\label{concrete-parameter}

We rewrite the expression $\rmm_\alpha = \rmm_\alpha ( \{ \trmm_\beta \} )$ in terms of the metric, $g_{\lambda \mu} = g_{\lambda \mu} ( \{ \tg_{\nu \rho} \} )$. With taking into account $g_{0 \lambda} = \tg_{0 \lambda}$, this reduces to $g_{\alpha \beta} = g_{\alpha \beta} ( \{ \tg_{\gamma \delta} \} )$. $\trmm$ is expressed through $\tg$ in the same way as $\rmm$ is expressed through $g$,
\begin{equation}                                                           
\trmm_1^2 = \tg_{2 2} \tg_{3 3} - \tg_{2 3}^2 , \quad 2 ~ \mbox{perm}(123) .
\end{equation}

\noindent We take $\tg^{(0 )}_{\alpha \beta} = \delta_{\alpha \beta}$, $\tg_{\alpha \beta} = \delta_{\alpha \beta} + \tw_{\alpha \beta}$. Thus $\rmm$ can be expressed in terms of $\tw$,
\begin{equation}                                                           
\rmm_1 = \frac{ k }{ \pi } + \sqrt{ \frac{ k }{ 2 \pi } } \left( \tw_{2 2} + \tw_{3 3} + \tw_{2 2} \tw_{3 3} - \tw_{2 3}^2 \right) + \dots .
\end{equation}

\noindent The expansion up to bilinear terms in $\tw_{\alpha \beta}$ is shown. Thus we find $g_{\alpha \alpha}$ using
\begin{equation}                                                           
g_{1 1} = \frac{ \rmm_2 \rmm_3 }{ \rmm_1 } \frac{ \left( 1 - \tg_{\htwo \hthr }^2 \right)^{1 / 2} }{ \left( 1 - \tg_{\hthr \hone }^2 \right)^{1 / 2} \left( 1 - \tg_{\hone \htwo }^2 \right)^{1 / 2} } , \dots
\end{equation}

\noindent then
\begin{eqnarray}                                                           
& & \hspace{-15mm} \frac{ \pi }{ k } g_{1 1} = 1 + \sqrt{ \frac{ 2 \pi }{ k } } \tw_{1 1} + \frac{ 1 }{ 2 } \left( \tw_{3 1}^2 + \tw_{1 2}^2 - \tw_{2 3}^2 \right) \nonumber \\ & & \hspace{-15mm} + \sqrt{ \frac{ \pi }{ 2 k } } \left( \tw_{3 3} \tw_{1 1} + \tw_{1 1} \tw_{2 2} - \tw_{2 2} \tw_{3 3} - \tw_{3 1}^2 - \tw_{1 2}^2 + \tw_{2 3}^2 \right) + \frac{ \pi }{ 2 k } \left( \tw_{1 1} - \tw_{2 2} \right) \left( \tw_{1 1} - \tw_{3 3} \right) .
\end{eqnarray}

Further, we take into account another adopted equations
\begin{equation}                                                           
g_{\halpha \no \hbeta} = \tg_{\halpha \no \hbeta} , \mbox{ then } \tg_{\alpha \no \beta} = O ( k^{- 1 / 2} ) .
\end{equation}

\noindent (Since $g_{\halpha \no \hbeta} = O ( b_{\rm s}^{- 1} ) = O ( k^{- 1 / 2} ) $ from (\ref{ww=O(bb)}).) This prompts us to redefine components of $\tg$ as
\begin{equation}                                                           
\tg_{\alpha \no \beta} \Rightarrow \sqrt{ \frac{ 2 \pi }{ k } } \tg_{\alpha \no \beta} .
\end{equation}

\noindent This leaves the measure proportional to the Lebesgue one $\d^{1 0} \tg_{\lambda \mu}$; at the same time, the linear part of $g_{\alpha \beta} ( \{ \tg_{ \gamma \delta } \} )$ is $\propto \delta_{\alpha \beta} + \mathrm{const} \cdot \tw_{\alpha \beta}$. As a result, we get
\begin{eqnarray}                                                           
& & \frac{ \pi }{ k } g_{1 1} = 1 + \sqrt{ \frac{ 2 \pi }{ k } } \left[ \tw_{1 1} + \frac{ 1 }{ 2 } \left( \tw_{3 3} \tw_{1 1} + \tw_{1 1} \tw_{2 2} \right. \right. \left. - \tw_{2 2} \tw_{3 3} \right) \nonumber \\ & & + \frac{1}{2} \sqrt{ \frac{ 2 \pi }{ k } } \left( 1 - \sqrt{ \frac{ 2 \pi }{ k } } \right) \left( \tw_{3 1}^2 + \tw_{1 2}^2 - \tw_{2 3}^2 \right) \left. + \frac{ 1 }{ 4 } \sqrt{ \frac{ 2 \pi }{ k } } \left( \tw_{1 1} - \tw_{2 2} \right) \left( \tw_{1 1} - \tw_{3 3} \right) \right], \nonumber \\ & & \frac{ \pi }{ k } g_{2 3} = \sqrt{ \frac{ 2 \pi }{ k } } \left[ \tw_{2 3} - \frac{ 1 }{ 2 } \left( 1 - \sqrt{ \frac{ 2 \pi }{ k } } \right) \left( \tw_{2 2} + \tw_{3 3} \right) \tw_{2 3} \right] , \nonumber \\ & & 2 ~ \mbox{perm}(123) .
\end{eqnarray}

\noindent Thus we get the parametrization in the form
\begin{equation}\label{g=1+u}                                              
\frac{ \pi }{ k } g_{\alpha \beta} = \delta_{\alpha \beta} + \sqrt{ \frac{ 2 \pi }{ k } } u_{\alpha \beta} \left( \{ \tw_{\gamma \delta} \} \right) ~ {\mbox or} ~ g_{\alpha \beta} = b_{\rm s}^2 \delta_{\alpha \beta} + \sqrt{2} b_{\rm s} u_{\alpha \beta} , ~ u_{\alpha \beta} = \tw_{\alpha \beta} + V_{\alpha \beta} ,
\end{equation}

\noindent where $V_{\alpha \beta} = V_{\alpha \beta} \left( \{ \tw_{\gamma \delta} \} \right)$ denotes bilinear and higher order terms in $\tw$. Here $u_{\alpha \beta} = O ( 1 )$ with respect to $k$ (and therefore $\eta$) in the sense of its correlator with any component of the metric, since $\tw_{\alpha \beta} = O ( 1 )$.

\subsection{Vertex scales and expansion parameters}\label{vertex-scales}

\subsubsection{Disappearance of the diagrammatic contribution of the gauge-fixing term at \texorpdfstring{$\varepsilon \to 0$}{ε → 0}}\label{diagr-from-gauge-fix}

By substituting the parametrization of the metric $g_{\alpha \beta} ( \{ \tw_{\gamma \delta} \} )$ (\ref{g=1+u}) into the action $\cS_{\rm g}^\prime [ g , J ]$ (\ref{cS_g'}), we can obtain a propagator for $\tw_{\lambda \mu}$ and vertices. The deviation of the gauge-fixing term ($\ccF [ \{ w_{\lambda \mu} \} ( - g )^\ialpha ]$ with nonzero $\ialpha$ when scaling $w_{0 \lambda} \Rightarrow \rgamma^{ - \ialpha } w_{0 \lambda}$) from the bilinear form in the original metric variables was discussed in subsection \ref{replace}. The largest coefficient, $O ( \varepsilon^{- 2} )$, appears in front of some bilinear form of the variable $w_{0 \lambda}$, $\varepsilon^{- 2} w_{0 \lambda} \dots w_{0 \mu}$. The $w_{\alpha \beta}$ components appear as part of the term with the largest coefficient $O ( \varepsilon^0 ) = O ( 1 )$, $w_{0 \lambda} \dots [ w_{\alpha \beta} ( - g )^\ialpha ]$. This term consists of the bilinear $w_{0 \lambda} \dots w_{\alpha \beta}$ and 3-linear and higher parts $w_{0 \lambda} \dots \{ w_{\alpha \beta} [ ( - g )^\ialpha - 1 ] \}$. Here, the dots represent certain (non-local) operators. These terms contain the field $w_{0 \lambda}$, which is $O ( \varepsilon )$ in the sense of its correlator with any component of the metric. Therefore, such a vertex gives this smallness to any diagram containing it.

Now the parametrization affects precisely these terms. We obtain the terms of the type of $w_{0 \lambda} \dots [ w_{\alpha \beta} ( \{ \tw_{\gamma \delta} \} ) ( - g ( \{ \tw_{\gamma \delta} \} ) )^\ialpha ]$ and just some additional nonlinearity contained in $w_{\alpha \beta} = w_{\alpha \beta} ( \{ \tw_{\gamma \delta} \} )$. This term is linear over the field $w_{0 \lambda}$, and any corresponding vertex gives the smallness $O ( \varepsilon )$ to any diagram containing it. Thus, the diagrammatic effect in the calculation of any particular amplitudes caused by the vertices contained in the gauge-fixing term disappears as $\varepsilon$ tends to 0, also when the metric is parameterized, and for the same reasons.

\subsubsection{Scales of interaction terms}\label{scales}

Thus, for vertices it is sufficient to use the action without the gauge-fixing term, $\cS_{\rm g}$ (\ref{cSg}). As for the propagator components for $\tw$, they are proportional to the propagator components for $w$. The propagator components that do not vanish at $\varepsilon \to 0$ are the propagator components for the spatial-spatial metric components,
\begin{equation}                                                           
\langle w_{\alpha \beta} w_{\gamma \delta} \rangle = 2 b_{\rm s}^2 \langle \tw_{\alpha \beta} \tw_{\gamma \delta} \rangle .
\end{equation}

\noindent According to subsection \ref{scale}, the parametrically largest (for a large $b_{\rm s}$, $b_{\rm s} >> b_{\rm t}$) part of the propagator  of $\tw$ is
\begin{equation}\label{<twtw>}                                             
- i \langle \tw_{\alpha \beta} \tw_{\gamma \delta} \rangle \propto \frac{b_{\rm t}^{- 1} b_{\rm s}^{- 1} ( \tdelta_{\alpha \gamma} \tdelta_{\beta \delta} + \dots )}{ b_{\rm t}^{- 2} \sin^2 \frac{p_0 }{2 } - b_{\rm s}^{- 2} \sum_{\alpha = 1}^3 \sin^2 \frac{p_\alpha }{2 } + i 0 } + \dots .
\end{equation}

Continuing with the vertices, the action is a sum of terms of the general form
\begin{eqnarray}                                                           
& & \sqrt{- g} g^{\lambda_1 \lambda_2} g^{\lambda_3 \lambda_4} g^{\lambda_5 \lambda_6} \left( \Delta^\#_{\mu_5} g_{\mu_1 \mu_2} \right) \left( \Delta^\#_{\mu_6} g_{\mu_3 \mu_4} \right), \quad \Delta^\# \stackrel{\rm def}{=} \Delta \mbox{ or } \Delta^{(s ) } ,
\end{eqnarray}

\noindent where the superscripts should be contracted with the subscripts in certain ways. In the synchronous gauge limit $\varepsilon \to 0$, in the changing part of the metric, only the spatial-spatial components survive in the sense of their correlators with any components of the metric. Among the contravariant metric components, only the spatial-spatial and temporal-temporal components survive. So we are left with the terms
\begin{eqnarray}\label{dbdb-term}                                       
& & \sqrt{- g} g^{\alpha_1 \alpha_2} g^{\alpha_3 \alpha_4} g^{\alpha_5 \alpha_6} \left( \Delta^\#_{\beta_5} g_{\beta_1 \beta_2} \right) \left( \Delta^\#_{\beta_6} g_{\beta_3 \beta_4} \right), \\ & & \label{d0d0-term} \sqrt{- g} g^{\alpha_1 \alpha_2} g^{\alpha_3 \alpha_4} g^{0 0} \left( \Delta_0 g_{\beta_1 \beta_2} \right) \left( \Delta_0 g_{\beta_3 \beta_4} \right) .
\end{eqnarray}

\noindent Eq. (\ref{d0d0-term}) has a largeness $g^{0 0} = - b_{\rm t}^{- 2} + O ( \varepsilon^2 )$. It includes the time derivative $\Delta_0$ (and not $\Delta^{(s )}_0$) from our refined action. At $\varepsilon \to 0$, it is proportional to
\begin{equation}                                                           
\sqrt{- g} g^{\alpha_1 \alpha_2} g^{\alpha_3 \alpha_4} \left( b_{\rm s} b_{\rm t}^{- 1} \Delta_0 u_{\beta_1 \beta_2} \right) \left( b_{\rm s} b_{\rm t}^{- 1} \Delta_0 u_{\beta_3 \beta_4} \right) .
\end{equation}

\noindent Here in $u_{\beta_1 \beta_2}$, $u_{\beta_3 \beta_4}$ we single out arbitrary monomials of $\{ \tw_{\gamma \delta} \}$, the corresponding contribution to this expression is proportional to
\begin{eqnarray}                                                           
& & \sqrt{- g} g^{\alpha_1 \alpha_2} g^{\alpha_3 \alpha_4} \left[ b_{\rm s} b_{\rm t}^{- 1} \Delta_0 \left( \tw_{\gamma_n \delta_n} \dots \tw_{\gamma_{m + 1} \delta_{m + 1}} \right) \right] \left[ b_{\rm s} b_{\rm t}^{- 1} \Delta_0 \left( \tw_{\gamma_m \delta_m} \dots \tw_{\gamma_1 \delta_1} \right) \right] .
\end{eqnarray}

With taking into account the parametrically largest part of the propagator of $\tw$ (\ref{<twtw>}), the correlator of this value with some another $n$ fields $\tw$ is proportional to
\begin{eqnarray}\label{d0d0-term'}                                         
& & \frac{ b_{\rm s} }{ b_{\rm t} } \Delta_0 \left( \sum_{ j = m + 1 }^n p_j \right) \frac{ b_{\rm s} }{ b_{\rm t} } \Delta_0 \left( \sum_{ j = 1 }^m p_j \right) \prod_{j = 1}^n \frac{b_{\rm t}^{- 1} b_{\rm s}^{- 1} }{ b_{\rm t}^{- 2} \sin^2 \frac{p_{j 0} }{2 } - b_{\rm s}^{- 2} \sum_{\alpha = 1}^3 \sin^2 \frac{p_{j \alpha} }{2 } + i 0 }
\end{eqnarray}

\noindent in the momentum representation; $\Delta_0 ( p ) \equiv \exp ( i p_0 ) - 1 $. Here $p_j$ are the quasi-momenta of $n \geq 2$ lines of $\tw$. The propagators of the $\tw$ fields that appear when substituting the expression $g_{\alpha \beta} = b_{\rm s}^2 \delta_{\alpha \beta} + \sqrt{2} b_{\rm s} u_{\alpha \beta} ( \{ \tw_{\gamma \delta} \} ) $ into $\sqrt{- g} g^{\alpha_1 \alpha_2} g^{\alpha_3 \alpha_4}$, expanding in $\tw$, and contracting with other $\tw$ fields in the diagram are not yet written out here. Further, as in subsection \ref{scale}, we will consider various regions of change of quasi-momenta $p_{j 0}$. Not all quasi-momenta are independent, in particular, they can depend on the external quasi-momenta. Let us assume that we begin the calculation of a given diagram by integrating over the temporal components of the quasi-momenta by calculating the residues at the poles of the integrand. These poles are located at $p_{j 0} = O ( b_{\rm t} / b_{\rm s} )$. This integration is performed over independent quasi-momenta. If the dependent quasi-momenta have the same order, $p_{j 0} = O ( b_{\rm t} / b_{\rm s} )$ (this is the normal order of the quasi-momentum for a distribution close to continuous at $b_{\rm t} \to 0$ and characterized by the true momentum $\tp_{j 0} = b_{\rm t}^{- 1} p_{j 0}$), then
\begin{eqnarray}                                                           
& & \sum_{j = m + 1}^n p_{j 0} = O ( b_{\rm t} / b_{\rm s} ) , ~ \sum_{j = 1}^m p_{j 0} = O ( b_{\rm t} / b_{\rm s} ) , \nonumber \\ & & \frac{ b_{\rm s} }{ b_{\rm t} } \Delta_0 \left( \sum_{j = m + 1}^n p_j \right) = O ( 1 ), \quad \frac{ b_{\rm s} }{ b_{\rm t} } \Delta_0 \left( \sum_{j = 1}^m p_j \right) = O ( 1 ) .
\end{eqnarray}

Then consider the case when there is at least one dependent momentum $p_{j 0} = O ( 1 )$ for some $j$, $j = m + 1 , \dots , n$, whereas $p_{j 0} = O ( b_{\rm t} / b_{\rm s} ) \forall j = 1, \dots , m$ as before. Then we have $b_{\rm s} b_{\rm t}^{- 1} \Delta_0 \left( \sum_{j = 1}^m p_j \right) = O ( 1 )$ as before. For $p_{j 0} = O ( 1 )$ for some $j = m + 1 , \dots , n$, the corresponding propagator contains a largeness $\sim b_{\rm t}^{- 2}$ in the denominator and therefore a smallness $\sim ( b_{\rm t} / b_{\rm s} )^2$ compared to this propagator for $p_{j 0} = O ( b_{\rm t} / b_{\rm s} )$. The propagator connects two vertices, and we can conditionally assign the square root of this additional factor, $[ ( b_{\rm t} / b_{\rm s} )^2 ]^{1 / 2} = b_{\rm t} / b_{\rm s}$, to each vertex, namely, multiply $\Delta_0$ by it: $b_{\rm s} b_{\rm t}^{- 1} \Delta_0 \left( \sum_{j = m + 1}^n p_j \right) \times b_{\rm t} b_{\rm s}^{- 1} = \Delta_0$ $\left( \sum_{j = m + 1}^n p_j \right) = O ( 1 )$. Moreover, if some large quasi-momentum flows in/out of a vertex, some another large quasi-momentum flows out/into this vertex. Here it must be the quasi-momentum of a line of $\tw$ in $\sqrt{- g} g^{\alpha_1 \alpha_2} g^{\alpha_3 \alpha_4}$. It introduces one else factor $b_{\rm t} / b_{\rm s}$ to the given vertex, which further suppresses it. Analogously, we consider the case where we have a large dependent momentum $p_{j 0} = O ( 1 )$ for some $j = 1, \dots , m$, whereas $p_{j 0} = O ( b_{\rm t} / b_{\rm s} ) \forall j = m + 1, \dots , n$.

Finally, large dependent momenta $p_{j 0} = O ( 1 )$ can be both at some $j = 1, \dots , m$ and at some $j = m + 1 , \dots , n$. Again, additional factors $b_{\rm t} / b_{\rm s}$ multiply both $\Delta_0$ in (\ref{d0d0-term'}), (\ref{d0d0-term}).

As a result, the estimates of the scales of the vertices provided by the term (\ref{d0d0-term}) in the action are the same as those in the case of the term (\ref{dbdb-term}). For the latter, we substitute $g_{\alpha \beta} = b_{\rm s}^2 \delta_{\alpha \beta} + \sqrt{2} b_{\rm s} u_{\alpha \beta}$ there and find for the scale of this term
\begin{eqnarray}                                                           
& & \hspace{-15mm} \sqrt{- g} g^{\alpha_1 \alpha_2} g^{\alpha_3 \alpha_4} g^{\alpha_5 \alpha_6} \left( \Delta^\#_{\beta_5} g_{\beta_1 \beta_2} \right) \left( \Delta^\#_{\beta_6} g_{\beta_3 \beta_4} \right) \sim b_{\rm t} b_{\rm s}^{- 1} \left( \Delta^\#_{\beta_5} u_{\beta_1 \beta_2} \right) \left( \Delta^\#_{\beta_6} u_{\beta_3 \beta_4} \right) \sim b_{\rm t} b_{\rm s}^{- 1}
\end{eqnarray}

\noindent in the leading order in $u$, in which this term is bilinear in $u$. Higher orders in $u$ follow by expanding $\sqrt{- g} g^{\alpha_1 \alpha_2} g^{\alpha_3 \alpha_4} g^{\alpha_5 \alpha_6}$ over $u$. The relative contribution of $u$ is of the order of $b_{\rm s}^{- 1} u$ there, and each subsequent order in $u$ is accompanied by an additional factor $O ( b_{\rm s}^{- 1} )$ in the coefficient. Thus we can write for the scale of the coefficients of the vertices
\begin{eqnarray}\label{sumC(n)u...ududu}                                   
& & \sqrt{- g} g^{\alpha_1 \alpha_2} g^{\alpha_3 \alpha_4} g^{\alpha_5 \alpha_6} \left( \Delta^\#_{\beta_5} g_{\beta_1 \beta_2} \right) \left( \Delta^\#_{\beta_6} g_{\beta_3 \beta_4} \right) = \nonumber \\ & & \sum_{n \geq 2} C_{(n )}^{\alpha_1 \dots \alpha_6 \gamma_1 \delta_1 \dots \gamma_{n - 2} \delta_{n - 2}} u_{\gamma_1 \delta_1} \dots u_{\gamma_{n - 2} \delta_{n - 2}} \left( \Delta^\#_{\beta_5} u_{\beta_1 \beta_2} \right) \left( \Delta^\#_{\beta_6} u_{\beta_3 \beta_4} \right) , \nonumber \\ & & C_{(n )}^{\alpha_1 \dots \alpha_6 \gamma_1 \delta_1 \dots \gamma_{n - 2} \delta_{n - 2}} = O ( b_{\rm t} b_{\rm s}^{1 - n} ) .
\end{eqnarray}

Besides, in the considered case—theories with an interaction that is quadratic in derivatives (namely, in $\Delta^\#_0$)—there are non-pole contributions. They are obtained by deforming the integration path $[ - \pi, \pi ]$ in the complex quasi-momentum plane $p_0$ towards large $\Im p_0$: into $[ - \pi, - \pi + i L ] \cup [ - \pi + i L, \pi + i L ] \cup [ \pi + i L, \pi ]$, $L \to \infty$. They are given by rather special terms in diagrams—vertices of the type (\ref{d0d0-term}) forming a loop where neighboring $\Delta_0 u$ are contracted—and can be analyzed (subsections \ref{contrib-new}\&\ref{Smatrix}). Here, too, by setting $\Im p_0 \to \infty$ for a loop quasi-momentum, we find that such a vertex in such a loop has the scale $O ( b_{\rm s}^{2 - n} )$ which, as $O ( b_{\rm t} b_{\rm s}^{1 - n} )$ (\ref{sumC(n)u...ududu}) in a pole diagram, involves negative powers of $b_{\rm s}$, not $b_{\rm t}$.

\subsubsection{Correspondence with the continuum diagrams}\label{continuum-correspond}

Considering first $u_{\alpha \beta} = \tw_{\alpha \beta}$, we neglect the effect of the non-Lebesgue form of the original measure. Parameterizing $g_{\lambda \mu}$ at $u_{\alpha \beta} = \tw_{\alpha \beta}$ by some scaling of $g_{\lambda \mu}$ and redefining $u_{\lambda \mu}$ reduces to some parametrization standard for the continuum theory, which we write in ordinary units as $g_{\lambda \mu} = \eta_{\lambda \mu} + l_{\rm Pl} u_{\lambda \mu}$ (in the corresponding synchronous gauge), where $l_{\rm Pl}$ is simultaneously gravitational coupling constant. The effect of the elementary edge length scale $b_{\rm s}$ in the continuum notations consists in the occurrence of the momenta cutoff at $\sim l_{\rm Pl}^{- 1 } \eta^{- 1 / 2}$. If we trace the maximal divergence of loop integrations, then, for example, adding a new graviton line between two vertices of a diagram leads to the appearance of the two graviton couplings $l_{\rm Pl}$, the graviton propagator $\sim (\mbox{momentum})^{- 2}$ and integration over this new loop momentum. This integration gives the cutoff squared, and thus this addition of a loop to a diagram leads to a factor $\sim l_{\rm Pl}^2 ( l_{\rm Pl}^{- 1 } \eta^{- 1 / 2} )^2 = \eta^{- 1}$. $\eta^{- 1}$ is a small parameter of this expansion.

In the discrete theory, we get a finite-difference form of the standard perturbative expansion of the continuum GR. Now, $n$-graviton $\tw$ vertices—for example, in pole diagrams—have an effective scale of the order of $b_{\rm t} b_{\rm s}^{1 - n}$, $n \geq 3$ (the $n = 2$ terms define the propagator, not the vertices). It can be seen that when adding new vertices to a given diagram and/or increasing the number of lines in any of them, the scale of the diagram is parametrically suppressed by a negative power of $b_{\rm s}$. That is, the perturbative expansion occurs over the small parameter $b_{\rm s}^{- 1}$.

Thus, the non-renormalizability of the continuum GR (large degrees of the UV cutoff that arise when calculating diagrams) simultaneously means large degrees of the small parameter $b_{\rm s}^{- 1}$ and better convergence of the perturbative series in the discrete analogue of the theory.

The continuum analogue diagram arises when passing from quasi-momenta $p_\lambda$ to momenta $\tp_\lambda$, $p_\alpha = b_{\rm s} \tp_\alpha$, $p_0 = b_{\rm t} \tp_0$:
\begin{eqnarray}                                                           
& & b_{\rm s}^{- 1} \d p_\alpha \Rightarrow \d \tp_\alpha , \quad b_{\rm s}^{- 1} \Delta^{(s )}_\alpha \Rightarrow i \tp_\alpha , \quad b_{\rm s}^{- 1} \Delta_\alpha \Rightarrow i \tp_\alpha , \nonumber \\ & & b_{\rm t}^{- 1} \d p_0 \Rightarrow \d \tp_0 , \quad b_{\rm t}^{- 1} \Delta^{(s )}_0 \Rightarrow i \tp_0 . \quad b_{\rm t}^{- 1} \Delta_0 \Rightarrow i \tp_0 ,
\end{eqnarray}

\noindent This diagram may diverge in the formal limit $b_{\rm s} \to 0$, with the divergence being proportional to $b_{\rm s}^{- n}$ multiplied by a specific structure of interest, and we have an expansion over a small parameter $\eta^{- 1}$.

In particular, if a continuum diagram (or a certain structure in it) is finite, then we obtain a lattice approximation for it, and for habitual external momenta or distances much larger than the typical length scale, this diagram is reproduced by its discrete counterpart with high accuracy. Of course, the continuum and discrete versions of this diagram should be considered in corresponding versions of the same soft synchronous gauge in the principal value prescription.

\subsubsection{New vertices}\label{new-vertex}

Also we should take into account the nonlinear terms $V_{\alpha \beta} \left( \{ \tw_{\gamma \delta} \} \right)$ in $u_{\alpha \beta} = \tw_{\alpha \beta} + V_{\alpha \beta}$. In the $n$th order over $u$ (\ref{sumC(n)u...ududu}), some or all of these $u$ are replaced now by monomials of $\tw$ in various combinations, not by simply $\tw$, as in the lowest order $n$ in $\tw$ above. As a result, the number of $\tw$ lines grows. But the scale of these new vertices remains the same as the scale $O ( b_{\rm t} b_{\rm s}^{1 - n} )$ of the $n$-graviton $\tw$ vertex for $u = \tw$ since the order of $\tw$ is $O ( 1 )$ in the sense of its correlator with any component of the metric. This leads to subseries of diagrams with the growing number of loops and no parametrical diminishing of their scales; in the resulting expansion over the number of loops, only a numerical decrease in their scales is expected, including due to loop smallness.

In particular, the set of 3-graviton vertices consists of (the finite-difference version of) the standard vertex (that is, for $u = \tw$), a combination of
\begin{equation}\label{wdwdw}                                              
\tw_{\gamma_1 \delta_1} \left( \Delta^\#_{\mu_5} \tw_{\beta_1 \beta_2} \right) \left( \Delta^\#_{\mu_6} \tw_{\beta_3 \beta_4} \right) ,
\end{equation}

\noindent and the vertex due to the bilinear term in $\tw$ in $u$, a combination of
\begin{equation}\label{dwwdw}                                              
\left[ \Delta^\#_{\mu_5} \left( \tw_{\gamma_1 \delta_1} \tw_{\gamma_2 \delta_2} \right) \right] \left( \Delta^\#_{\mu_6} \tw_{\beta_3 \beta_4} \right) .
\end{equation}

\noindent These vertices arise at the 1-loop level, for example in diagrams describing corrections to the Newtonian potential. The ratio of the contributions of these vertices depends on the specific momentum dependence of the corresponding diagrams and structures, but at least for quasi-momenta of order 1 (i.e. for extremely large momenta of order $b_{\rm s}^{- 1}$), the vertex combined from (\ref{dwwdw}) can dominate since it has the scale $O ( b_{\rm t} b_{\rm s}^{- 1} )$ that should be compared with the scale $O ( b_{\rm t} b_{\rm s}^{- 2} )$ of the standard vertex combined from (\ref{wdwdw}), as we discussed above, equation (\ref{sumC(n)u...ududu}).

\section{The Newtonian potential}\label{Newton}

\subsection{Mutual cancellation of some contributions}\label{cancel}

Illustrating the above with the example of the Newtonian potential or the scattering of two matter particles with the participation of gravitons, we can take as the matter field a (discrete) scalar field $\phi$ with an action in the form of a sum over sites, $(- 1 / 2) \sum_{\rm sites} [ g^{\lambda \mu} ( \Delta_\lambda \phi )$ $\! \! \! \!$ $( \Delta_\mu \phi ) + m^2 \phi^2 ] \sqrt{ - g }$. The 2-matter-1-graviton vertex arises from the term $\propto \phi \oDelta_\gamma u_{\alpha \beta} ( \delta_{\alpha \gamma} \delta_{\beta \delta} - \frac{1}{2} \delta_{\alpha \beta} \delta_{\gamma \delta} ) \Delta_\delta \phi $ ($g_{\alpha \beta} = b_{\rm s}^2 \delta_{\alpha \beta} + \sqrt{2} b_{\rm s} u_{\alpha \beta}$) by replacing $u_{\alpha \beta} = \tw_{\alpha \beta} + V_{\alpha \beta} ( \tw )$ with $\tw_{\alpha \beta}$.

\begin{figure}[h]
	\centering
	\includegraphics[scale=1]{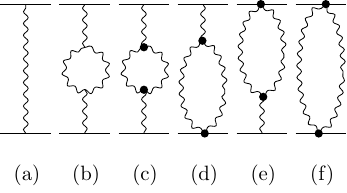}
	\caption{The main (a), some standard correction (b) and some potentially enhanced correction diagrams (c,d,e,f). The bolded points are new vertices.}
	\label{f2}
\end{figure}

The above discretized standard 3-graviton vertex (\ref{wdwdw}) arises from the $u ( \Delta^\# u ) ( \Delta^\# u )$-term in the action by substituting $\tw_{\alpha \beta}$ for $u_{\alpha \beta}$; the vertex (\ref{dwwdw}) arises from the bilinear term by substituting $\tw_{\alpha \beta}$ for one $u_{\alpha \beta}$ and the bilinear part of $V_{\alpha \beta} ( \tw ) \sim \tw \tw + \dots$ for another $u_{\alpha \beta}$. As mentioned, the last vertex has scale $O ( b_{\rm t} b_{\rm s}^{- 1} )$ compared to $O ( b_{\rm t} b_{\rm s}^{- 2} )$ of the first, that is, it has an additional large factor $\sqrt{ \eta }$. Consequently, the diagram in Fig.~\ref{f2}(c) with two new 3-graviton vertices of the $( \Delta^\# \tw ) [ \Delta^\# ( \tw \tw ) ]$ type, denoted by thick dots contains the large parameter $\eta$ compared to the analogous diagram in Fig.~\ref{f2}(b) with standard vertices of the $\tw ( \Delta^\# \tw ) ( \Delta^\# \tw )$ type (denoted by ordinary dots). That is, if the relative contribution of the standard diagram in Fig.~\ref{f2}(b) to the Newtonian potential of the skeleton diagram in Fig.~\ref{f2}(a) is $O ( l_{\rm Pl}^2 r^{- 2} )$, then the contribution of the new diagram in Fig.~\ref{f2}(c) should be $O ( b_{\rm s}^2 r^{- 2} )$,
\begin{equation}                                                           
U_{( a )} + U_{( b )} + U_{( c )} = - \frac{ G m_1 m_2 }{ r } \left[ 1 + O ( 1 ) \frac{ l_{\rm Pl}^2 }{ r^2 } + O ( 1 ) \frac{ b_{\rm s}^2 }{ r^2 } \right] .
\end{equation}

\noindent A priori, this is quite expected from a physical point of view, since the theory contains the scale $b_{\rm s}$. However, such a diagram is mutually cancelled with somewhat similar diagrams with new 2-matter-2-graviton vertices shown as bold dots in Fig.~\ref{f2}(d,e,f). These vertices are obtained from the above term $\propto \phi \oDelta_\gamma u_{\alpha \beta} ( \delta_{\alpha \gamma} \delta_{\beta \delta} - \frac{1}{2} \delta_{\alpha \beta} \delta_{\gamma \delta} ) \Delta_\delta \phi $ by replacing $u_{\alpha \beta} = \tw_{\alpha \beta} + V_{\alpha \beta} ( \tw )$ with $V_{\alpha \beta} = V_{\alpha \beta} ( \tw )$. To clarify the mechanism of the cancellation, let us slightly detail the form of the new 3-graviton term in the action. It is contained in
\begin{equation}\label{VMw}                                                
\sum_{\rm sites} ( b_{\rm s}^2 V_{\alpha \beta} \ccM^{\alpha \beta \gamma \delta} \tw_{\gamma \delta} + b_{\rm s}^2 \tw_{\alpha \beta} \ccM^{\alpha \beta \gamma \delta} V_{\gamma \delta} ) = \sum_{\rm sites} 2 b_{\rm s}^2 V_{\alpha \beta} \ccM^{\alpha \beta \gamma \delta} \tw_{\gamma \delta}
\end{equation}

\noindent (up to boundary terms; keeping in mind the Hermitianity of $\ccM$), where the bilinear part of $V_{\alpha \beta}$, $\propto \tw \tw$, should be taken. Here $\ccM$ is the above considered matrix of the bilinear form of the action. As for the gauge-fixing term $\ccF [ \{ w_{\lambda \mu} \} ( - g )^\ialpha ]$, we found earlier (subsection \ref{diagr-from-gauge-fix}) that the vertices it generates have a vanishing diagrammatic effect at $\varepsilon \to 0$. In particular, we can say the same about the vertices generated (as a result of the parametrization) by its bilinear part $\ccF [ \{ w_{\lambda \mu} \} ]$, namely, $2 b_{\rm s}^2 V_{\alpha \beta} \Delta \ccM^{\alpha \beta \gamma \delta} \tw_{\gamma \delta}$, and for convenience we can add this part back to (\ref{VMw}):
\begin{equation}\label{VG^(-1)w}                                           
2 b_{\rm s}^2 V_{\alpha \beta} \ccM^{\alpha \beta \gamma \delta} \tw_{\gamma \delta} \Rightarrow 2 b_{\rm s}^2 V_{\alpha \beta} ( \ccM + \Delta \ccM )^{\alpha \beta \gamma \delta} \tw_{\gamma \delta} = 2 b_{\rm s}^2 V_{\alpha \beta} ( \cG^{- 1} )^{\alpha \beta \gamma \delta} \tw_{\gamma \delta}
\end{equation}

\noindent since $\cG = ( \ccM + \Delta \ccM )^{- 1} $ by definition. When averaging this term in a product with other fields, when contracting $\tw_{\gamma \delta}$ with such a field $\tw_{\epsilon \zeta}$ at another site, we get the following expression:
\begin{equation}\label{G^(-1)ww}                                           
\langle ( \cG^{- 1} )^{\alpha \beta \gamma \delta} \tw_{\gamma \delta} ( x ) \quad \tw_{\epsilon \zeta} ( y ) \rangle = ( \cG^{- 1} )^{\alpha \beta \gamma \delta} \frac{ i }{ 2 b_{\rm s}^2 } \cG_{\gamma \delta \epsilon \zeta} ( x, y ) = \frac{ i }{ 2 b_{\rm s}^2 } \left( \delta^\alpha_\epsilon \delta^\beta_\zeta + \delta^\alpha_\zeta \delta^\beta_\epsilon \right) \delta_{x , y}.
\end{equation}

\begin{figure}[h]
	\centering
	\includegraphics[scale=1]{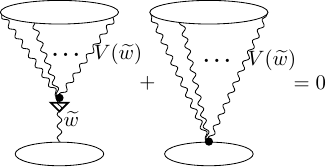}
	\caption{Cancellation of some (blocks of) vertices inside a diagram. The arrow at the vertex $V_{\alpha \beta} ( \cG^{- 1} )^{\alpha \beta \gamma \delta} \tw_{\gamma \delta}$ points out the $\tw_{\gamma \delta}$ line.}
	\label{f3}
\end{figure}

\noindent Here $\delta_{x , y}$ is the Kronecker delta, which requires the identity of the sites $x$, $y$ (two quadruples of integer coordinates), which are the arguments of the corresponding fields (also $\cG_{\gamma \delta \epsilon \zeta} ( x, y )$ is actually a function of $x - y$). That is, the field line $\tw$ on the diagram can be considered as compressed to a point. Suppose we have to average a product of field monomials, in one of which some instance of the field $u_{\epsilon \zeta}$ is singled out. Then, firstly, we can replace $u_{\epsilon \zeta}$ with $\tw_{\epsilon \zeta}$ and simultaneously take into account to the first order the interaction term in the action (\ref{VG^(-1)w}) and contract $\tw_{\gamma \delta}$ and $\tw_{\epsilon \zeta}$; secondly, we can replace $u_{\epsilon \zeta}$ with $V_{\epsilon \zeta}$ without taking into account the interaction. Then the corresponding contributions to the considered correlator cancel each other (Fig.~\ref{f3}):
\begin{eqnarray}\label{VG^(-1)ww+V=0}                                      
& &
\rlap{$\phantom{\langle \dots \quad 2 i b_{\rm s}^2 V_{\alpha \beta} ( \cG^{- 1} )^{\alpha \beta \gamma \delta} \,} \overbracket[0.75pt]{\phantom{\tw_{\gamma \delta} \quad \,}} $}
\langle \dots \quad 2 i b_{\rm s}^2 V_{\alpha \beta} ( \cG^{- 1} )^{\alpha \beta \gamma \delta} \tw_{\gamma \delta} \quad \tw_{\epsilon \zeta} \dots \rangle + \langle \dots \quad V_{\epsilon \zeta} \dots \rangle = \langle \dots \quad 2 i b_{\rm s}^2 \frac{ i }{ 2 b_{\rm s}^2 } V_{\epsilon \zeta} \dots \rangle \nonumber \\ & & + \langle \dots \quad V_{\epsilon \zeta} \dots \rangle = 0 .
\end{eqnarray}

\noindent The contraction of two fields ($\tw_{\gamma \delta}$ and $\tw_{\epsilon \zeta}$) according to Wick's theorem when averaging is denoted by a bracket above.

In order to contribute to the $\ln \bq^2$ dependence of interest to us (where $\bq$ is the momentum transfer), the Fourier transform of which is the $\propto 1 / r^3$ potential, each 3-graviton vertex $\propto V_{\alpha \beta} ( \cG^{- 1} )^{\alpha \beta \gamma \delta} \tw_{\gamma \delta}$ must have its field line $\tw_{\gamma \delta}$ as external to the loop containing this vertex. Then the rule (\ref{VG^(-1)ww+V=0}) (Fig.~\ref{f3}) leads to a pairwise cancellation of the diagrams in Fig.~\ref{f2}(c,d,e,f). Moreover, such a pairwise reduction of the diagrams also occurs in the case where the 3-graviton vertices in these diagrams are new $\propto V_{\alpha \beta} ( \cG^{- 1} )^{\alpha \beta \gamma \delta} \tw_{\gamma \delta}$, for which the $\tw_{\gamma \delta}$ line does not belong to the loop containing this vertex, only on one side, the upper or lower in Fig.~\ref{f2}(c,d,e), while on the other side such a vertex is standard or of the $ V_{\alpha \beta} ( \cG^{- 1} )^{\alpha \beta \gamma \delta} \tw_{\gamma \delta}$ type, but with the $\tw_{\gamma \delta}$ line belonging to the loop containing this vertex.

Analogously, we can perform a reduction of more complex vertices encountered at a greater number of loops. We denote the n-graviton vertex as $u_{\gamma_1 \delta_1} u_{\gamma_2 \delta_2} \dots u_{\gamma_n \delta_n}$, $n \geq 3$, where the finite differences for some two instances of the $u$ field are not shown, but the position of each $u$ is assumed to be meaningful. For any $u = \tw + V ( \tw )$ in this vertex $\overbrace{u \dots u}^n$ we can, firstly, take $\tw$, secondly, $V ( \tw )$, thirdly, take that $\tw$ and take into account the interaction vertex $\propto V_{\alpha \beta} ( \cG^{- 1} )^{\alpha \beta \gamma \delta} \tw_{\gamma \delta}$ in the first order and contract its field $\tw_{\gamma \delta}$ with that $\tw$. According to the rule (\ref{VG^(-1)ww+V=0}), the last two contributions cancel each other out, and we are left with $\tw$ instead of $u$. In the suggestion of the independency of such a reduction over each $u$, it is natural to expect that finally $u \dots u$ will be reduced to the standard vertex $\tw \dots \tw$. Indeed, in the general case of $k$ instances of $u$ substituted by $V$, it is convenient to think of these $k$ instances of $V$ as occupying the $k$ first places, the corresponding part of the vertex being $V_{\gamma_1 \delta_1} \dots V_{\gamma_k \delta_k} \tw_{\gamma_{k + 1} \delta_{k + 1}} \dots \tw_{\gamma_n \delta_n}$. This configuration can be obtained if we take any term in $u_{\gamma_1 \delta_1} \dots u_{\gamma_n \delta_n}$ containing $k - j$ instances of $V$ among the first $k$ places ($C^k_j$ ways) and contract the rest $j$ instances of $\tw$ with the interaction term $V \cG^{- 1} \tw$ taken in the $j$th order in $\frac{1}{ j! } (i \cS_{\rm int} )^j$ from the perturbative expansion of the exponential in the functional integral. When averaging, the denominator $j!$ cancels out with the number of permutations of $j$ fields $\tw$ to be contracted with other $j$ fields $\tw$. Each such a contraction leads to $(- 1)$ times the considered expression with $\tw$ replaced by $V$ (\ref{VG^(-1)ww+V=0}), and the overall sign factor is $(- 1)^j$. Thus, the effective vertex with $k$ instances of $V$ (at certain $k$ places) acquires a factor that is obtained by summing $(- 1)^j C^k_j$ from $j = 0$ to $j = k$:
\begin{equation}                                                           
\sum^k_{j = 0} (- 1)^j C^k_j = (1 - 1)^k = 0.
\end{equation}

\noindent Here $1 \leq k \leq n$, but the case $k = 0$ (standard vertex) with the result 1 is also covered by this formula. Thus, the standard $n$-graviton vertex upon the parametrization effectively reduces to the original non-parameterized vertex,
\begin{equation}\label{u...utow...w}                                       
u_{\gamma_1 \delta_1} u_{\gamma_2 \delta_2} \dots u_{\gamma_n \delta_n} \Rightarrow \tw_{\gamma_1 \delta_1} \tw_{\gamma_2 \delta_2} \dots \tw_{\gamma_n \delta_n}.
\end{equation}

\begin{figure}[h]
	\centering
	\includegraphics[scale=1]{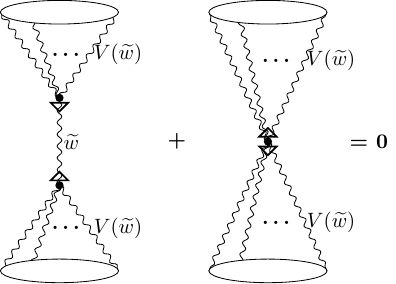}
	\caption{Cancellation of some (blocks of) vertices inside a diagram. The arrows at the vertex $V_{\alpha \beta} ( \cG^{- 1} )^{\alpha \beta \gamma \delta} V_{\gamma \delta}$ point out its components, the sets of lines $V ( \tw )$.}
	\label{f4}
\end{figure}

The case $n = 2$ has the peculiarity that a standard 2-graviton vertex to which $u_{\gamma_1 \delta_1} u_{\gamma_2 \delta_2}$ supposedly might be reduced must not exist since it would be contribution to the bilinear form of the action. In reality, the interaction part of the $uu$ term is effectively reduced to the $V_{\gamma_1 \delta_1} \tw_{\gamma_2 \delta_2}$ (or $\tw_{\gamma_1 \delta_1} V_{\gamma_2 \delta_2}$), that is, $V \cG^{- 1} \tw$ term. The effective $VV$ part vanishes. Indeed, a contribution to the effective $VV$ part follows either from taking interaction in a diagram in the first order $i \cS_{\rm int}$ and replacing both $u$ by $V$ in the $uu$ vertex there or from taking interaction in the second order $\frac{1}{ 2! } (i \cS_{\rm int} )^2$ and contracting the $\tw$ fields in the $V \tw$ terms there with each other. These contributions cancel each other out (Fig.~\ref{f4}):
\begin{equation}\label{VG^(-1)wwG^(-1)V+VG^(-1)V=0}                        
\rlap{$\phantom{\frac{1}{ 2! } \langle \dots \quad 2 i b_{\rm s}^2 V \cG^{ - 1 } \, \,} \overbracket[0.75pt]{\phantom{\tw \quad \,}} $}
\frac{1}{ 2! } \langle \dots \quad 2 i b_{\rm s}^2 V \cG^{ - 1 } \tw \quad \tw \cG^{ - 1 } V \cdot 2 i b_{\rm s}^2 \quad \dots \rangle + \langle \dots \quad \frac{ 1 }{ 2 } \cdot 2 i b_{\rm s}^2 V \cG^{ - 1 } V \quad \dots \rangle = 0 .
\end{equation}

\noindent This looks as a particular case of (\ref{VG^(-1)ww+V=0}), it was only necessary to clarify the coefficients. Thus, we are left effectively with the $V \cG^{- 1} \tw$ part of the $uu$ vertex.

\begin{figure}[h]
	\centering
	\includegraphics[scale=1]{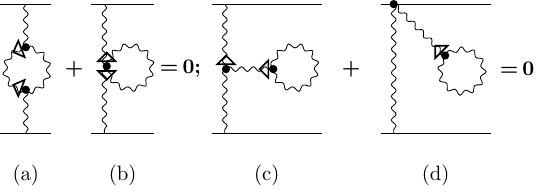}
	\caption{Cancellation of some tadpole diagrams.}
	\label{f5}
\end{figure}

The aforementioned vertex $\propto V_{\alpha \beta} ( \cG^{- 1} )^{\alpha \beta \gamma \delta} V_{\gamma \delta}$ includes at least a 4-graviton term, and at the one-loop level it leads to tadpole type diagrams. Then the considered rule (\ref{VG^(-1)wwG^(-1)V+VG^(-1)V=0}) leads to the cancellation of such diagrams, Fig.~\ref{f5}(a,b). The new 3-graviton vertex also can lead to tadpoles, and some of them are cancelled with taking into account the rule (\ref{VG^(-1)ww+V=0}), Fig.~\ref{f5}(c,d).

Here an attention should be paid to the issue of normal ordering. If interaction terms are given in the normal form, then when performing Gaussian averaging of a product of such terms and fields to calculate, for example, a Green function, the contractions by Wick's theorem should not be performed for the pairs of the field variables referred to the same vertex. If the gravity action is given in the normal ordered form with respect to the original metric field variables $g_{\lambda \mu}$ or $w_{\lambda \mu}$, then this will not be the normal ordered form with respect to new field variables ($\tw_{\lambda \mu}$) nonlinearly related to $w_{\lambda \mu}$. If we introduce normal ordering with respect to the new variables, the above mutual cancellation of some new diagrams will be violated as far as tadpoles are concerned. In this case, the tadpole of Fig.~\ref{f5}(b) is set to be zero while the tadpole of Fig.~\ref{f5}(a) remains the same, namely, equal to $( - 1 )$ times the tadpole of Fig.~\ref{f5}(b) without normal ordering. Thus, introducing normal ordering here leads to arising rather than vanishing a tadpole contribution and seems to be not appropriate to the problem at hand. That is, the reason is that in the theory with no less than quadratic in derivatives interaction some formally $n$-point loops, $n \geq 2$, can be non-distinguishable from tadpoles. Further, in subsection \ref{normal}, we discuss normal ordering with respect to the original variables $u$, which may prove useful.

\subsection{Contribution of the new vertices}\label{contrib-new}

From the above it follows that the only way to obtain a nonzero diagram involving new vertices is to contract the field $\tw$ from the vertex $V \cG^{- 1} \tw$ with any $\tw$ from $V( \tw )$ from another vertex $V \cG^{- 1} \tw$, from this last one - $\tw$ with any $\tw$ from $V( \tw )$ from the third vertex $V \cG^{- 1} \tw$, etc.

To be more accurate, a diagram with the interaction vertex $V \cG^{- 1} \tw$ is possible where the graviton $\tw$ line is external. Such a diagram gives a non-cancelled contribution to the corresponding Green function, but a zero contribution to the S-matrix, since amputating this $\tw$ line on-shell gives zero.

Thus, for a given diagram with a finite number of lines and vertices to give a nonzero S matrix contribution, the end field $\tw$ will sooner or later have to contract with $\tw$ from some $V( \tw )$ from the same chain, forming a closed chain. Without loss of generality, we can assume that in the chain $ \rlap{$\phantom{V \cG^{- 1} \,} \overbracket[0.75pt]{\phantom{\tw \,}} $}
\rlap{$\phantom{V \cG^{- 1} \tw V \cG^{- 1} \tw \dots V \cG^{- 1} \,} \overbracket[0.75pt]{\phantom{\tw \,}} $} V \cG^{- 1} \tw V \cG^{- 1} \tw \dots V \cG^{- 1} \tw V \cG^{- 1} \tw $ the terminal field $\tw$ is contracted with some $\tw$ from the terminal $V( \tw )$, and other unclosed chains (possibly with branches) can pair with the resulting closed chain through their terminal $\tw$ and $\tw$ from the factors $V( \tw )$ in the closed chain,
\begin{eqnarray}                                                          
\rlap{$\phantom{V \cG^{- 1} \,} \overbracket[0.75pt]{\phantom{\tw \,}} $}
\rlap{$\phantom{V \cG^{- 1} \tw V \dots \cG^{- 1} \,} \overbracket[0.75pt]{\phantom{\tw \quad }} $}
\rlap{$\phantom{V \cG^{- 1} \tw V \dots \cG^{- 1} \tw \quad \:} \overbracket[0.75pt]{\phantom{V \cG^{- 1} \tw V \cG^{- 1} \tw V \dots \cG^{- 1} }} $}
\rlap{$\phantom{V \cG^{- 1} \tw V \dots \cG^{- 1} \tw \quad V \cG^{- 1} \,} \overbracket[0.75pt]{\phantom{\tw \,}} $}
\rlap{$\phantom{V \cG^{- 1} \tw V \dots \cG^{- 1} \tw \quad V \cG^{- 1} \tw V \cG^{- 1} \,} \overbracket[0.75pt]{\phantom{\tw }} $}
\rlap{$\phantom{V \cG^{- 1} \tw V \dots \cG^{- 1} \tw \quad V \cG^{- 1} \tw V \cG^{- 1} \tw \:} \overbracket[0.75pt][0.60em]{\phantom{V \dots \cG^{- 1} \tw \quad }} $}
\rlap{$\phantom{V \cG^{- 1} \tw V \dots \cG^{- 1} \tw \quad V \cG^{- 1} \tw V \cG^{- 1} \tw V \dots \cG^{- 1} \tw \quad \tw \cG^{- 1} \,} \overbracket[0.75pt]{\phantom{\tw \,}} $}
V \cG^{- 1} \tw V \dots \cG^{- 1} \tw \quad V \cG^{- 1} \tw V \cG^{- 1} \tw V \dots \cG^{- 1} \tw \quad \tw \cG^{- 1} V \tw \cG^{- 1} \dots \nonumber \\
\rlap{$\phantom{\dots \, \;} \overbracket[0.75pt]{\phantom{V \tw \cG^{- 1} V \quad \,}} $}
\rlap{$\phantom{\dots \; \;} \overbracket[0.75pt]{\phantom{V }} $}
\rlap{$\phantom{\dots V \tw \cG^{- 1} V \quad \tw \cG^{- 1} \dots \; \;} \overbracket[0.75pt]{\phantom{V }} $}
\dots V \tw \cG^{- 1} V \quad \tw \cG^{- 1} \dots V \tw \cG^{- 1} V .
\end{eqnarray}

\begin{figure}[h]
	\centering
	\includegraphics[scale=1]{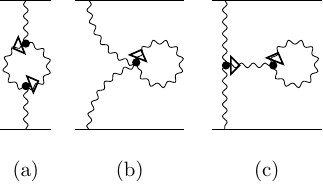}
	\caption{Diagrams with new vertices. In (b) the $[ \tw ]^3$-part of $V$ at the vertex $V \cG^{- 1} \tw$ acts.}
	\label{f6}
\end{figure}

\noindent Loop integration over such a closed chain gives a local contribution to the effective action. This integration is non-pole and real, that is, this contribution is imaginary. Taking into account the above unclosed chains (possibly with branches) results again in a local imaginary term in the action. Diagrams with such subdiagrams at the one-loop level are shown in Fig.~\ref{f6}(a,b,c).

Note that the contraction of $\tw$ with a function of $\tw$, in particular $V( \tw )$, means differentiation of this function, in particular
\begin{equation}                                                          
\rlap{$\phantom{V_A ( \cG^{- 1} )^{A B} \,} \overbracket[0.75pt]{\phantom{\tw_B \quad \,}} $}
\rlap{$\phantom{V_A ( \cG^{- 1} )^{A B} \tw_B \quad V_C ( \tw ) = V_A ( \cG^{- 1} )^{A B} \,} \overbracket[0.75pt]{\phantom{\tw_B \quad \,}} $}
V_A ( \cG^{- 1} )^{A B} \tw_B \quad V_C ( \tw ) = V_A ( \cG^{- 1} )^{A B} \tw_B \quad \tw_D \frac{ \partial V_C ( \tw ) }{ \partial \tw_D } = \frac{ i }{ b_{\rm s}^2 } V_A \frac{ \partial V_C ( \tw ) }{ \partial \tw_A } ,
\end{equation}

\noindent where $A$, $B$, $C$, \dots denote pairs of Lorentz indices like $\alpha \beta$; the Kronecker delta at the locations of the quantities being contracted is implied on the right-hand side. The basic imaginary contribution to the effective action is given by the sum of closed chains with different numbers $n$ of vertices,
\begin{eqnarray}                                                          
& & i \sum_{n = 1}^\infty \frac{ ( - 1 )^{ n + 1 } }{ n } \sum_{\rm sites} \frac{ \partial V_{ A_1 } }{ \partial \tw_{ A_n } } \frac{ \partial V_{ A_2 } }{ \partial \tw_{ A_1 } } \dots \frac{ \partial V_{ A_{ n - 1} } }{ \partial \tw_{ A_{ n - 2 } } } \frac{ \partial V_{ A_n } }{ \partial \tw_{ A_{ n - 1 } } } = i \Tr \ln \left\| \delta^B_A + \frac{ \partial V_A ( \tw ) }{ \partial \tw_B } \right\| \nonumber \\ & & = i \ln \frac{ D ( u ) }{ D ( \tw ) } .
\end{eqnarray}

\noindent This is ready to be rewritten in terms of $u = \tw + V ( \tw )$ as independent variables as
\begin{equation}                                                          
- i \ln \frac{ D ( f ( u ) ) }{ D ( u ) } ,
\end{equation}

\noindent where $f ( u )$ is the inverse function to $u = u ( \tw ) = \tw + V ( \tw )$. We can find that the contribution of $V ( \tw )$ from $u$ cancels out with the contribution of the vertex $V \cG^{- 1} \tw$ pairing its $\tw$ with $\tw$ from $u$ (that is, the contribution of the above unclosed chains of new vertices paired with closed chains cancels out), only $\tw$ from $u = \tw + V ( \tw )$ acts at the vertices given by this expression, just as we have found in subsection (\ref{cancel}) that only $\tw$ from $u$ acts at the vertices given by the gravity action $\cS_{\rm g}$ and the interaction terms with matter fields. We can rename the dummy integration variable $\tw$ back to $u$. Then we get the original non-parameterized gravity action, including interaction terms with matter fields, and the original measure
\begin{equation}                                                          
\frac{ D ( f ( u ) ) }{ D ( u ) } D u = \frac{ D ( \tw ( u ) ) }{ D ( u ) } D u .
\end{equation}

\noindent This demonstrates the formal consistency of the approach, allowing us to return to the original field variables at the diagrammatic level. At any finite order of the perturbative expansion, we have a contribution to the effective action that is different from that corresponding to the original measure, but still remains a local imaginary quantity.

Let us return for a moment to the previous subsection, \ref{cancel}: there, we considered subdiagrams in which $\tw$ from $V( \tw )$ is not contracted with anything—that is, $V$ behaves like an external field and can be absorbed by $\tw$ during functional integration over $\tw$: $\tw + V \Rightarrow \tw$. This explains quite simply why the contribution of the field $V$, arising from the diagrams considered there, vanishes, but it was important to handle this contribution diagrammatically, on an equal footing with the case of the dependence of $V$ on $\tw$ considered in the present subsection.

\subsection{The S matrix}\label{Smatrix}

Thus, new vertices (in fact, $V \cG^{- 1} \tw$) lead to the emergence of loops, the integration of which leads to local imaginary contributions to the effective action. Direct use of these vertices in the calculation of S-matrix elements would lead to non-unitarity of the S matrix. This is a somewhat trivial violation of the unitarity, associated with the transition between the bases of the fields $w$ and $\tw$ and the appearance/removal of a functional measure factor. The considered diagram contributions are non-pole ones; in the continuum theory they arise from large momenta and small distances (in the discrete theory, from large imaginary quasi-momenta) and depend on the definition of the products of field operators at close and coinciding points encountered in the definition of the S matrix. As discussed in the general theory of the S matrix \cite{BogShir} (we will take some points from that book below), the theory allows some freedom (or rather requires further definition) in defining such products when the arguments coincide, which can be used to fix the form of the S matrix based on the physical conditions of
\begin{enumerate}
\vspace{-5mm}
\item[a)] causality,
\vspace{-5mm}
\item[b)] unitarity,
\vspace{-5mm}
\item[c)] relativistic covariance,
\vspace{-5mm}
\item[d)] correspondence.
\vspace{-5mm}
\end{enumerate}
(In the discrete theory, one can consider analogues of these conditions; in particular, relativistic covariance can be satisfied to the leading order in finite differences.)

For effectively developing such a theory, the S matrix is considered depending on a function $g ( x )$ of “switching on” and “switching off” the interaction. It has values in the segment $[0, 1]$, and the interaction Lagrangian density $\cL_{\rm int} ( x )$ is replaced by $\cL_{\rm int} ( x ) g ( x )$. The S matrix is considered as an operator-valued functional $\hS ( g )$ of $g ( x )$ so that the ordinary S matrix is its target value of interest for $g ( x ) = 1$ everywhere, $\hS = \hS ( 1 )$. The functional expansion in powers of g(x) is considered,
\begin{equation}                                                          
\hS ( g ) = 1 + \sum_{n \geq 1} \frac{1}{ n! } \int \hS_n ( x_1 , \dots , x_n ) g ( x_1 ) \dots g ( x_n ) \d^4 x_1 \dots \d^4 x_n .
\end{equation}

Based on considerations of correspondence, it was established that
\begin{equation}                                                          
\hS_1 ( x ) = i \cL_{\rm int} ( x ) .
\end{equation}

\noindent The conditions of causality and unitarity made it possible to recurrently express the coefficient functions of subsequent terms of the expansion $\hS_n ( x_1 , \dots , x_n )$, $n > 1$, through the preceding $\hS_j$, $j = 1, 2, \dots, n - 1$, up to anti-Hermitian operators $i \Lambda_n ( x_1 , \dots , x_n )$ (that is, $\Lambda_n$ are Hermitian), which are equal to zero everywhere except at the point $x_1 = x_2 = \dots = x_n$ (so-called quasi-local operators; this is imposed by the causality conditions). As a result, $\hS_n ( x_1 , \dots , x_n )$, $n > 1$, depends on $\Lambda_j$, $j = 2, 3, \dots, n$. For example,
\begin{equation}\label{S2(x1x2)}                                          
\hS_2 ( x_1 , x_2 ) = i^2 T \{ \cL_{\rm int} ( x_1 ) \cL_{\rm int} ( x_2 ) \} + i \Lambda_2 ( x_1 , x_2 ) .
\end{equation}

\noindent This condenses into the expression
\begin{equation}\label{hS(g)}                                             
\hS ( g ) = T \left\{ \exp \left( i \int \cL_{\rm int} ( x; g ) d^4 x \right) \right\} ,
\end{equation}

\noindent where
\begin{eqnarray}\label{L(x;g)}                                            
& & \cL_{\rm int} ( x; g ) = \cL_{\rm int} ( x ) g ( x ) + \sum_{j \geq 2} \frac{ 1 }{ j! } \int \Lambda_j ( x, x_1, \dots , x_{j - 1} ) g ( x ) g ( x ) \dots g ( x_{j - 1} ) \nonumber \\ & & \cdot d^4 x_1 \dots d^4 x_{j - 1} ,
\end{eqnarray}

\noindent so that the target $\cL_{\rm int} ( x; g )$ at $g = 1$ is
\begin{equation}\label{L(x;1)}                                            
\cL_{\rm int} ( x; 1 ) = \cL_{\rm int} ( x ) + \sum_{j \geq 2} \frac{ 1 }{ j! } \int \Lambda_j ( x, x_1, \dots , x_{j - 1} ) d^4 x_1 \dots d^4 x_{j - 1} .
\end{equation}

In QED and other renormalizable theories, $\Lambda_j$s serve to compensate for divergences at coincident points when performing the standard renormalization procedure. In the present non-renormalizable theory we use a restricted version of this procedure that does not renormalize any real coupling constants, but prevents the appearance of imaginary constants that are not suitable for the unitary S matrix.

First, we consider the contribution to imaginary constants due to the new vertices.

For simplicity, we denote by the same letters some continuum analogue of the parametri\-zation $u = \tw + V ( \tw )$ of the discrete metric variables $u$ by variables $\tw$. The total Lagrangian density $\cL$ is the sum of the free Lagrangian density $\cL_0$ and the interaction Lagrangian density $\cL_{\rm int}$, $\cL [ u ] = \cL_0 [ u ] + \cL_{\rm int} [ u ]$. The square brackets here indicate a functional dependency, that is, for example, $\cL [ u ( x ) ] = \cL ( x )$. After parametrization, the Lagrangian interaction density becomes $\cL_0 [ \tw + V ] - \cL_0 [ \tw ] +\cL_{\rm int} [ \tw + V ]$. This should be compared with the case $u = \tw$, that is, with $\cL_{\rm int} [ \tw ]$. The difference is $\cL [ \tw + V ] - \cL [ \tw ]$. The interaction can contain the gauge-fixing term (more precisely, that part of this term that is related to the interaction). This term is non-local, but its contribution to the diagram technique is $O ( \varepsilon )$ and disappears as the synchronous gauge softening parameter $\varepsilon \to 0$. The continuum analogue of the arguments in subsections \ref{cancel}\&\ref{contrib-new} leads to a peculiar situation where the new interaction terms $\cL [ \tw + V ] - \cL [ \tw ]$ lead to diagrams (and only to them) that either vanish upon amputating some external graviton line on the mass shell and hence do not contribute to the S matrix, or contain new vertices via subdiagrams that contain only new vertices and represent anti-Hermitian local imaginary contributions to the effective action. Such contributions violate the unitarity of the S matrix and manifest themselves already at the level of $\hS_2 ( x_1 , x_2 )$ (\ref{S2(x1x2)}) due to the imaginary graviton self-energy insertions, as in Fig.~\ref{f6}, which in the continuum case also diverge as $i \delta^4 ( 0 )$. The only way to remove exactly the local imaginary part of the effective action (without making any other changes) is to cancel in $\cL_{\rm int} ( x; 1 )$ (\ref{L(x;1)}) exactly the new terms $\cL [ \tw + V ] - \cL [ \tw ]$ in $\cL_{\rm int}$ by an appropriate choice of (Hermitian) $\Lambda_j$s such that
\begin{equation}\label{Lambda=L(w)-L(w+V)}                                
\sum_{j \geq 2} \frac{ 1 }{ j! } \int \Lambda_j ( x, x_1, \dots , x_{j - 1} ) d^4 x_1 \dots d^4 x_{j - 1} = \cL [ \tw ( x ) ] - \cL [ \tw ( x ) + V ( \tw ( x ) ) ] .
\end{equation}

\noindent This can be done in an infinite number of ways. In the present context, the specific choice of $\Lambda_j$s does not matter to us. For example, we can take for a given $n$
\begin{equation}                                                          
\frac{ 1 }{ n! } \Lambda_n ( x, x_1, \dots , x_{n - 1} ) = \{ \cL [ \tw ( x ) ] - \cL [ \tw ( x ) + V ( \tw ( x ) ) ] \} \delta^4 ( x - x_1 ) \dots \delta^4 ( x - x_{n - 1} )
\end{equation}

\noindent and $\Lambda_j = 0$ for $j \neq n$.

The result of such a construction of the S matrix can be immediately transferred directly to the discrete version: the amplitude is evaluated based on the original non-parameterized action, and diagrams with new vertices (like Fig.~\ref{f6} at the one-loop level) are absent.

Symmetry in RC should be restored by averaging over simplicial structures. But already now, by removing the new terms, we get rid of the trace of measure non-covariance left in the parametrization $g_{\lambda \mu} = g_{\lambda \mu} ( \{ \tg_{\nu \rho} \} )$, and since the discrete action is covariant in the leading order of finite differences, we have a covariance-like symmetry in this order if we consider $\tg_{\lambda \mu}$ as a true metric.

Secondly, we note that anti-Hermitian local imaginary contributions to the effective action can also arise in diagrams containing only standard vertices. In the continuum case, such contributions are infinite and proportional to $i \delta^4 ( 0 )$, and in Ref.~\cite{FraVil} it was proposed to choose a local measure in the functional integral in such a way as to compensate for such infinities. In the discrete case, such contributions are finite, but imaginary and violate the unitarity of the matrix S, so we factor out a part of the functional measure to compensate for them. In our case of the soft synchronous gauge, the effective ghost action or the Faddeev-Popov determinant is calculated in closed form if the synchronous gauge softening parameter $\varepsilon \to 0$, is reduced to a power of $(- g)$ ($( - g )^{1 / 2}$) and is already included in the measure. It remains to analyze the integration over quasi-momenta in graviton loops.

In a graviton loop, if we start with the continuum case and begin integrating over momentum from $p_0$, then non-standard $i$-ness occurs at the maximum divergence of the integral over $p_0$ (linear), which occurs when parts of all involved vertices contain $p_0^2$. As a result, $p_0^{2 n}$ in the numerator obtained from $n$ vertices cancels out with $p_0^{2 n}$ in the denominator obtained from $n$ propagators for large $p_0$. Together with the subsequent integration over $\d^3 \bp$ of the sum of these diagrams with different numbers of vertices $n$, this gives a certain local metric function multiplied by an infinite imaginary number $i \delta^4 ( 0 )$ as a contribution to the effective action. That is, this gives a factor in the functional measure which, acting in the spirit of Ref.~\cite{FraVil}, can be found through the determinant of the quadratic form of the time derivatives in the action as
\begin{equation}\label{|d2S/((d0g)(d0g))|^(-1/2)}                         
\det \left\| \frac{ \delta^2 S }{ \delta ( \partial_0 g_{ \lambda \mu } ) \delta ( \partial_0 g_{ \nu \rho } ) } \right\|^{- 1 / 2} .
\end{equation}

\noindent In the considered synchronous gauge, such a form in the Lagrangian density $\cL ( x )$ is proportional to
\begin{equation}                                                          
( g^{\alpha \gamma} g^{\beta \delta} - g^{\alpha \beta} g^{\gamma \delta} ) ( \partial_0 g_{\alpha \beta} ) ( \partial_0 g_{\gamma \delta} ) \sqrt{ \rgamma } ,
\end{equation}

\noindent while $g_{0 \lambda}$ are fixed. As a result, the corresponding factor in the measure is found by limiting the space-time indices in (\ref{|d2S/((d0g)(d0g))|^(-1/2)}) by the spatial indices and is proportional to
\begin{equation}                                                          
\rgamma^{ 1 / 2 } .
\end{equation}

\begin{figure}[h]
	\centering
	\includegraphics[scale=1]{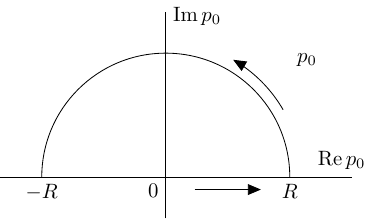}
	\caption{The loop momentum integration prescription (the continuum case): the integration interval $p_0 \in ( - R, R )$, $ R \to \infty $ is extended to a closed contour by adding a semicircle.}
	\label{f6-5}
\end{figure}

\noindent The appearance of this factor is equivalent to adding
\begin{equation}\label{id40ln}                                            
- \frac{ i }{ 2 } \delta^4 ( 0 ) \ln \rgamma
\end{equation}

\noindent to the effective Lagrangian density. To compensate for this term, we introduce the counterterm
\begin{equation}                                                          
\frac{ i }{ 2 } \delta^4 ( 0 ) \ln \rgamma
\end{equation}

\noindent into the Lagrangian density. Since (\ref{id40ln}) is a non-pole contribution, the same compensation effect can be achieved without adding a counterterm, but using Feynman diagrams to find the S matrix, in which the integration over $p_0 \in ( - \infty , \infty )$ in each loop is extended to the closed contour integration over the segment $p_0 \in [ - R , R )$ and the semicircle $p_0 = R \exp ( i \varphi )$, $\varphi \in [0 , \pi )$, $R \to \infty$, see Fig.~\ref{f6-5}. This only takes into account the contribution of the poles. This definition is consistent: the result does not depend on which loop momenta are chosen as independent when integrating over them and in what order the integrations are performed.

Furthermore, when choosing quasi-local operators $\Lambda_j$s to compensate for the new terms $\cL [ \tw + V ] - \cL [ \tw ]$ according to (\ref{Lambda=L(w)-L(w+V)}) in constructing the unitary S matrix, we see that such compensation is not needed since the contribution of the new vertices is non-pole and in any case disappears when using the above prescription in Fig.~\ref{f6-5}. Conversely, this prescription excludes only non-pole contributions, which are exhausted by the contributions of the diagrams considered above: firstly, graviton loops consisting of $p_0^2$-parts of standard vertices; secondly, graviton loops consisting of (continuous analogues of) the new vertices $V \cG^{- 1} \tw$.

Now, in the discrete case, the non-pole contributions to the effective action are not infinite, but imaginary. To reduce them, we single out the required factor $\rgamma^{ - 1 / 2 }$ in the original measure in the path integral (\ref{funct-int-full}), rewriting the power-of-volume type factor in the measure (\ref{(-g)^(e/2)}) as
\begin{equation}                                                          
( - g )^\frac{\eta - 8}{2} \propto \rgamma^\frac{\eta - 8}{2} = \rgamma^\frac{\eta - 7}{2} \rgamma^\frac{ - 1}{2} ,
\end{equation}

\noindent and convert this factor into the counterterm
\begin{equation}\label{counterterm}                                       
\frac{ i }{ 2 } \sum_{\rm sites} \ln \rgamma
\end{equation}

\noindent to the action. The remaining measure and subsequent formulas look as if $\eta$ were enhanced by 1. In particular,
\begin{equation}                                                          
b_{\rm s} = \sqrt{ \frac{ \eta - 8 }{ 2 \pi } } .
\end{equation}

\begin{figure}[h]
	\centering
	\includegraphics[scale=1]{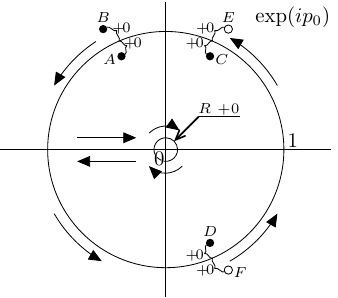}
	\caption{The prescription for the integration contour over $p_0$ in the discrete case in terms of the variable $\exp ( i p_0 )$ most clearly reflecting the periodicity at $p_0 \Rightarrow p_0 + 2 \pi$. To the standard counterclockwise circle of radius 1 centered at the origin, an infinitesimal clockwise concentric circle is added. The bold points denote typical physical poles $A, B$ or pairs of the nonphysical ones $C, D$ or $E, F$ (of an expression to be summed with the analogous expression with the conjugated such poles $E, F$ or $C, D$, respectively, according to the principal value prescription for softening the synchronous gauge).}
	\label{f7}
\end{figure}

Again, instead of the counterterm (\ref{counterterm}) one can use the prescription for any loop quasi-momentum integration, according to which the integration over $p_0 \in [ - \pi, \pi ]$ is extended to the integration over the closed loop $[ - \pi, \pi ] \cup [ \pi, \pi + i L ] \cup [ \pi + i L, - \pi + i L ] \cup [ - \pi + i L, - \pi ]$, $L \to \infty$, Fig.~\ref{f7}.

This prescription also results in the contribution of new vertices being zero; in general, it precisely prevents all unwanted non-pole terms from appearing in Feynman diagrams.

In particular, as far as the one-loop level is concerned, the diagrams in Fig.~\ref{f6} disappear. The same result would have been obtained if we had not introduced the above $\Lambda_j$s (in the continuum) and had not used any prescriptions, but had used the diagrams to estimate the real part of the amplitude. This means finding the physical amplitude itself in a sufficiently spacelike momenta region, where it just should be real. We could then analytically continue it to all (quasi-)momenta.

It is worth noting that in any habitual field theory with interactions less than quadratic in derivatives, the result of such a prescription is no different from the standard integration over $p_0 \in ( - \infty , \infty )$ or $p_0 \in [ - \pi , \pi ]$ in the continuum or discrete case, respectively, since the contribution of large momenta (or large imaginary quasi-momenta) is zero, as can be seen when any integration over the loop (quasi-)momenta starts from $p_0$.

Thus, we can estimate the amplitude using the prescription for integration over loop (quasi-)momenta (Figs.~\ref{f6-5},\ref{f7} for the continuum or discrete case, respectively) and confining ourselves to the standard vertices of the original non-parameterized action: some diagrams with new vertices cancel each other out, and the remaining diagrams with new vertices disappear when calculated according to this prescription. However, in the discrete case, such a prescription is made only for the S matrix. The non-pole contributions also make sense and allow one to calculate, as found in subsection \ref{contrib-new}, the functional measure. This is where a significant difference from the continuum theory emerges. There, such contributions are infinite like $i \delta^4 ( 0 )$ and the requirement for their reduction actually sets a functional measure (\cite{FraVil}). In the discrete case, the measure has an independent meaning, can be introduced from the outside and serve to fix the elementary length as a discretization parameter, which at the same time describes the optimal starting point of the perturbative expansion, which is one of the issues considered in the present paper.

\subsubsection{Normal ordering}\label{normal}

Returning to the topic of normal ordering, we noted at the end of subsection \ref{cancel} that such ordering with respect to the new field variables $\tw$ is undesirable, as it disrupts the cancellation of tadpoles. But normal ordering with respect to the original variables $u$ may prove useful. By Wick's theorem, the T-ordered product of $n$ instances of a free field $u$ is equal to the sum of the normal products of all possible chronological contractions of $u$, including the normal product with no contractions. Based on this, normal products can be recursively expressed in terms of T-ordered products and chronological contractions. Then the procedure of passing from the ordinary time-ordered product of free fields to the normal-ordered product consists in adding to that ordinary product a combination of such products with all possible partial and complete contractions, for example
\begin{eqnarray}                                                          
uuuu \Rightarrow : u u u u : & = &
\rlap{$\phantom{uuuu - \; \; } \overbracket[0.75pt]{\phantom{u }} $}
\rlap{$\phantom{uuuu - uuuu - \; \; } \overbracket[0.75pt]{\phantom{u u }} $}
\rlap{$\phantom{uuuu - uuuu - uuuu - \; \; } \overbracket[0.75pt]{\phantom{u u u }} $}
\rlap{$\phantom{uuuu - uuuu - uuuu - uuuu - u \, } \overbracket[0.75pt]{\phantom{u }} $}
\rlap{$\phantom{uuuu - uuuu - uuuu - uuuu - uuuu - u \, } \overbracket[0.75pt]{\phantom{u u }} $}
\rlap{$\phantom{uuuu - uuuu - uuuu - uuuu - uuuu - uuuu - u u \, } \overbracket[0.75pt]{\phantom{u }} $}
uuuu - uuuu - uuuu - uuuu - uuuu - uuuu - uuuu \nonumber \\ & &
\rlap{$\phantom{+ \; \; } \overbracket[0.75pt]{\phantom{u }} $}
\rlap{$\phantom{+ u u \; \; } \overbracket[0.75pt]{\phantom{u }} $}
\rlap{$\phantom{+ uuuu + \; \; \; \, } \overbracket[0.75pt]{\phantom{u u }} $}
\rlap{$\phantom{+ uuuu + u \; \; } \overbracket[0.75pt][0.60em]{\phantom{u u }} $}
\rlap{$\phantom{+ uuuu + uuuu + u \; \; } \overbracket[0.75pt]{\phantom{u }} $}
\rlap{$\phantom{+ uuuu + uuuu + \; \; \; \, } \overbracket[0.75pt][0.60em]{\phantom{u u u }} $}
+ uuuu + uuuu + uuuu
\end{eqnarray}

\noindent (symbolically, omitting indices). Here $u$ can mean either a functional integral field variable or a field operator. Correspondingly, $\rlap{$\phantom{\, } \overbracket[0.75pt]{\phantom{u }} $} u u$ is a functional integral average of two fields or their chronological contraction. The symbol for $T$-ordering of field products is omitted for brevity. This procedure keeps the degree of each monomial—in particular, the separation into the main (bilinear) part of the action and the interaction part. The bilinear part in $u$ transforms into the bilinear part in $\tw$ upon renaming $u$ to $\tw$. Therefore, $\rlap{$\phantom{\, } \overbracket[0.75pt]{\phantom{u }} $} u u$ for the free fields $u$ and $\rlap{$\phantom{\, } \overbracket[0.75pt]{\phantom{\tw }} $} \tw \tw$ for the free fields $\tw$ coincide. As for the fields $u$ in the expression for $: u \dots u :$ that do not participate in these contractions with one another, it follows from the discussion in subsection \ref{cancel} that, at the interaction vertices, these fields effectively reduce to $\tw$. In overall, reducing the parameterized vertex $u \dots u$ to the non-parameterized one $\tw \dots \tw$ (\ref{u...utow...w}) can also be made under the normal-ordering sign:
\begin{equation}\label{:u...u:to:w...w:}                                  
: u_{\gamma_1 \delta_1} u_{\gamma_2 \delta_2} \dots u_{\gamma_n \delta_n} : \Rightarrow : \tw_{\gamma_1 \delta_1} \tw_{\gamma_2 \delta_2} \dots \tw_{\gamma_n \delta_n} : .
\end{equation}

\noindent Again, among the new vertices in the diagrams containing them, the vertex $V \cG^{- 1} \tw$ (an ordinary one—i.e., without normal ordering) effectively acts. The consideration of the present subsection shows that the diagrams for the S-matrix once again feature precisely the non-parameterized vertices—this time with normal ordering, (\ref{:u...u:to:w...w:}).

An advantage of the discrete theory is that diagram contributions are finite even in the absence of normal ordering, whereas the introduction of normal ordering is precisely motivated by the possibility of making the Hamiltonian eigenvalues finite for physical states. However, if necessary, we can use the normal-ordered form of the original action—that is, disregard contractions of pairs of field variables belonging to the same vertex.

It should be noted that such contractions do not contribute to the $\ln \bq^2$- or $r^{-3}$-type dependence of the potential correction at the one-loop level discussed in the literature.

\section{Conclusion}

Thus, if the starting point of the perturbative series is far enough from the maximum point of the measure, the series becomes a large parameter series and should explode. This can probably be interpreted as uncontrolled perturbations, whereas the starting point located in some neighborhood or coinciding with the maximum point of the measure is most favorable for the perturbative series. This looks as a dynamic mechanism for establishing this starting point, in particular an elementary length scale that is finite and, more importantly, nonzero, so as not to obtain large-parameter series like (\ref{(1+eta x)^(1/eta)}), thereby ensuring discreteness (arbitrarily small lengths would mean a return to the continuum).

The S matrix is defined by the discrete analogues of standard continuum Feynman diagrams, the prescription for computing which is to restrict oneself to the pole contribution at each loop integration. This prescription may consist in using a somewhat modified integration contour for each loop over $p_0$ (Fig.~\ref{f7}), a discrete version of Fig.~\ref{f6-5}. If necessary, we can use the normal-ordered form of the action—that is, disregard field lines that begin and end at the same vertex.

An alternative way to construct $\hS$ may be to extract some unitary part from the non-unitary S matrix operator $\hS_0$ based on the “as is” diagrams; for example, $\exp [ ( \ln \hS_0 - \ln \overline{\hS}_0 ) / 2 ]$ or $( \hS_0 \overline{\hS}_0 )^{- 1 / 2} \hS_0 = \hS_0  ( \overline{\hS}_0 \hS_0 )^{- 1 / 2}$. Obviously, this can be done by an infinite number of ways. This can be made unambiguous by requiring that it does not contain the infinite value $i \delta^4 ( 0 )$ in the continuum that comes with non-pole contributions. At the lowest order of interaction, all these definitions are equivalent and do not contain $i \delta^4 ( 0 )$, but at higher orders the only option is to discard the non-pole contributions. In the discrete case we have a finite, but still abnormally large value instead of $i \delta^4 ( 0 )$. Thus, we arrive at the above prescription for $\hS$ (Figs.~\ref{f6-5},\ref{f7}), singling out the pole parts from the loop integrations.

Thus we have a subtle mechanism for fixing the elementary length, such that the details of the metric parametrization of $V ( \tw )$ do not enter into the amplitudes, but the specific definition of $V ( \tw )$ was required to show the overall consistency of the approach.

The diagrams for the physical amplitude form a series over the small parameter $\eta^{- 1}$. The discrete counterparts of the finite continuum diagrams for usual non-Planckian external momenta reproduce them with high accuracy. In particular, one-loop corrections to the Newtonian potential in the continuum formulation are finite and have been found in the literature; in the present approach, they form the leading order in such a series.

Regarding the convergence of diagrams for physical amplitudes and the expansion parameter, an analogy with standard continuum QED is observed: the renormalizability and small expansion parameter $\alpha$ in QED correspond to the discreteness and small expansion parameter $\eta^{-1}$ in the case under consideration.

The standard (non-parameterized) action is invariant under the diffeomorphism group in the leading order in finite differences. A given amplitude on its basis will be invariant in the leading order if the corresponding diagrams are dominated by small quasi-momenta, that is, if their continuum counterparts are UV finite. Otherwise, if these continuum analogues are UV divergent, the discrete diagrams are dominated by extreme quasi-momenta and invariance is violated. In this case, it is important to stick to the RC strategy, which implies summing/averaging over all possible simplicial structures, and this should restore symmetry. In particular, this should lead to gauge independence. The simplest such averaging can be made over the orientation of the hypercubic axes or (for higher loop corrections to the Newtonian potential, if the corresponding continuum diagrams are UV divergent) over orientations of interacting bodies relative to the hypercubic axes.

For the actual calculation of discrete diagrams (including both pole and non-pole contributions), all aspects considered implied the discrete version of the soft synchronous gauge together with the refined finite-difference form of the action. In this framework, the analytical structure of discrete diagrams corresponds to the structure of continuous diagrams. The gravitational propagator for the refined finite-difference form of the action $\cS_{\rm g}$ (\ref{cSg}) is quite bulky, but in most cases we can use its effective form $G^{\rm eff}$ (\ref{Geff}) for small quasi-momenta or the one restricted to the spatial-spatial metric components $\cG_{\alpha \beta \gamma \delta}$ (\ref{Gabgd}), (\ref{Gabgd-explicit}) neglecting terms of normal order $O ( \varepsilon^2 )$. More exactly, the actual principal value type propagator is the half-sum (\ref{Gnn+Gn*n*}) of two such expressions with poles with respect to $p_0$ that in the neighborhood of zero are approximately $- i \varepsilon$ and $i \varepsilon$, respectively.

\section*{Acknowledgments}

The present work was supported by the Ministry of Education and Science of the Russian Federation.


\begin{thebibliography}{99}
\bibitem{Regge}
 T. Regge, General relativity theory without coordinates, {\it Nuovo Cimento} {\bf 19}, 558 (1961).
\bibitem{Fein}
 G. Feinberg, R. Friedberg, T. D. Lee, and M. C. Ren, Lattice gravity near the continuum
 limit, {\it Nucl. Phys. B} {\bf 245}, 343 (1984).
\bibitem{CMS}
 J. Cheeger, W. M\"{u}ller, and R. Shrader, On the curvature of the piecewise flat spaces,
 {\it Commun. Math. Phys.} {\bf 92}, 405 (1984).
\bibitem{HamWil1}
 H. W. Hamber and R. M. Williams, Newtonian Potential in Quantum Regge Gravity, {\it Nucl.Phys. B} {\bf 435}, 361 (1995); arXiv:hep-th/9406163.
\bibitem{HamWil2}
 H. W. Hamber and R. M. Williams, On the Measure in Simplicial Gravity, {\it Phys. Rev. D} {\bf 59},  064014 (1999); arXiv:hep-th/9708019.
\bibitem{Ham1}
 H. W. Hamber, {\it Quantum Gravitation: The Feynman Path Integral Approach}, (Springer Berlin Heidelberg 2009). doi:10.1007/978-3-540-85293-3.
\bibitem{cdt}
 J. Ambjorn, A. Goerlich, J. Jurkiewicz, and R. Loll,  Nonperturbative Quantum Gravity, {\it Physics Reports} {\bf 519}, 127 (2012); arXiv:1203.3591[hep-th].
\bibitem{cdt1}
 R. Loll, Quantum gravity from causal dynamical triangulations: A review, {\it Class. Quantum Grav.} {\bf 37}, 013002 (2020); arXiv:1905.08669[hep-th].
\bibitem{Mik}
 A. Mikovi\'{c} and M. Vojinovi\'{c}, Quantum gravity for piecewise flat spacetimes, Proceedings of the MPHYS9 conference, 2018; arXiv:1804.02560[gr-qc].
\bibitem{hooft} G. 't Hooft and M. Veltman, One-loop divergencies in the theory of gravitation, {\it Ann. Inst. H. Poincare} {\bf A20}, 69 (1974).
\bibitem{don} J. F. Donoghue, Leading Quantum Correction to the Newtonian Potential, {\it Phys. Rev. Lett.} {\bf 72}, 2996 (1994); arXiv:gr-qc/9310024.
\bibitem{don1} J. F. Donoghue, General relativity as an effective field theory: The leading quantum corrections, {\it Phys. Rev.} {\bf D50}, 3874 (1994); arXiv:gr-qc/9405057.
\bibitem{don2} J. F. Donoghue, Introduction to the effective field theory description of gravity, in \textit{Advanced School on Effective Theories}, ed. by F. Cornet and M.J. Herrero (World Scientific, Singapore, 1996); arXiv:gr-qc/9512024.
\bibitem{don3} J. F. Donoghue, Perturbative dynamics of quantum general relativity, in \textit{Proceedings of the Eighth Marcel Grossmann Meeting on General Relativity}, ed. by Tsvi Piran and Remo Ruffini (World Scientific, Singapore, 1999); arXiv:gr-qc/9712070.
\bibitem{muz} I. J. Muzinich and S. Vokos, Long Range Forces in Quantum Gravity, {\it Phys. Rev.} {\bf D52}, 3472 (1995); arXiv:hep-th/9501083.
\bibitem{akh} A. Akhundov, S. Belucci, and A. Shiekh, Gravitational interaction to one loop in effective quantum gravity, {\it Phys. Lett.} {\bf B395}, 16 (1998); arXiv:gr-qc/9611018.
\bibitem{hamliu1} H. Hamber and S. Liu, On the Quantum Corrections to the Newtonian Potential, {\it Phys. Lett.} {\bf B357}, 51 (1995); arXiv:hep-th/9505182.
\bibitem{kk} G. G. Kirilin and I. B. Khriplovich, Quantum power correction to the Newton law, {\it Zh. Eksp. Teor. Fiz.} {\bf 109}, 1139 (2002); [{\it Sov. Phys. JETP} {\bf 95}, 981 (2002)]; arXiv:gr-qc/0207118.
\bibitem{BDH1}
 N. E. J. Bjerrum-Bohr, J. F. Donoghue and B. R. Holstein, Quantum Corrections to the Schwarzschild and Kerr Metrics, {\it Phys. Rev.} {\bf D 68}, 084005 (2003) [Erratum-ibid. {\bf D 71}, 069904 (2005)]; arXiv:hep-th/0211071.
\bibitem{BDH2}
 N. E. J. Bjerrum-Bohr, J. F. Donoghue and B. R. Holstein, Quantum Gravitational Corrections to the Nonrelativistic Scattering Potential of Two Masses, {\it Phys. Rev.} {\bf D 67}, 084033 (2003) [Erratum-ibid. {\bf D 71}, 069903 (2005)]; arXiv:hep-th/0211072.
\bibitem{kk1} G. G. Kirilin and I. B. Khriplovich, Quantum long-range interactions in general relativity, {\it Zh. Eksp. Teor. Fiz.} {\bf 125}, 1219 (2004); [{\it Sov. Phys. JETP} {\bf 98}, 1063 (2004)]; arXiv:gr-qc/0402018.
\bibitem{Frob} M. B. Fr\"{o}b, C. Rein and R. Verch, Graviton corrections to the Newtonian potential using invariant observables, {\it JHEP} {\bf 01}, 180 (2022); arXiv:2109.09753[hep-th].
\bibitem{Tiber} Tib'{e}rio de Paula Netto, Leonardo Modesto and Ilya L. Shapiro, Universal leading quantum correction to the Newton potential, {\it Eur. Phys. J. C} {\bf 82}, 160 (2022); arXiv:2110.14263[hep-th].
\bibitem{HamLiu}
 H. W. Hamber and S. Liu, Feynman Rules for Simplicial Gravity, {\it Nucl.Phys. B} {\bf 472}, 447 (1996); ({\it Preprint} arXiv:hep-th/9603016).
\bibitem{our1}
 V. M. Khatsymovsky, On the non-perturbative graviton propagator, {\it Int. J. Mod. Phys. A} {\bf 33}, 1850220 (2018); arXiv:1804.11212[gr-qc].
\bibitem{Fro}
 J. Fr\"{o}hlich, Regge calculus and discretized gravitational functional integrals, in {\it Nonperturbative Quantum Field Theory: Mathematical Aspects and Applications, Selected Papers} (World Scientific, Singapore, 1992), p. 523, IHES preprint 1981 (unpublished).
\bibitem{Kha}
 V. M. Khatsymovsky, Tetrad and self-dual formulations of Regge calculus, {\it Class. Quantum Grav.} {\bf 6}, L249 (1989).
\bibitem{Ash}
 A. Ashtekar, New Variables for Classical and Quantum Gravity, {\it Phys. Rev. Lett.} {\bf 57}, 2224 (1986).
\bibitem{RovSmo}
 C. Rovelli and L. Smolin, Discreteness of Area and Volume in Quantum Gravity, {\it Nucl. Phys.} {\bf B442}, 593 (1995); erratum, ibid. {\bf B456}, 753 (1995); arXiv:gr-qc/9411005.
\bibitem{ADM1}
 R. Arnowitt, S. Deser, and C. W. Misner, The Dynamics of General Relativity, in {\it Gravitation: an introduction to current research, Louis Witten ed.} (Wiley, 1962), chapter 7, p. 227; arXiv:gr-qc/0405109[gr-qc].
\bibitem{Kha3}
 V. M. Khatsymovsky, Defining integrals over connections in the discretized gravitational functional integral, {\it Mod. Phys. Lett. A} {\bf  25}, 1407 (2010); arXiv:1005.0060[gr-qc].
\bibitem{our}
 V. M. Khatsymovsky, On the discrete version of the Kerr-Newman solution, {\it Int. J. Mod. Phys. A} {\bf 38}, 2350025 (2023); arXiv:2212.13547[gr-qc].
\bibitem{RocWil}
 M. Rocek and R. M. Williams, The quantization of Regge calculus, {\it Z. Phys. C} {\bf 21}, 371 (1984).
\bibitem{khat}
 V. M. Khatsymovsky, On the gravitational diagram technique in the discrete setup, {\it Int. J. Mod. Phys. A} {\bf 38}, 2350143 (2023); arXiv:2306.11531[gr-qc].
\bibitem{Ste}
 F. Steiner, A new improved temporal gauge: the soft temporal gauge, {\it Phys. Lett.} {\bf B173}, 321 (1986).
\bibitem{Land} P. V. Landshoff, The propagator in axial gauge, {\it Phys. Lett.} {\bf 169B}, 69 (1986).
\bibitem{khat0}
 V. M. Khatsymovsky, Soft synchronous gauge in the perturbative gravity; {\it Int. J. Mod. Phys. A} {\bf 39}, 2450114 (2024); arXiv:2312.17119[hep-th].
\bibitem{khat0.5}
 V. M. Khatsymovsky, Soft synchronous gauge in gravity: Principal value prescription; {\it Int. J. Mod. Phys. A} {\bf 40}, 2550080 (2025); arXiv:2407.00713[hep-th].
\bibitem{khat1}
 V. M. Khatsymovsky, Towards the consistent perturbative expansion in discrete gravity; arXiv:2601.02181[gr-qc].
\bibitem{BogShir}
 N. N. Bogoliubov; D. V. Shirkov (1980). Introduction to the Theory of Quantized Field. John Wiley \& Sons Inc. ISBN 0-471-04223-4. (3rd edition)
\bibitem{HonMee}
 J. Honerkamp and K. Meetz, Chiral-invariant perturbation theory, {\it Phys. Rev.} {\bf D 3}, 1996 (1971).
\bibitem{GJLW}
 I. S. Gerstein, R. Jackiw, S. Weinberg and B. W. Lee, Chiral loops, {\it Phys. Rev.} {\bf D 3}, 2486 (1971).
\bibitem{don4}
 John F. Donoghue, The cosmological constant and the use of cutoffs, {\it Phys. Rev.} {\bf D 104}, 045005 (2021); arXiv:2009.00728[hep-th].
\bibitem{FraVil} E. S. Fradkin and G. A. Vilkovisky, S Matrix for Gravitational Field. II. Local Measure; General Relations; Elements of Renormalization Theory, {\it Phys. Rev.} {\bf D8}, 4241 (1973).
\bibitem{Holst}
 S. Holst, Barbero's Hamiltonian Derived from a Generalized Hilbert-Palatini Action, {\it Phys. Rev. D} {\bf 53}, 5966 (1996); arXiv:gr-qc/9511026.
\bibitem{Fat}
 L. Fatibene, M. Francaviglia, and C.Rovelli, Spacetime Lagrangian For\-mu\-la\-ti\-on of Bar\-be\-ro-Immirzi Gravity, {\it Class. Quantum Grav.} {\bf 24}, 4207 (2007); arXiv:0706.1899[gr-qc].
\bibitem{Loll}
 R. Loll, Discrete approaches to quantum gravity in four dimensions, {\it Living Rev.Rel.}{\bf 1}, 13 (1998); arXiv:gr-qc/9805049.
\bibitem{our2}
 V. M. Khatsymovsky, On the discrete Christoffel symbols, {\it Int. J. Mod. Phys. A} {\bf 34}, 1950186 (2019); arXiv:1906.11805[gr-qc].
\end{thebibliography}
\end{document}